\documentclass[manuscript,screen]{acmart}

\usepackage{listings}
\usepackage{xcolor}
\usepackage{color}
\usepackage{float}
\usepackage{algorithm2e}
\usepackage[export]{adjustbox}
\usepackage{subfig}
\usepackage{multirow}
\usepackage{tcolorbox}

\definecolor{commentcolor}{RGB}{0,128,0}
\definecolor{stringcolor}{RGB}{163,21,21}
\definecolor{keywordcolor}{RGB}{0,0,255}

\lstdefinestyle{pythonstyle}{
    language=Python,
    basicstyle=\footnotesize\ttfamily,
    keywordstyle=\color{keywordcolor}\bfseries,
    commentstyle=\color{commentcolor}\itshape,
    stringstyle=\color{stringcolor},
    showstringspaces=false,
    breaklines=false,
    frame=none,
    captionpos=b,
    showlines=true
}

\lstdefinestyle{jsonstyle}{
    language=JSON,
    basicstyle=\footnotesize\ttfamily,
    stringstyle=\color{stringcolor},
    showstringspaces=false,
    breaklines=false,
    frame=none,
    captionpos=b
}
\lstdefinelanguage{json}{
    basicstyle=\footnotesize\ttfamily,
    stepnumber=1,
    showstringspaces=false,
    breaklines=false,
    frame=none,
    captionpos=b,
    string=[s]{"}{"},
    comment=[l]{:\ "},
    morestring=[b]',
    literate=
     *{0}{{{\color{blue}0}}}{1}
      {1}{{{\color{blue}1}}}{1}
      {2}{{{\color{blue}2}}}{1}
      {3}{{{\color{blue}3}}}{1}
      {4}{{{\color{blue}4}}}{1}
      {5}{{{\color{blue}5}}}{1}
      {6}{{{\color{blue}6}}}{1}
      {7}{{{\color{blue}7}}}{1}
      {8}{{{\color{blue}8}}}{1}
      {9}{{{\color{blue}9}}}{1}
      {:}{{{\color{black}:}}}{1}
      {,}{{{\color{black},}}}{1}
      {\{}{{{\color{black}{\{}}}}{1}
      {\}}{{{\color{black}{\}}}}}{1}
      {[}{{{\color{black}[}}}{1}
      {]}{{{\color{black}]}}}{1},
}

\lstdefinelanguage{yaml}{
  keywords={true,false,null,y,n},
  keywordstyle=\color{keywordcolor}\bfseries,
  basicstyle=\ttfamily\footnotesize,
  comment=[l]{\#},
  commentstyle=\color{commentcolor}\itshape,
  stringstyle=\color{stringcolor},
  moredelim=[l][\color{black}]{-},
  morestring=[b]',
  morestring=[b]",
  literate=
   *{:}{{{\color{black}{:}}}}{1}
    {,}{{{\color{black}{,}}}}{1}
    {\{}{{{\color{black}{\{}}}}{1}
    {\}}{{{\color{black}{\}}}}}{1}
    {[}{{{\color{black}{[}}}}{1}
    {]}{{{\color{black}{]}}}}{1}
}

\lstdefinestyle{yamlstyle}{
    language=yaml,
    basicstyle=\ttfamily\footnotesize,
    keywordstyle=\color{keywordcolor}\bfseries,
    commentstyle=\color{commentcolor}\itshape,
    stringstyle=\color{stringcolor},
    showstringspaces=false,
    breaklines=false,
    frame=none,
    captionpos=b,
    xleftmargin=0.05\textwidth,
    xrightmargin=0.05\textwidth,
}

\AtBeginDocument{%
  \providecommand\BibTeX{{%
    \normalfont B\kern-0.5em{\scshape i\kern-0.25em b}\kern-0.8em\TeX}}}

\setcopyright{acmlicensed}
\acmJournal{TOSEM}
\acmYear{2024} \acmVolume{1} \acmNumber{1} \acmArticle{1} \acmMonth{1}\acmDOI{10.1145/3702980}

\begin{document}

\title[Detecting Code Inclusion In Language Models Trained on Code]{\underline{Tra}ined \underline{Wi}thout My \underline{C}onsent: Detecting Code Inclusion In Language Models Trained on Code}

\author{Vahid Majdinasab}

\email{vahid.majdinasab@polymtl.ca}
\orcid{0000-0003-4411-0810} 
\affiliation{%
  \institution{Polytechnique Montreal}
  \streetaddress{}
  \city{Montreal}
  \state{Quebec}
  \country{Canada}
  \postcode{}
}

\author{Amin Nikanjam}
\orcid{0000-0002-0440-6839}
\affiliation{%
  \institution{Polytechnique Montreal}
  \streetaddress{P.O. Box 1212}
  \city{Montreal}
  \state{Quebec}
  \country{Canada}
  \postcode{43017-6221}}
\email{amin.nikanjam@polymtl.ca}

\author{Foutse Khomh}
\orcid{0000-0002-5704-4173}
\affiliation{\institution{Polytechnique Montreal}
  \streetaddress{}
  \city{Montreal}
  \state{Quebec}
  \country{Canada}
  \postcode{}
}
\email{foutse.khomh@polymtl.ca}

\renewcommand{\shortauthors}{Majdinasab et al.}
\
\begin{abstract}
    Code auditing ensures that the developed code adheres to standards, regulations, and copyright protection by verifying that it does not contain code from protected sources. The recent advent of Large Language Models (LLMs) as coding assistants in the software development process poses new challenges for code auditing. The dataset for training these models is mainly collected from publicly available sources. This raises the issue of intellectual property infringement as developers' codes are already included in the dataset. Therefore, auditing code developed using LLMs is challenging, as it is difficult to reliably assert if an LLM used during development has been trained on specific copyrighted codes, given that we do not have access to the training datasets of these models. Given the non-disclosure of the training datasets, traditional approaches such as code clone detection are insufficient for asserting copyright infringement. To address this challenge, we propose a new approach, TraWiC; a model-agnostic and interpretable method based on membership inference for detecting code inclusion in an LLM's training dataset. We extract syntactic and semantic identifiers unique to each program to train a classifier for detecting code inclusion. In our experiments, we observe that TraWiC is capable of detecting 83.87\% of codes that were used to train an LLM. In comparison, the prevalent clone detection tool NiCad is only capable of detecting 47.64\%. In addition to its remarkable performance, TraWiC has low resource overhead in contrast to pair-wise clone detection that is conducted during the auditing process of tools like CodeWhisperer reference tracker, across thousands of code snippets.
\end{abstract}



\received{January 2024}
\received[revised]{2024}
\received[accepted]{2024}

\maketitle

\section{Introduction}\label{sec:introduction}
    Machine Learning (ML) models have been used in various industrial sectors ranging from healthcare to finance \cite{mckingseyreport2023}. These models allow for easier and faster data analysis and decision-making. A significant development in ML is the recent rise of Large Language Models (LLM) which are now being used for various Software Engineering (SE) tasks such as software development, maintenance, and deployment. LLMs are reported to increase developers' productivity and reduce software development time \cite{chen2021evaluating,anil2023palm}. These models are trained on code collected from publicly available sources such as GitHub which contains licensed code. However, the inaccessibility of these models' training datasets raises the important challenge of auditing the generated code. 
Code clone detection techniques are one of the more prominent approaches used to safeguard against using copyrighted code during software development \cite{ralhan2021study,saini2018code}. However, to do so, having access to the codes used for training the models (i.e., original codes) is required which is not possible when analyzing codes generated by an LLM as the training datasets of these models are not made publicly available.

Previous works have shown that LLMs can re-create instances from their training data \cite{carlini2022quantifying, carlini2021extracting, chang2023speak}. This is known as the \textit{memorization issue} in which ML models re-create their training dataset instead of generalizing \cite{zhang2021understanding}. By exploiting memorization, Membership Inference Attacks (MIA) \cite{shokri2017membership} have shown to be effective at both extracting information from ML models and inferring the presence of specific instances in a model's training dataset \cite{long2018understanding,salem2018ml,he2021node,melis2019exploiting,leino2020stolen}. Inspired by these approaches, we present TraWiC: A model-agnostic, interpretable approach that exploits the memorization ability of LLMs trained on code to detect whether codes from a project (collection of codes) were included in a model's training dataset. To assess whether a given project was included in the training dataset of a model $M$, TraWic parses the codes in the project to extract unique textual elements (variable names, comments, etc.) from each of them. Afterwards, these elements are masked and the model is queried to predict what the masked element is. Finally, the generated outputs of the model $M$ are compared with the original masked elements to determine if the code under analysis was included in the model's training dataset.

To evaluate our approach, we constructed two datasets of projects which were used to train three distinct LLMs. These datasets will act as the ground truth for inclusion detection and are comprised of over 10,700 code files from 314 projects. Our results indicate that TraWiC achieves an accuracy of 83.87\% for dataset inclusion detection. As code clone detectors are traditionally used for code auditing \cite{ralhan2021study, codewhisperer}, we compare TraWiC against two of the widely used open-source code clone detectors, NiCad \cite{roy2008nicad} and JPlag \cite{wise1993string}, to have an appropriate baseline for our evaluations.
Our analysis shows that NiCad is only capable of achieving an accuracy of 47.64\% for detecting dataset inclusion in an LLM's training dataset while JPlag achieves an accuracy of 55\%. Furthermore, unlike code clone detection approaches, TraWiC does not require pair-wise comparison between the codes to detect inclusion which makes it more computationally efficient compared to clone detection. 

Briefly, this paper makes the following contributions:
\begin{itemize}
    \item We propose a model-agnostic, interpretable, and efficient approach for dataset inclusion detection for LLMs.
    \item Our approach outperforms the widely used code clone detection tools NiCad and JPlag.
    \item We conduct thorough error and sensitivity analyses on TraWiC's performance. We demonstrate how targeting MIAs at non-syntactical parts of code (strings, documentation) can help detect dataset inclusion even in the presence of data obfuscation. Furthermore, we show how conducting MIAs on less prevalent textual elements within the code such as infrequent variable names, can be useful for dataset inclusion detection. 
    \item We demonstrate our TraWiC's effectiveness on multiple LLMs. Namely, SantaCoder, Llama-2, and Mistral.
    \item We release TraWiC's code alongside our data to be used by other researchers and studies \cite{reppackage}.
\end{itemize}

The rest of the paper is organized as follows: In Section \ref{sec:background}, we present background concepts and the related literature necessary for understanding the work presented in this paper. Section \ref{sec:trawic} introduces our methodology and details of TraWiC's implementation. We present our experiment design in Section \ref{sec:exp_design} and our results in Section \ref{sec:results}. We review the related works in section \ref{sec:related_works}, discuss the limitations of our approach in Section \ref{sec:limitations}, and discuss the threats to our work's validity in Section \ref{sec:threats_to_validity}. Finally, we conclude the paper in Section \ref{sec:conclusion}.

\section{Background}\label{sec:background}
    In this section, we will review the necessary background and concepts related to our approach. TraWiC is designed to work with LLMs trained on code, therefore, we will first review the literature on large language models. Afterward, we will go over code clone detection as such approaches can be used for dataset inclusion detection in training datasets of LLMs (albeit with high compute costs). Finally, we will review the current works on MIAs.

\subsection{Large Language Models}\label{subsec:background__llms}
    One of the highly active fields of research in ML is Natural Language Processing (NLP). Here, research is focused on designing approaches that are capable of understanding language and generating responses that are contextually relevant, accurate, and coherent. Neural Language Models (NLMs) were trained in order to predict the probability of the next word (token) in a sequence by considering the previous tokens \cite{qiu2020pre}. Models such as ELMo \cite{peters2018deep} and BERT \cite{devlin2018bert} were introduced by training a model on a large dataset of unlabeled corpora to learn word representations based on the context (other tokens in the sequence). This process is known as \textit{pre-training}; a model is trained on an unlabeled dataset to learn word representations, language structure, syntax, and semantics \cite{petroni2019language, feldman2019commonsense}. Pre-trained models can be further \textit{fine-tuned} for specific tasks by continuing their training on a labeled dataset designed for the task at hand. By incorporating the Transformer architecture and using the self-attention mechanism \cite{vaswani2023attention}, models such as GPT-2 \cite{radford2019language} were trained using the pre-training/fine-tuning paradigm alongside introducing new architectures. Models such as GPT-3 \cite{brown2020language} were trained by mainly using similar architectures but scaling the size of the model alongside its training data. As the number of these models' parameters is extremely large, they are called \textit{Large Language Models}. Code is also used in the training dataset of LLMs that are not fine-tuned for program synthesis as it has been shown to increase a model's capability in reasoning \cite{chen2021evaluating, austin2021program}. LLMs fine-tuned on large datasets of code have displayed high performance in coding tasks. 
    Currently, enterprises are offering LLMs fine-tuned for coding tasks such as GitHub's Copilot (based on Codex \cite{chen2021evaluating}), Amazon's CodeWhisperer, and Google's Codey. Alongside these enterprise services, many open-source models such as SantaCoder \cite{allal2023santacoder}, WizardCoder \cite{luo2023wizardcoder}, CodeLlama \cite{rozière2023code}, etc., are available. These LLMs were trained using different approaches, are based on various architectures, and are designed to achieve different objectives (program synthesis, code summarization, comment generation, etc.). However, regardless of their intended use, all these models require a large corpus of high-quality code (code that contains documentation explaining the functionality of its different parts) during their pre-training and fine-tuning phases. The primary resources for collecting such data are GitHub and StackOverflow, which contain vast amounts of open-source codes, and most of the open-source datasets online (e.g., TheStack \cite{Kocetkov2022TheStack}) are just an accumulation of data collected from these resources.  
    
\subsection{Code Clone Detection}\label{subsec:background__code_clone_detection}
    Code Clone Detection (CCD) is a widely researched area and is an important aspect of software quality assurance. A fault/bug/vulnerability that is present in one project can exist in other projects that contain similar codes and correction is required for all clones. Other software engineering activities such as plagiarism detection, program understanding, code compaction, etc. require identifying codes that are similar to each other either semantically or syntactically as well \cite{roy2009comparison}. CCD tools such as NiCad \cite{roy2008nicad}, SorcererCC \cite{sajnani2016sourcerercc}, Simian \cite{simian_ccd}, etc., are available as either open-source or enterprise tools to facilitate CCD activities. In CCD literature, code clones are categorized into four categories \cite{roy2007survey}: 
    
    \begin{itemize}
        \item \textit{Type-1:} An exact copy without modifications (except for whitespace and comments).
        \item \textit{Type-2:} A syntactically identical copy with differences in variable/type/function identifiers.
        \item \textit{Type-3:} Similar code fragments with extra addition or deletion of statements.
        \item \textit{Type-4:} Code fragments with different syntax, but similar semantics. 
    \end{itemize}
    
    By exploiting the high learning capacity of Deep Learning (DL), a large amount of research has been conducted on using DL-based approaches for CCD. To detect clones, these models use various code representations such as Abstract Syntax Trees (AST), Control Flow Graphs (CFG), code metrics (number of functions, number of variable calls, etc.), and raw code itself. DL-based approaches such as CoCoNut \cite{lutellier2020coconut}, AST-NN \cite{zhang2019novel}, FCCA \cite{hua2020fcca}, etc. have been able to achieve high performance on clone detection benchmarks. However, there exist some problems with using CCD models for dataset inclusion detection. Namely, the high cost of training/re-training/inference of large DL models, and the computational infeasibility of attempting clone detection in a pair-wise manner for all of the codes that exist in a dataset.

\subsection{Membership Inference Attacks}\label{subsec:background__mia}
    Memorization is a long-standing problem in ML wherein an ML model reproduces instances from its training data instead of generalizing \cite{zhang2021understanding}. 
    By exploiting the memorization problem, MIAs \cite{shokri2017membership} have shown to be able to extract information from ML models regarding their training data. Formally, given access to query on a model $M$, and a data record $D$, an MIA is considered successful if an attacker can successfully identify $D$ being in $M$'s training data or not. Shokri et al. \cite{shokri2017membership} studied how an attacker can identify the presence of a record in a model's training dataset by analyzing the model's output probabilities. Based on Shokri et al.'s work, many approaches have been proposed for both attacks on a model for extracting information about its training dataset and defenses in order to prevent the model from recreating sensitive records \cite{shokri2017membership, jia2019memguard, hu2021ear, hui2021practical}. Carlini et al. \cite{carlini2021extracting} attempted to extract the training data of GPT-2 by initializing the model with a sequence start token and having the model generate tokens until it generates an end-of-sequence token. They generate a large number of sequences in this fashion and remove samples that contain low likelihood. Their approach is based on the idea that samples with high likelihood are generated either because of trivial memorization or repeated sub-strings in their dataset. By doing so, they were able to extract information about individuals' contact information, news pieces, tweets, etc. In another work, Carlini et al. \cite{carlini2022quantifying} study the quantity of memorization in LLMs. They show that by increasing the number of parameters of a model, the amount of memorized data increases as well. Chang et al. \cite{chang2023speak} attempt to probe ChatGPT and GPT-4 by using ``name cloze'' membership inference queries. In their approach, they query ChatGPT/GPT-4 by giving it a context (which in their work are texts extracted from novels), mask the token which contains a \textit{name}, and inspect the output's similarity to the original token. The underlying idea is that without the model seeing the input during its training, the name should be impossible to predict. ``Counterfactual Memorization'' \cite{zhang2021counterfactual} is another approach proposed for studying the memorization issue in LLMs. Here, the changes in a model's predictions are characterized by omitting a particular document during its training. Even though this approach shows promise in studying a model's degree of memorization, it is very computationally expensive. 

\section{TRAWIC}\label{sec:trawic} 
    In this section, we first define the concepts and terminology used throughout the rest of this paper to describe TraWiC, discuss the motivating example behind TraWiC's design, and then show an example of how code is processed for dataset inclusion detection in our approach. We will then explain TraWiC's dataset inclusion detection pipeline in detail.

\subsection{Definitions}\label{subsec:definitions}
    As explained in Section \ref{subsec:background__mia}, memorization can be leveraged in MIAs to investigate the presence of data records in a model's training dataset. Previous works have used \textit{exact memorization} \cite{tirumala2022memorization} and \textit{name cloze} \cite{chang2023speak} approaches to analyze the outputs of a model for detecting dataset inclusion. In both works, parts of the input (i.e., token(s)) to the model are \textbf{masked}, and the model is tasked with predicting the masked tokens given the rest of the input as context. Afterwards, the model's outputs are compared to the original masked token. In these approaches, the model should not be able to predict the exact token (given that it does not exist anywhere else in the input and is unique enough) unless it has seen the input during its training. In our work, we follow a similar rationale with modifications for code. We consider code as consisting of two distinct parts: \textbf{syntax} and \textbf{documentation}; with syntax being the code itself (which follows the programming language's structural rules) and documentation being an explanation of the syntax for future reference. From here on, we use the word \textit{``script''} to denote a single file in a \textit{``project''} (which can consist of multiple scripts) that contains the code alongside its documentation. We formally define masking as follows:

    \begin{tcolorbox}[colback=gray!5,colframe=black]
            \textbf{Definition 1 (masking):} Given input $I$ which consists of tokens $[t_1, t_2, ..., t_n]$, and a chosen token $T$, masking $T$ entails replacing all the tokens in $I$ which match $T$ with another token named, $MASK$. The model $M$ is queried to predict what $MASK$ is, given $I$ as input.
    \end{tcolorbox}
    
    We also define exact and partial matching of masked tokens as follows:
    
    \begin{tcolorbox}[colback=gray!5,colframe=black]
            \textbf{Definition 2 (exact match):} Let $I$ be an input composed of a sequence of tokens $[t_1, t_2, ..., t_n]$ with a specific token $T$ masked. An \underline{exact match} is detected if the output of model $M$ for the masked token $T$ is identical to the actual token $T$:
            \begin{center}
                $M(I)=T$
            \end{center}
    \end{tcolorbox}
    
    \begin{tcolorbox}[colback=gray!5,colframe=black]
            \textbf{Definition 3 (partial match):} Let $I$ be an input composed of a sequence of tokens $[t_1, t_2, ..., t_n]$ with a specific token $T$ masked. Given a similarity function $S$, a \underline{partial match} is detected if the similarity score $S(M(I), T)$, between the output of the model $M$ and the masked token $T$ surpasses a specified threshold, $L$.
    \end{tcolorbox}
    
    In alignment with established coding standards, such as those described in \cite{martin2009clean}, variable/function/class names must be selected in a manner that reflects their purpose and context within the codebase. These identifiers, beyond the common names that are universally used (e.g., using ``set'' for accessors or ``get'' for mutators), are unique to the script and its corresponding project, and reflect the developers' individual coding style. Additionally, documentation, which is an important part of the program, is closer to natural language as it is not constrained to the programming language's syntax and as such, allows for a greater expression of the developers' individual styles \cite{ying2005source, fluri2007code}. Therefore, in order to detect whether a project  was used in a model's training dataset, we break each script in the project into 2 different identifier groups and compare the model's generations with the inputs as follows:
    
    \begin{itemize}
        \item \textbf{Syntactic identifiers} are names or symbols used to represent various elements in programming languages. These identifiers are used to name variables, functions, classes, and other entities within the code. Excluding code reuse (i.e., using code written for one script in another) within similar projects, these identifiers are generally unique to their codebases. In fact, Feitelson et al. \cite{feitelson2020developers} have shown that there is only a 6.9\% probability that two developers choose the same name for the same variable. Therefore, given the uniqueness of such identifiers, we look for exact matches between the masked tokens in the original script and the model's outputs. Specifically, we extract the following identifiers from each script:
            \begin{enumerate}
                \item Variable names
                \item Function names
                \item Class names
            \end{enumerate}
            
        \item \textbf{Semantic identifiers} are expressions that are either used to explain the logic of code into human-readable, natural language terms (i.e., documentation) or contain natural language elements (e.g., strings). Documentation serves to clarify code for future reference or other developers, while strings are a specific data type used within the code. As Aghajani et al.~\cite{aghajani2019software} show, developers have different standards for ``what'' to document and ``how'' to convey the underlying information. They report that different developers apply their own terminology and style based on experience, individual style, and the codebase's context. Therefore, these identifiers are not strictly uniform between different projects developed by different developers. As such, we look for partial matches between the original script's identifiers and the model's outputs, given the individual and contextual nature of how developers use both documentation and strings.

            \begin{enumerate}   
                \setcounter{enumi}{3}
                    \item Strings: data structures that are used to handle textual data. For example, in Listing \ref{listing:motivating_example_code} the phrase ``Hello World!'' is a string.
                
                    \item Statement-level documentation (e.g., comments): explanation of single or multiple statements' functionalities.
                
                    \item Method/Class-level documentation (e.g., Docstrings, Javadocs): explanations of a method's or class's functionality. Some programming languages like Python or Lisp follow specific conventions and syntax for separating method/class-level documentation from statement-level documentation \cite{pep8, lisp_style_guide}, while other languages such as C or Java only establish a style guide for how they should be written and do not provide a separate syntax for them. Regardless of the programming language, established best practices require developers to thoroughly document the inputs, operations, and outputs of a method/class \cite{aghajani2020software} and they are longer and more detailed than statement-level documentation \cite{haouari2011good}. Therefore, we categorize method/class-level documentation in a different group than statement-level documentation.
                    
            \end{enumerate}
        \end{itemize}
        
    From here on, we use the word \textit{``element''} to denote a single extracted syntactic/semantic identifier. Outside of these two identifier groups and their constituent elements, what remains in the code is either related to the programming language's syntax or operations that execute the logic of the code. 
    
    Finally, We define the prefix and suffix of an element as follows: 
    
    \begin{tcolorbox}[colback=gray!5,colframe=black]
            \textbf{Definition 4 (prefix and suffix):} Let $S$ be a script composed of a sequence of tokens $[t_1, t_2, ..., t_n]$ with a specific token $T$ masked. We consider all the tokens in $S$ that come before $T$ from $t_1$ as \underline{prefix} and all the tokens that immediately come after $T$ until $t_n$ as \underline{suffix}.
    \end{tcolorbox}
    
    To determine whether a script was included in the training data of a given model, we apply the \textit{Fill-In-the-Middle} (FIM) technique which is commonly used in both pre-training and fine-tuning of LLMs \cite{bavarian2022efficient}. This technique consists of breaking a script into three parts: prefix, masked element, and suffix. The model is then tasked to predict the masked element given the prefix and suffix as input. In the next section, we present an example of how an incoming script is processed for detecting dataset inclusion. 

\subsection{Motivating Example}
    In this section, we present how TraWiC differs from and improves the previous MIA approaches for code. Consider model $M$ to be a model trained on code with project $P$ that contains the script presented in Listing \ref{listing:motivating_example_code} ($C$) being included in $M$'s training dataset. Detecting both $P$ and $C$'s inclusion in $M$'s training dataset using the previous approaches (exact memorization \cite{tirumala2022memorization}, name cloze \cite{chang2023speak}) would require the following steps:
    
    \begin{itemize}
        \item If the task is next token prediction using exact memorization or other similar approaches, then all scripts in $P$ would be broken down into multiple, separate parts. For each generated part, the model would be tasked with predicting \underline{only the next token}.
        
        \item If the task is name cloze detection or other similar approaches, each script in $P$ would be broken down into multiple prompts, and for each prompt, the model will be tasked with predicting the \underline{masked token}.
    \end{itemize}
    
    Both approaches introduce multiple points of failure for inclusion detection and require a large number of costly inference calls on the LLM. We further describe the details in the rest of this section.
    
    \subsubsection{Next token prediction} As an example, let us consider the exact memorization approach. Here, we would be required to break down each script in $P$ into $n$ different parts with $n$ being the number of tokens in the script, and call the model $n$ times in order to get the models' predictions for each token. After collecting the outputs of $M$ for each input, one needs to check for exact matches for each of the predicted tokens. While this approach might be useful for models that are only trained on non-code textual inputs (e.g. blogs, books, forums, etc.), this would not be useful on code, especially on semantic elements given how developers follow different styles for writing them \cite{aghajani2019software}. Furthermore, given the number of parameters of the LLM under study, inference on $M$ can become extremely costly and time-consuming. Following the running example in Listing \ref{listing:motivating_example_code}, consider the docstring for the function \texttt{add\_variables}, if $M$ generates outputs that differ from the target tokens, the exact memorization approach would return a negative response. However, $M$ has predicted a token for a \underline{semantic identifier} which is closer to natural language, can be worded in similar ways without breaking the code, and is close to what was included in the original script. Furthermore, employing next token prediction results in losing all the context that comes after the token that is to be predicted by $M$. Therefore, for tokens that are at the beginning and near the middle of the code, the input to the model will not contain the entirety of the code, and therefore $M$ will be prone to generating tokens that are not similar to what is present in the training data and result in false negatives.
    
    \subsubsection{Name cloze prediction} Using name cloze prediction, similar to exact memorization, would entail breaking down the code into $n$ parts, constructing $n$ prompts, and calling the model on each prompt. While this approach can be useful for detecting rare names in the model and provides context about what comes after the target token prediction, it is based on an instruct prompt and therefore sensitive to how the prompt is designed. As shown by Srivastava et al.\cite{srivastava2022beyond}, LLMs are sensitive to changes in prompts. Such sensitivity may cause the model to output false negatives or positives. Furthermore, similar to the exact memorization approach, name cloze approaches check for exact matches between what the model predicts and the target token and therefore face the same challenge of resulting in false negatives when $M$’s outputs differ slightly in wording compared to what was originally observed during training.
    
    In comparison, TraWiC benefits from the best aspects of both approaches. Specifically, TraWiC is not dependent on the prompt similar to exact matching approaches while providing the context of the entire code to the model similar to name cloze approaches. Furthermore, TraWiC also introduces a new method for ensuring the minimization of false negatives as much as possible during the checking process and performs fewer calls on the LLM, making it more efficient. Following the running example presented in Listing \ref{listing:motivating_example_code}, TraWiC requires only \underline{13} calls on LLMs in contrast to \underline{82} calls of the previous approaches (one call for each token). Specifically, by looking for exact similarities for parts of the input that are required to be exactly the same (syntactic elements) and fuzzy matching for parts of the inputs that can differ (semantic elements) we check the model’s outputs for code inclusion on 6 different levels. Furthermore,  previous approaches rely solely on the number of matches without employing any comparison of the obtained results with previous observations. Therefore, by using a separate classifier trained on previous observations, we add another level of accuracy compared to exact matching models’ outputs by identifying patterns and variations that may not be captured through exact matching alone. As a result, TraWiC is much more efficient, accurate, and suited for detecting code inclusion in a model’s training dataset compared to the previously proposed approaches.

\subsection{End-to-End Data Processing Example}\label{subsec:motivating_example}
    In this section, we present an example of how TraWiC processes the scripts in a project for dataset inclusion detection. Listing \ref{listing:motivating_example_code}, shows an example of a Python script. This script consists of 2 functions, 3 variable declarations, and corresponding docstrings and comments inside each function. Listing \ref{listing:motivating_example_identifiers} displays all the syntactic and semantic identifiers extracted from \ref{listing:motivating_example_code}. Following the example in Listing \ref{listing:motivating_example_code}, Listings \ref{listing:motivating_example_prefix} and \ref{listing:motivating_example_suffix} show the constructed prefix and suffix for the variable \underline{\texttt{dummy\_variable}}, respectively. Note that we keep all the information from the original script and only mask the element itself. This process is repeated for every semantic and syntactic element extracted from the input script.
    \begin{figure}[h]
    \centering
    \captionsetup[listing]{position=bottom}
    \begin{minipage}[b]{0.48\textwidth}
        \centering
        \begin{lstlisting}[style=pythonstyle, caption={An example of a Python script and its extracted identifiers.}, label={listing:motivating_example_code}]
def print_input_string(input_string):
    """
    This function prints the input string.
    Args:
        input_string(str): the string to be printed
    """
    # store the input string in a variable
    dummy_variable  = input_string
    # print the input string
    print(input_string)

def add_variables(a, b):
    """
    This function adds two variables.
    Args:
        a (float): first variable
        b (float): second variable
    Returns:
        float: result of the addition
    """
    # add the two variables and 
    # store the result in a new variable
    c = a + b
    # return the result
    return c

print_input_string("hello World!")

result = add_variables(a=1, b=2)
        \end{lstlisting}
    \end{minipage}
    \hfill
    \begin{minipage}[b]{0.48\textwidth}
        \centering
        \begin{lstlisting}[style=jsonstyle, caption={JSON representation of extracted identifiers and documentation}, label={listing:motivating_example_identifiers}]
{
  "Variable names": [
    "dummy_variable",
    "a",
    "b",
    "c",
    "result"
  ],
  "Function names": [
    "print_input_string",
    "add_variables"
  ],
  "Class names": [],
  "Strings": [
    "'Hello World!'"
  ],
  "Statement-level documentation": [
    "store the input string in a variable",
    "print the input string",
    "add the two variables and store
    the result in a new variable",
    "return the result"
  ],
  "Method/Class-level documentation": [
    "This function prints the input string.\n\nArgs:\n    input_string (str): the string to be printed",
    "This function adds two variables.\n\nArgs:\n    a (float): first variable\n    b (float): second variable\n\nReturns:\n    float: result of the addition"
  ]
}
        \end{lstlisting}
    \end{minipage}
    \end{figure}

\begin{figure}[h]
    \centering
    \captionsetup[listing]{position=bottom}
    \begin{minipage}[b]{0.48\textwidth}
        
        \begin{lstlisting}[style=pythonstyle, caption={Prefix generated for variable \protect\underline{\texttt{dummy\_variable}}}, label={listing:motivating_example_prefix}]
def print_input_string(input_string):
    """
    This function prints the input string.
    Args:
        input_string(str): the string to be printed
    """
    # store the input string in a variable








            
        \end{lstlisting}

    \end{minipage}
    \hfill
    \begin{minipage}[b]{0.48\textwidth}
        \vspace{0pt}  
        \begin{lstlisting}[style=pythonstyle, caption={Suffix generated for variable \protect\underline{\texttt{dummy\_variable}}}, label={listing:motivating_example_suffix}]
    = input_string
    # print the input string
    print(input_string)

def add_variables(a, b):
    """
    This function adds two variables.
    Args:
        a (float): first variable
        b (float): second variable
    Returns:
        float: result of the addition
    """
    # add the two variables and store 
    # the result in a new variable
    c = a + b
    # return the result
    return c

print_input_string("hello World!")

result = add_variables(a=1, b=2)
        \end{lstlisting}
        \vspace{\fill}  
    \end{minipage}
\end{figure}

    By breaking each script into multiple (prefix, suffix) pairs, we aim to leverage the memorization capability of LLMs. More specifically, as reported by Carlini et al. \cite{carlini2022quantifying}, as the capacity (i.e., number of parameters) of a language model increases, so does its ability to memorize its training dataset instead of generalizing. Therefore, by using a large part of the script that was present in the model's training dataset as input, and only masking a small piece (i.e., a single element) to be used with the FIM technique as defined in Section \ref{subsec:definitions}, the model would likely be pushed in the direction of generating an output that is the same or highly similar to the original masked token if the script was included in its training dataset \cite{tirumala2022memorization}.
    
    Once we have collected all the model's outputs for the extracted elements of a script,  we look for the degree of similarity between the generated outputs and the original masked elements to determine dataset inclusion which we will explain in detail in Section \ref{subsec:methdology__comparison}. After obtaining the results of similarity comparison, a classification model predicts whether a project was included in a model's training dataset given the results of similarity comparison as input.

\subsection{Methodology}\label{subsec:trawic_methodology}
\begin{figure}[t]
        \centering
        \includegraphics[width=0.9\textwidth]{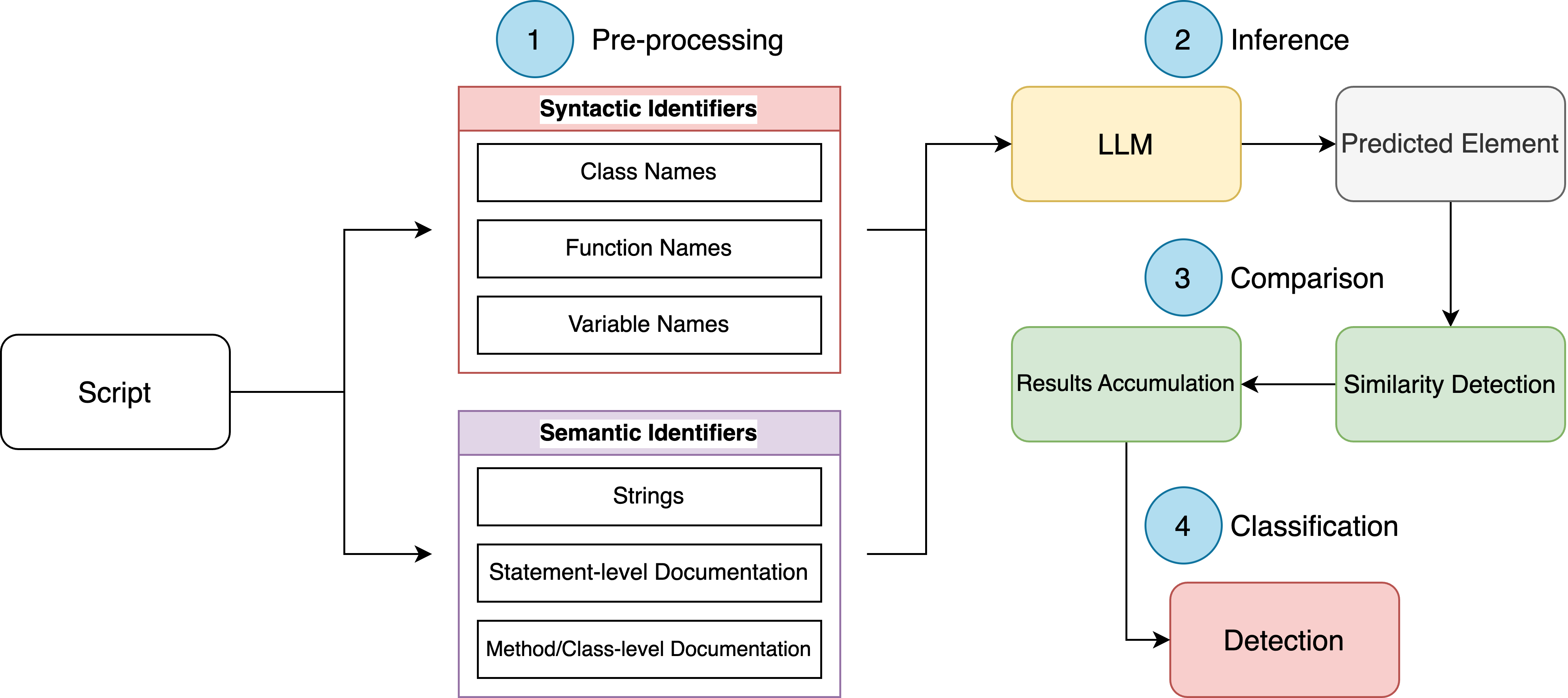}
        \caption{TraWiC's dataset inclusion detection pipeline}
        \label{fig:methodology_overview}
    \end{figure}

    In this section, we will explain how TraWiC can be used for the task of dataset inclusion detection in detail. TraWiC's pipeline consists of four stages: pre-processing, inference, comparison, and classification. Figure \ref{fig:methodology_overview} provides an overview of our approach. 

    \subsubsection{Pre-processing}\label{subsec:methodology__preprocessing}
        As stated in Section \ref{subsec:definitions}, we break each script into its six constituent elements (i.e., syntactic/semantic identifiers). From each script, we extract the elements and for each extracted element, we break down the script into two parts: prefix and suffix. By creating a prefix and suffix pair for each extracted element, we will be able to query the model to predict what comes in between the prefix and suffix. 
        Algorithm \ref{algo:algo_preprocessing} describes this pre-processing step in detail. 
        \RestyleAlgo{ruled}
        
        \begin{algorithm}[htb]
            \caption{Pre-processing}
            \small
            \SetKwFunction{ExtractIdentifiers}{ExtractIdentifiers}
            \SetKwFunction{GeneratePrefixAndSuffix}{GeneratePrefixAndSuffix}
            \SetKwComment{Comment}{/* }{ */}
            
            \KwIn{SCRIPT \Comment*[r]{SCRIPT is an entire script from a project. A project can contain many scripts.}}
            \KwOut{ProcessedScript \Comment*[r]{A list containing all the (prefix, suffix) tuples generated for each element from the input script.}}
            
            \SetAlgoLined
            \tcp{Extract all syntactic and semantic identifiers from SCRIPT}
            identifiers $\leftarrow$ \ExtractIdentifiers{SCRIPT}\;
            
            \tcp{Generate prefix and suffix tuple for each identifier}
            \ForEach{$i \in identifiers$}{
                $(p_i, s_i) \leftarrow \GeneratePrefixAndSuffix(i)$\;
                store $(p_i, s_i, i)$ in ProcessedScript\;
            }
            
            \KwRet{ProcessedScript}\;
        
            \label{algo:algo_preprocessing}
        \end{algorithm}

    \subsubsection{Inference}\label{subsec:methodology__inference}
        After breaking the script into (prefix, suffix, masked element) tuples as shown in Algorithm \ref{algo:algo_preprocessing}, the given model is queried to predict the masked element given prefix and suffix as input. From this tuple, the prefix and suffix are used for inference on the model while the masked element will be used in the next step for comparing the model's outputs to the original masked element. This process is repeated for all of the tuples extracted from a script. Algorithm \ref{algo:algo_inference} describes this process. Once we have collected all the model's predictions for each masked element, we move to the next step in which we compare the model's predictions with the original elements. 
        
        \begin{algorithm}[htb]
            \caption{Inference}
            \small
            \SetKwFunction{Predict}{Predict}
            \SetKwComment{Comment}{/* }{ */}
            
            \KwIn{ProcessedScript \Comment*[r]{Output of Algorithm \ref{algo:algo_preprocessing}.}}
            \KwOut{ModelPredictions \Comment*[r]{A list containing all of the given model's predictions for the masked element.}}
            
            \tcp{Predict the masked element given prefix and suffix for the corresponding element as input}
            \ForEach{$(p_i, s_i, i) \in  ProcessedScript$}{
                $o_i \leftarrow \Predict(p_i,s_i)$\;
                store $(o_i, i, type_i)$ in ModelPredictions\;
            }
            \KwRet{ModelPredictions}\;
            \label{algo:algo_inference}
        \end{algorithm}

    \subsubsection{Comparison}\label{subsec:methdology__comparison}
        Given the outputs of the model for each element, we define two different comparison criteria given an element's type:
            \begin{itemize}
                \item For syntactic identifiers (Variable/Function/Class names), we look for an \textbf{exact} match. If the model generates the exact same token(s), we count it as a \textit{“hit”}, which is a binary flag that indicates whether the generated output matches the desired output.
                
                \item For semantic identifiers (strings/documentation), to assess the similarity between two sequences (series of tokens), we use the Levenshtein distance (edit distance) score \cite{levenshtein_distance}. This metric measures the number of single-character edits that are required to change one sequence to the other. These elements are closer to natural language and do not require the syntactic restrictions required for code (incorrect/inaccurate generations will not result in syntax errors). Therefore, we compare the generated output with the original elements and calculate the edit distance between the two. Here, we define a threshold, and edit distances higher than this threshold will be considered as a \textit{hit}. The final threshold was chosen empirically to provide a balance between semantic similarity and performance gains. We present the results with different thresholds in Section \ref{sec:results}.
                
                The other metrics that can be used for this objective are based on semantic similarity where a semantic score is calculated between the vector representations of two strings (e.g., cosine similarity, TF-IDF distance, etc.) \cite{wang2020measurement}. We choose the edit distance metric as we do not aim to inspect whether the model's output is similar to the original masked element in meaning but in syntactic structure. Therefore, analyzing the semantic distance would not be helpful for this task.
            \end{itemize}
        
        After processing the model's outputs as described above, we normalize the hit numbers by dividing the number of hits against the total number of checks for each type of identifier. Algorithm \ref{algo:algo_comparison} shows the comparison process in detail. Table \ref{table:ds_sample} shows a sample of the dataset that is constructed for dataset inclusion detection as an output of the comparison process for the example presented in Section \ref{subsec:motivating_example}.
        
        \begin{algorithm}
            \caption{Comparison}
            \small
            \SetKwFunction{Predict}{Predict}
            \SetKwFunction{SemanticComparison}{SemanticComparison}
            \SetKwComment{Comment}{/* }{ */}
            \SetKwFunction{AccumulateResults}{AccumulateResults}
            \SetKwData{TypeMap}{TypeMap}
            \SetKwData{UpdateHits}{UpdateHits}
            
            \KwIn{ModelPredictions \Comment*[r]{Output of Algorithm \ref{algo:algo_inference}.}}
            \KwOut{(Hits)}
            
            \tcp{Function to update hits based on the result}
            \SetKwProg{Fn}{Function}{}{}
            \Fn{\UpdateHits{$Hits, key, result$}}{
                $Hits[key] \leftarrow Hits[key] + result$\;
            }
            
            \tcp{Initialize Hits as a dictionary for each identifier type}
            $Hits \leftarrow \{VariableHits: 0, FunctionHits: 0, ClassHits: 0, StringHits: 0, CommentHits: 0, DocStringHits: 0\}$\;
            \BlankLine
            \tcp{Compare the given model's outputs $(o_i)$ to the masked element $(i)$ depending on its $type$ (i.e., variable/function/class name or string/comment/doscstring)}
            \ForEach{$(o_i, i, type_i) \in  ModelPredictions$}{
                \If{$type_i$ is variable/function/class name}{
                    \eIf{$ o_i = i$ }
                        {$result \leftarrow 1$\;}
                        {$result \leftarrow 0$\;}
                }
                \If{$type_i$ is string/comment/docstring}{
                    \eIf{($SemanticComparison(o_i, i) \geq SemanticThreshold$} 
                        {$result \leftarrow 1$\;} 
                        {$result \leftarrow 0$\;}
                }
                \UpdateHits{$Hits, type_i, result$}\;
            }
            
            \KwRet{Hits}\;
            \label{algo:algo_comparison}
        \end{algorithm}
             
        \begin{table}[H]
            \centering
            \caption{Data representation sample for the example presented in Section \ref{subsec:motivating_example}}
            \begin{adjustbox}{max width=\textwidth}
                \begin{tabular}{lrrrrrr}
                \toprule
                Script Name & Class Hits & Function Hits & Variable Hits & String Hits & Comment Hits & Docstring Hits \\
                \midrule
                sample\_project/sample.py & 0 & 1.0 & 0.08 & 0.4 & 0.66 & 0.0 \\
                \bottomrule
                \end{tabular}
            \end{adjustbox}
            
            \label{table:ds_sample}
        \end{table}

    \subsubsection{Classification}
            With our extracted features, the classifier's task is to detect whether a script was included in the LLM's training dataset; which makes this a binary classification problem. Any classification method, such as decision trees or Deep Neural Networks (DNN), can be used based on the desired performance. We experiment with different classification methods as explained in Section \ref{subsubsec:results__classifier}
            
            \begin{algorithm}[H]
                \caption{Classification}
                \small
                \SetKwFunction{ClassificationModel}{ClassificationModel}
                \SetKwComment{Comment}{/* }{ */}
                
                \KwIn{Hits \Comment*[r]{Output of Algorithm \ref{algo:algo_comparison}.}}
                \KwOut{ClassificationResults \Comment*[r]{A binary variable with 1 indicating inclusion in the dataset and 0 otherwise.}}
                
                \tcp{Predict dataset inclusion}
                $ClassificationResults \leftarrow \ClassificationModel(hitType, hitCount)$\;
                
                \KwRet{ClassificationResults}\;
                \label{algo:algo_classification} 
            \end{algorithm}

\section{Experimental Design}\label{sec:exp_design}
    In this section, we explain our experimental design including the construction of the dataset used as the ground truth for validating our approach, the LLM under study, the classifier used for detecting dataset inclusion, and details of TraWiC's performance comparison with a state of the art CCD approach. Specifically, we aim to answer the following Research Questions (RQs):

\begin{itemize}
    \item \textbf{RQ1 [Effectiveness]}
    
        \begin{itemize}
            \item {\textbf{RQ1a}}: What is TraWiC's performance on the dataset inclusion detection task?
            \item {\textbf{RQ1b}: How does TraWiC compare against traditional CCD approaches for dataset inclusion detection?}
            \item {\textbf{RQ1c}: What is the effect of using different classification methods on TraWiC's performance?}
        \end{itemize}
    
    \item \textbf{RQ2 [Sensitivity Analysis]} 

        \begin{itemize}
            \item {\textbf{RQ2a}}: How robust is TraWiC against data obfuscation techniques?
            \item {\textbf{RQ2b}}: What is the importance of each feature in detecting dataset inclusion?
        \end{itemize}
    
    \item \textbf{RQ3 [Error Analysis]} How does the model under study make mistakes in predicting the masked element?

\end{itemize}

\subsection{Models Under Study}\label{subsubsec:results__model_under_study}

    For evaluating our approach, without loss of generality, we have selected three distinct LLMs. Namely, SantaCoder \cite{allal2023santacoder}, Mistral 7B \cite{jiang2023mistral}, and Llama-2 \cite{touvron2023llama}. We expand more on each model in the following.
    
    \subsubsection{SantaCoder}
        SantaCoder is an LLM trained by HuggingFace on TheStack for program synthesis and supports Python, Java, and JavaScript programming languages. This model was chosen as one of the LLMs under study for the following reasons:
            \begin{itemize}
                \item TheStack is publicly available. Therefore, we can inspect the data and confirm the presence of a script in SantaCoder's training dataset.
                
                \item SantaCoder's data cleaning procedure is clearly explained and replicable with different cleaning criteria being applied \cite{allal2023santacoder}. To clean the data for SantaCoder's training dataset, its developers have used:  1) ``GitHub stars'' which filters repositories based on the number of their stars, 2) ``Comment-to-code-ratio'' which considers the script's inclusion based on the ratio of comment characters to code characters, and 3) ``more near-deduplication'' which employs multiple different deduplication settings. Hence, given that we have access to the original dataset, we can reproduce the dataset that the model was trained on by following the same procedures.
                \item SantaCoder is trained for code completion, using the FIM objective. As such, it natively supports the FIM task that is used for TraWiC.
            \end{itemize}
        
        HuggingFace has released multiple versions of SantaCoder. Each model is trained on data that were constructed using different data-cleaning criteria. For our experiments, we chose the \underline{\texttt{comment\_to\_code}} model\footnote{https://huggingface.co/bigcode/santacoder} because the comment-to-code ratio for this model is calculated by ``using the \texttt{ast} and \texttt{tokenize} modules to extract docstrings and comments from Python files'' \cite{allal2023santacoder}; which allows us to accurately reproduce the data. Following the data cleaning steps outlined in \cite{allal2023santacoder}, the training dataset for this version was constructed by filtering any script in TheStack that has a comment-to-code ratio between 1\% and 80\% on a character level after the standard data cleaning steps (removing files from opt-out requests, near-deduplication, PII-reduction, etc).
    
    \subsubsection{Llama-2}
        Llama-2 is a family of LLMs developed by Meta \cite{touvron2023llama}. Llama-2 was pre-trained on 2 trillion tokens of data, has a context length of 4096 tokens, and uses Grouped-Query Attention (GQA) \cite{ainslie2023gqa}. This family of LLMs comes in different sizes (7B, 13B, 34B, and 70B), and both the pre-trained and instruct models have been released on HuggingFace\footnote{https://huggingface.co/collections/meta-llama/llama-2-family-661da1f90a9d678b6f55773b}. The pre-trained base versions provide a foundation for further fine-tuning according to specific downstream tasks while the fine-tuned versions allow for enhanced performance in conversational applications. For our experiments, we use the pre-trained version with 7 billion parameters and fine-tune it on a dataset that we have constructed.
        
    \subsubsection{Mistral 7B}
        Mistral 7B \cite{jiang2023mistral} is a 7-billion-parameter LLM developed by the Mistral research group, trained for language modeling tasks such as chat, code completion, and complex reasoning. This model uses GQA \cite{ainslie2023gqa} similar to Llama-2 and additionally uses Sliding Window Attention (SWA) \cite{beltagy2020longformer}, which allows for faster inference speed and reduced computational costs. Alongside its fast inference, Mistral also outperforms Llama-2 13B on all evaluation benchmarks \cite{jiang2023mistral}. Similar to Llama-2, Mistral 7B has been released in two versions:  A pre-trained base, and fine-tuned version. For our experiments, we chose the pre-trained version\footnote{https://huggingface.co/mistralai/Mistral-7B-v0.1} for its flexibility and potential for fine-tuning.

    Both Mistral 7B and Llama-2 were selected for this study due to their widespread use as foundational models for fine-tuning \cite{minaee2024large}. Unlike models already optimized for specific downstream tasks, these pre-trained base models allowed us to demonstrate the effectiveness of our approach on versatile, general-purpose language models. Their selection also provided a balanced representation of different model architectures and training processes \cite{minaee2024large}. Even though the developers of both Llama-2 and Mistral provide some general descriptions about the datasets that were used to train the models, unlike SantaCoder, the datasets that were used to train these models are not publicly available. Moreover, unlike SantaCoder, neither Mistral nor Llama-2 support the FIM objective natively. Therefore, for our experiments, we first construct a dataset and fine-tune both models on the constructed dataset. We explain our dataset construction in Section \ref{subsubsec:results__dataset_construction} and our fine-tuning process in Section \ref{subsubsec:results__finetuning}.

\subsection{Dataset Construction}\label{subsubsec:results__dataset_construction}
    In order to assess our approach's correctness, we need to have access to the dataset that an LLM was trained on. By knowing which scripts/projects were used for training the model, we can construct a ground truth for training and validating the classifier for dataset inclusion detection. To do so, we use two datasets as follows.
    
    \subsubsection{TheStack}
        TheStack\footnote{https://huggingface.co/datasets/bigcode/the-stack} is a large dataset (3 terabytes) of code collected from GitHub by HuggingFace\footnote{https://huggingface.co/huggingface}. This dataset contains codes for 30 different programming languages and 137.36 million repositories \cite{Kocetkov2022TheStack}. We focus on the Python repositories in this dataset as the data cleaning process used for SantaCoder which we outline in Section \ref{subsubsec:results__model_under_study} is accurately replicable for Python. Given the large number of Python scripts and the data cleaning approach used for producing the training data for the LLM, some scripts within a project might be included in the LLM's dataset while others might be excluded. Therefore, we first filter the Python projects by considering a minimum of 10 and a maximum of 50 scripts per project. We focus on projects of such scale for the following reasons:
    
        \begin{itemize}
            \item Our aim is to increase the likelihood that some scripts within a project might be included in the training dataset of the LLM while some from the same project might not. This will allow us to have a more diverse dataset for both script-level and project-level dataset inclusion detection as explained in Section \ref{subsec:results_detecting_code_inclusion}.
            
            \item As our primary objective is to detect a project's inclusion in the training dataset of an LLM, by not including overly large projects that contain many scripts, we will have a larger number of projects in our dataset (e.g., instead of having a project with 200 scripts, we can have 5 projects with 40 scripts).
            
            \item As we outline in Section \ref{subsubsec:results__model_under_study}, we have chosen a data cleaning criterion based on documentation to code ratio. Therefore, to construct a balanced dataset where projects are not documented enough or too documented, we do not include overly small projects in our dataset. As reported by Mamun et al. \cite{mamun2017correlations}, there exists a correlation between the size of a software project and the amount of documentation written for it. Therefore, by not including overly small projects, we will increase the likelihood of having high-quality, well-documented scripts in our dataset.
        \end{itemize}
        It should be noted that SantaCoder was trained on all of the Python, Java, and JavaScript codes in TheStack after applying the data cleaning criterion. We then randomly sample from the remaining projects for constructing the dataset for inclusion detection for TraWiC, NiCad, and JPlag (as explained in Sections \ref{subsec:trawic_methodology} and \ref{subsec:methodology__comparison_with_nicad}).

    \subsubsection{GitHub repositories}

        As mentioned in Section \ref{subsubsec:results__model_under_study}, unlike SantaCoder, the datasets that were used to train Mistral and Llama-2 are not publicly available and the developers of both models do not explicitly declare what data was used to train their models. As such, in order to evaluate our approach on much larger and more capable models, we construct a dataset from GitHub repositories that were published \textbf{after} the release of both Mistral and Llama-2 (September 2023). Similar to the dataset used for SantaCoder, in order to have a fixed approach and compare the models methodologically, we focus on repositories that contain Python codes. As these repositories were published after these models' release, we can be certain that they were not used as the training data of the models under study. After collecting these repositories, we fine-tune the models on the dataset. As such, we will have a ground truth for both Mistral and Llama-2 which contains codes that the models have seen during their training and codes that were not included in their training dataset. The resulting dataset contains over 1355 Python scripts from 51 repositories out of which 20000 data points (which we will further explain in Section \ref{subsubsec:results__finetuning}) were constructed. 10000 data points were used for fine-tuning the models and the remaining 10000 were used for evaluation of TraWiC itself (therefore not included in the model's fine-tuning dataset). The list of repositories that were used to fine-tune the models is available in our replication package \cite{reppackage}.
    
\subsection{Finetuing}\label{subsubsec:results__finetuning}
    As described in Section \ref{subsubsec:results__model_under_study}, in contrast to SantaCoder, Mistral and Llama-2 are not trained to support the FIM objective natively. So, we use the dataset that we constructed from GitHub repositories as described in Section \ref{subsubsec:results__dataset_construction}, to fine-tune the models for the FIM objective. Doing so has two benefits:
    \begin{itemize}
        \item Constructing our own dataset allows for having a ground truth for determining whether a script/project was included in a model’s training dataset. Therefore, we can evaluate TraWiC on more capable models without the loss of generality.
        \item As the pre-trained versions of Mistral and Llama-2 do not support the FIM objective, using our own constructed dataset for fine-tuning the models allows for these models to support the FIM task.
    \end{itemize}
    
    In order to fine-tune the models for the FIM objective, we use the same prompt format as the one which was used to train SantaCoder. Listing \ref{listing:prompt_format} displays the prompt that was used to fine-tune the models with \texttt{prefix} and \texttt{suffix} being constructed as explained in Section \ref{subsec:definitions} and \texttt{infill} being the infilling objective which the model is tasked to predict.

        \begin{lstlisting}[style=yamlstyle, caption={The prompt format used for fine-tuning the models under study}, label={listing:prompt_format}]
            <fim-prefix>{prefix}<fim-suffix>{suffix}<fim-middle>{infill}<|endoftext|>
        \end{lstlisting}

    As an example, consider the script that was presented in Listing \ref{listing:motivating_example_code}. Listing \ref{listing:datapoint_example} presents a single data point extracted from this script after extracting the syntactic and semantic identifiers from this script as displayed in Listing \ref{listing:motivating_example_identifiers}.

        \begin{lstlisting}[style=yamlstyle, caption={The data point format used for fine-tuning the models under study}, label={listing:datapoint_example}]
            "prefix": "def print_input_string(input_string): \"\"\" ...\"\"\""
            "infill": "dummy_variable"
            "suffix": " = input_string \n # print the input ...."
        \end{lstlisting}

    Following this approach, we go over the collected dataset and generate the fine-tuning dataset. The models are only trained on Python scripts, and if codes from other languages exist in the dataset, they are not included in the fine-tuning dataset. In order to fine-tune the models on our dataset for the FIM objective, we use the Quantized Low-Rank Adaption (QLoRA) \cite{dettmers2024qlora} approach which is a parameter-efficient fine-tuning (PEFT) method \cite{ding2023parameter}. Furthermore, in order to have comparable results for both Mistral and Llama, both models were fine-tuned on the same dataset. The codes and the data that were used for training these models are available in our replication package \cite{reppackage} allowing for replication of our fine-tuned models and TraWiC's results. We present the details of fine-tuning both models in Section \ref{sec:appendix}.
    
\subsection{Classifier for Dataset Inclusion Detection}\label{subsubsec:results__classifier}
    We have chosen to use random forests as the classifier for our experiments for the following reasons:
    
    \begin{itemize}
        \item We emphasize the need for our approach to maintain a high degree of interpretability. By using multiple decision trees (which are inherently interpretable) that comprise the random forest, this type of classifier provides interpretability alongside high classification performance.

        \item Ensemble methods such as random forests are known for their superior performance over single predictors due to their ability to reduce overfitting and improve generalization. They are also more computationally efficient compared to the intensive resource demands of training and inference of DNN models \cite{lecun2015deep}.
    \end{itemize}

    To ensure a thorough comparative analysis of classifiers, we have also experimented with XGBoost (XGB) and Support Vector Machine (SVM) classifiers. For optimal classifier selection, we employed the grid search methodology which involves an exhaustive exploration of hyperparameters \cite{lecun2015deep}, where for each model, we evaluated various combinations of settings. We lay out the different sets of configurations used for grid search for each model in Appendix (Section \ref{sec:appendix}). The best models were selected based on the achieved F-score on the test dataset.

\subsection{Detecting Dataset Inclusion}\label{subsec:results_detecting_code_inclusion}
    Depending on the data cleaning criteria used for filtering the scripts for training the LLM as mentioned in Section \ref{subsubsec:results__model_under_study}, some scripts in a project may not be included in the model’s training dataset, e.g., out of 20 scripts in a project, some may not pass the data filtering criteria. Therefore, we outline two different levels of granularity for predicting whether a project was present in a model’s training dataset:
        
        \begin{itemize}
            \item File-level granularity: We determine whether a single script has been in the LLM’s training dataset.
                
            \item Repository-level granularity: We analyze all the scripts from a repository and consider that the repository has been included in the LLM’s training dataset if the classifier predicts a certain amount of scripts out of the total number of scripts in the repository as being in the training dataset.
        \end{itemize}
    
    To analyze the impact of edit distance on the classifier's performance, we define multiple levels of edit distance for a thorough sensitivity analysis. Our analysis for SantaCoder was conducted on 263 repositories extracted from TheStack, totaling 9,409 scripts with 1,866 scripts not included in SantaCoder’s training dataset. For both Llama-2 and Mistral, we used the same dataset that we fine-tuned the models on, designating half as the ground truth for scripts included in the training dataset, and the other half as scripts not included. This means the classifier was trained on 1,355 scripts from 51 repositories. Out of which, 25 repositories and their corresponding scripts were used for fine-tuning the models (thus seen by the model during training), while 26 repositories and their corresponding scripts were used as negative instances (not seen by the model during training).

\subsection{Comparison With other Clone Detection Approaches}\label{subsec:methodology__comparison_with_nicad}
    In software auditing, one of the approaches for detecting copyright infringement is using CCDs to check the code against a repository of copyrighted code \cite{ralhan2021study}. CCD is an active area of research and different approaches are proposed for the detection of clones, especially those of \textit{Type 3} and \textit{Type 4} \cite{ain2019systematic}. This process is carried out through a pair-wise comparison of the codes in a project against another dataset that contains copyrighted codes. As detailed in Section \ref{subsec:background__code_clone_detection}, many different approaches are proposed and available for CCD \cite{ain2019systematic}. Out of the proposed approaches, we choose NiCad \cite{roy2008nicad} and JPlag \cite{prechelt2002finding} for comparison with TraWiC on the dataset inclusion task, for the following reasons:

    \begin{itemize}
        \item \textbf{Python support:} Out of the proposed, available approaches, NiCad and JPlag are the only ones that support clone detection for Python scripts.
        
        \item \textbf{Availability:} A few of the proposed approaches are available for use. Furthermore, most of them require expensive re-training on new datasets. However, both NiCad and JPlag are open-source and do not require re-training (as opposed to the proposed ML-based solutions).
        
        \item \textbf{Syntactic errors in generated codes:} Most of the proposed approaches for CCD based on DNNs, extract information such as ASTs from the code which can only be done if the generated code is syntactically valid. Even though LLMs have shown a high capability in generating correct code, limiting CCD to codes that are syntactically valid would severely reduce the size of the dataset that can be used for this study.
    \end{itemize}

    NiCad employs a parser-based, language-specific approach using the TXL transformation system to extract and normalize potential code clones. It is capable of detecting code clones up to Type 3 \cite{roy2008nicad}. JPlag \cite{prechelt2002finding} compares the similarity of programs in two phases: converting programs into token strings and comparing these strings using the ``Greedy String Tiling'' \cite{wise1993string} algorithm. This algorithm identifies the longest common substrings between two token strings and marks them as tiles. The similarity value is determined by the percentage of tokens covered by these tiles. JPlag’s tokenization is language-dependent, and it prioritizes tokens that reflect a program’s structure while ignoring superficial aspects like whitespace and comments. For the purpose of this study, we use the latest version of JPlag\footnote{https://github.com/jplag/JPlag/releases/tag/v5.0.0}. As NiCad and JPlag are designed to detect clones between scripts, they operate differently compared to our approach where we inspect the similarity between one or multiple generated tokens. Therefore, to determine whether a code was in a model’s training dataset with NiCad/JPlag, we query the model to generate code snippets and compare the generated codes against each snippet that we know exists in the model’s training dataset. Algorithm \ref{algo:algo_nicad}, shows the steps for dataset inclusion detection using NiCad and JPlag. We generate code snippets instead of tokens since comparing scripts to each other using NiCad and JPalg will result in positive clone matches as a large portion of both scripts will be the same as shown in Figure \ref{fig:nicad_code_example}. Our detailed process is as follows:

\subsubsection{Pre-processing}\label{subsubsec:methodology__nicad__preprocessing}
\begin{figure}[b]
            \centering
            \includegraphics[width=\textwidth]{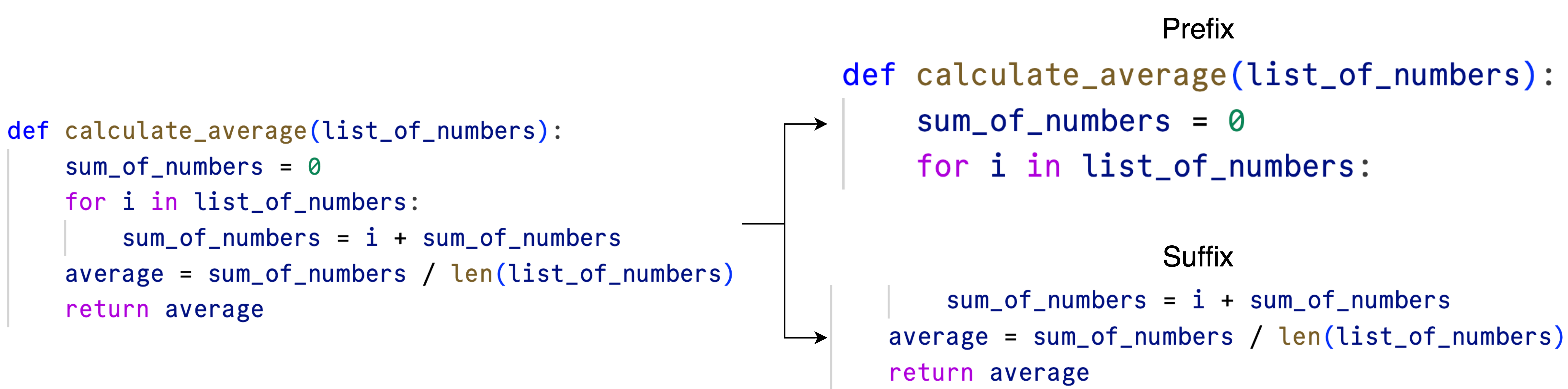}
            \caption{An example of how prefix and suffix are generated for NiCad and JPlag}
            \label{fig:nicad_code_example}
        \end{figure}

        We extract the functions and classes that are defined in a script by parsing its code. For example, the script in Listing \ref{listing:motivating_example_code}, has 2 functions. After doing so, we break each function/class into two parts based on its number of lines. Figure \ref{fig:nicad_code_example} displays an example of how this process is carried out. This process is repeated for every function and class that is extracted from the script. We consider the first half as the prefix and the second half as the suffix.
        
\subsubsection{Inference}\label{subsubsec:methodology__nicad__inference}
        After generating prefix and suffix pairs, we query the model with the prefix and have the model generate the suffix. Note that here, we do not restrict the model to predict the masked element by giving it the suffix. Instead, we allow the model to generate what comes after the prefix until the model itself generates an ``end-of-sequence'' token.

\subsubsection{Comparison}\label{subsubsec:methodology__nicad__comparison}
         TheStack dataset contains 106.91 million Python scripts \cite{Kocetkov2022TheStack}. Therefore, comparing each generated script by the model under study in a pair-wise manner against all the other scripts present in the training dataset is computationally infeasible. As such, we compare each generated script with those of multiple randomly sampled repositories from the model's training dataset to detect code clones across projects. Here, we aim to investigate whether the model generates codes from its training dataset when it is given a large part of a script as input. If NiCad/JPlag detect a clone between the generated script and another script from the sampled projects, we consider it as a \textit{hit} for detecting that the project was included in the model's training dataset. As we are sampling randomly, there exists the possibility that the model generates code from its training dataset but the generated code wasn't compared against the project that it was generated from. Therefore, to decrease the likelihood of such cases we do the following:
         \begin{itemize}
             \item We follow the same project selection criteria as described in Section \ref{subsubsec:results__dataset_construction} for both selecting the scripts to generate code from and comparing the generated codes against. By only sampling projects that contain a minimum of 10 and a maximum of 50 scripts, we aim to decrease the amount of missing \textit{hits}.
             \item As we expand upon in Section \ref{subsubsec:results_rq1b} we experiment with multiple random sampling ratios in order to have a comprehensive sensitivity analysis.
         \end{itemize}

        \begin{algorithm}
            \caption{Dataset Inclusion Detection Using NiCad/JPlag}
            \small
            \SetKwFunction{NiCad}{\textbf{NiCad/JPlag}}
            \SetKwFunction{Predict}{\textbf{Predict}}
            \SetKwFunction{ExtractPrefixAndSuffix}{\textbf{ExtractPrefixAndSuffix}}
            \SetKwComment{Comment}{/* }{ */}
            
            \KwIn{SCRIPT \Comment*[r]{SCRIPT is an entire script from a project. A project can contain many scripts.}}
            \KwOut{ClassificationResults \Comment*[r]{A binary variable with 1 indicating inclusion in the dataset and 0 otherwise.}}
            
            \tcp{Break the script into a prefix and suffix tuple based on the number of lines.}
            prefix, suffix $\leftarrow$ \ExtractPrefixAndSuffix{SCRIPT};
            
            \tcp{Generate the suffix by giving the model prefix as input.}
            GeneratedSuffix $\leftarrow$ \Predict{prefix};
            
            \tcp{Compare the generated suffix in a pair-wise manner with every script in the dataset.}
            \ForEach{$s \in \text{dataset}$}{
                r $\leftarrow$ \NiCad{s, GeneratedSuffix}\;
                Store $r$ in ClassificationResults\;
            }
            \KwRet{ClassificationResults};
            \label{algo:algo_nicad}
        \end{algorithm}

\section{Results and Analysis}\label{sec:results}
    In this section, we present our experimental results. We begin by laying out our experiment details, continue with presenting the results of our approach compared against a baseline, and conclude this section by conducting an error, feature importance, and sensitivity analysis. Our replication package is publicly accessible online \cite{reppackage}.

\subsection{\textbf{RQ1 [Effectiveness]}}\label{subsec:results__rq1}
    The results detailed in Tables \ref{table:edit_distance_sensitivity_results_repo_level}, \ref{table:edit_distance_sensitivity_results_repo_level_llama2}, and \ref{table:edit_distance_sensitivity_results_repo_level_mistral} showcase the performance of TraWiC's dataset inclusion detection when applied with varying edit distance thresholds. The ``Considering only a single positive'' criterion showcases the results on file-level granularity. We consider multiple inclusion criteria for predicting whether a repository (i.e., project) was included in the model's training dataset as outlined in ``Considering a threshold of 0.4/0.6''. ``Precision'' measures the percentage of correctly predicted positive observations out of all predicted positives, reflecting the classifier's accuracy in making positive predictions. ``Accuracy'' gives an overall assessment, showing the ratio of correctly predicted observations to the total predictions made. The ``F-score'' balances precision and sensitivity (recall), offering a comprehensive metric, especially for uneven datasets. ``Sensitivity'' assesses the model's capability to correctly distinguish actual positive cases, while ``Specificity'' measures its performance in distinguishing true negatives. We focus on the F-score as the main performance metric. After F-score, we focus on sensitivity as our goal is to detect dataset inclusion; therefore, the higher the classifier's ability to detect true positives, the more suited to our purpose it is.

    \subsubsection{RQ1a: What is TraWiC's performance for the dataset inclusion task?} \label{subsubsec:results_rq1a}
        We observe that the performance of our models varies based on the repository inclusion criterion and the edit distance threshold. Specifically, for SantaCoder, the best performance is achieved with an edit distance of 60 and a repository inclusion threshold of 0.4. In contrast, both fine-tuned versions of Mistral and Llama-2 perform best with an edit distance threshold of 20. However, there are differences in their best-performing conditions: Llama-2 achieves a higher F-score with an inclusion threshold of 0.4, while Mistral performs better in detecting the inclusion of single scripts. Nonetheless, the performance gap between Llama-2 and Mistral in terms of dataset inclusion detection is not significant when using an edit distance threshold of 20.
        For all 3 models under study, we can observe that precision, accuracy, and F-score generally decrease as the edit distance threshold increases. This shows that a higher threshold for edit distance may lead to less strict matching criteria, which in turn could result in more false positives and reduce TraWiC’s performance. Additionally, we observe that sensitivity marginally increases with the edit distance threshold for SantaCoder but marginally decreases for Mistral and Llama-2 when detecting single script inclusion. This suggests that while SantaCoder’s classifier becomes less precise, it captures more true positives at higher thresholds. However, the performance of Mistral and Llama-2 degrades as the threshold increases.
        
        \begin{table}
            \caption{Results of TraWiC's code inclusion detection with different edit distance thresholds on SantaCoder}
            \resizebox{\textwidth}{!}{
                \centering
                \begin{tabular}{c c r r r r r}
                    \toprule
                    Edit Distance Threshold & Repository Inclusion Criterion & Precision(\%) & Accuracy(\%) & F-score(\%) & Sensitivity(\%) & Specificity (\%)\\
                    \midrule
                        \multirow{3}{*}{\textbf{20}} & Considering only a \textbf{single} positive & 88.07 & 83.87 & \underline{88.57} & 89.08 & \underline{71.6}\\
                        & Considering a threshold of \textbf{0.4} & 76.52 & 75.1 & 84.82 & 95.13 & 20.63\\
                        & Considering a threshold of \textbf{0.6} & 80.46 & 71.26 & 80.31 & 80.15 & 47.08\\
                    \midrule
                        \multirow{3}{*}{\textbf{40}} & Considering only a \textbf{single} positive & 79.57 & 79.11 & 83.39 & 94.49 & 42.91\\
                        & Considering a threshold of \textbf{0.4} & 72.96 & 72.4 & 83.8 & 98.43 & 3.62\\
                        & Considering a threshold of \textbf{0.6} & 75.5 & 73.96 & 84.1 & 94.9 & 18.65\\
                    \midrule
                        \multirow{3}{*}{\textbf{60}} & Considering only a \textbf{single} positive & 78.09 & 77.46 & 85.46 & 94.36 & 37.7\\
                        & Considering a threshold of \textbf{0.4} & 71.28 & 71.23 & 82.95 & \underline{99.19} & 2.92\\
                        & Considering a threshold of \textbf{0.6} & 73.3 & 72.11 & 82.89 & 95.38 & 15.6\\
                    \midrule
                        \multirow{3}{*}{\textbf{70}} & Considering only a \textbf{single} positive & 78.47 & 78.08 & 85.85 & 94.77 & 38.81 \\
                        & Considering a threshold of \textbf{0.4} & 73.3 & 72.83 & 84.17 & 98.83 & 2.11\\
                        & Considering a threshold of \textbf{0.6} & 75.46 & 74.11 & 84.39 & 95.71 & 15.34\\
                    \midrule
                        \multirow{3}{*}{\textbf{80}} & Considering only a \textbf{single} positive & 77.79 & 77.47 & 85.55 & 95.02 & 36.17\\
                        & Considering a threshold of \textbf{0.4} & 71.22 & 70.83 & 82.84 & 99 & 1.47\\
                        & Considering a threshold of \textbf{0.6} & 72.42 & 70.98 & 82.41 & 95.6 & 10.3\\
                    \bottomrule
                \end{tabular}
            } 
            \label{table:edit_distance_sensitivity_results_repo_level}
        \end{table}
        
        Examining Table \ref{table:edit_distance_sensitivity_results_repo_level}, we observe diverse results based on different edit distance thresholds and repository inclusion criteria. The highest F-score, 88.57\%, is achieved at an edit distance threshold of 20 when considering a single script's inclusion. Specificity is highest, 71.6\%, with the same criteria, indicating a decent balance in predicting when a script was \textbf{not} included in the model's training dataset. For project inclusion detection, we report a sensitivity of 99.19\% and a specificity of 2.92\% while considering an edit distance threshold of 60. This means that TraWiC has a high capability in detecting whether a project was included in the training dataset. As mentioned in Section \ref{subsec:results_detecting_code_inclusion} after following the data cleaning criteria as laid out in \cite{allal2023santacoder} some scripts from a project may not be included in the training dataset of the given model. For example, out of the 20 scripts in a project, 10 may not have a comment-to-code ratio between 1\% to 80\% and therefore will not be used for training the model. Therefore, even though the project itself was included in the model's training dataset, a specific script from it may not be. As such, it should be noted that most of the projects in our dataset (in contrast to scripts) were included in SantaCoder's training dataset and a low specificity when it comes to \underline{project} inclusion detection (as opposed to script inclusion detection) is not alarming as other scripts from the project could have been used in the model's training. 

        Examining Tables \ref{table:edit_distance_sensitivity_results_repo_level_llama2} and \ref{table:edit_distance_sensitivity_results_repo_level_mistral}, we observe various performance metrics for TraWiC’s performance on fine-tuned versions of Llama-2 and Mistral across different edit distance thresholds and repository inclusion criteria. For Llama-2 (Table \ref{table:edit_distance_sensitivity_results_repo_level_llama2}), the highest F-score of 82.05\% is achieved at an edit distance threshold of 20 when considering a repository inclusion criterion of 0.4. The highest specificity, 87.5\%, is also achieved at an edit distance threshold of 20 with a repository inclusion criterion of 0.4. This indicates a good balance in predicting an entire project’s inclusion in the model’s training dataset. Sensitivity is highest at 84.23\% with an edit distance threshold of 20 and considering only a single positive. This shows that at this edit distance threshold, TraWiC can detect both a single script’s and a project’s inclusion in a model’s training dataset with high performance. For Mistral (Table \ref{table:edit_distance_sensitivity_results_repo_level_mistral}), the highest F-score, 85.37\%, is achieved at an edit distance threshold of 20 when considering a single script’s inclusion. Similarly, the highest sensitivity of 88.38\% is observed under the same conditions. The highest specificity, 85.71\%, is achieved with an edit distance threshold of 20 and a repository inclusion criterion of 0.4, indicating strong performance in predicting the absence of scripts in the training dataset. Therefore, we can observe that TraWiC shows high performance in code inclusion detection on LLMs trained on code. It should also be noted that the results presented in Tables  \ref{table:edit_distance_sensitivity_results_repo_level_llama2} and \ref{table:edit_distance_sensitivity_results_repo_level_mistral} are conducted on models that were fine-tuned on the dataset for 3 epochs. However, for both Llama-2 and Mistral models we can see that TraWiC’s performance starts to degrade as we increase the edit distance threshold. We discuss the effects of fine-tuning and the reasons behind TraWiC’s performance degradation on these models in RQ3.
        
        \begin{table}
            \caption{Results of TraWiC's code inclusion detection with different edit distance thresholds on Llama-2}
            \resizebox{\textwidth}{!}{
                \centering
                \begin{tabular}{c c r r r r r}
                    \toprule
                    Edit Distance Threshold & Repository Inclusion Criterion & Precision(\%) & Accuracy(\%) & F-score(\%) & Sensitivity(\%) & Specificity (\%)\\
                    \midrule
                        \multirow{3}{*}{\textbf{20}} & Considering only a \textbf{single} positive & 78.68 & 82.74 & 81.36 & \underline{84.23} & 81.54\\
                        & Considering a threshold of \textbf{0.4} & 80.0 & 86.27 & \underline{82.05} & 84.21 & \underline{87.5}\\
                        & Considering a threshold of \textbf{0.6} & 75.0 & 72.54 & 68.18 & 62.5 & 81.48\\
                    \midrule
                        \multirow{3}{*}{\textbf{40}} & Considering only a \textbf{single} positive & 75.98 & 80.93 & 79.09 & 82.47 & 79.73\\
                        & Considering a threshold of \textbf{0.4} & 75.0 & 75.6 & 75.0 & 75.0 & 76.19\\
                        & Considering a threshold of \textbf{0.6} & 66.66 & 70.0 & 70.0 & 73.68 & 66.66\\
                    \midrule
                        \multirow{3}{*}{\textbf{60}} & Considering only a \textbf{single} positive & 73.49 & 79.06 & 76.56 & 79.91 & 78.43\\
                        & Considering a threshold of \textbf{0.4} & 70.0 & 73.17 & 71.79 & 73.68 & 72.72\\
                        & Considering a threshold of \textbf{0.6} & 70.0 & 70.73 & 70.0 & 70.0 & 71.42\\
                    \midrule
                        \multirow{3}{*}{\textbf{70}} & Considering only a \textbf{single} positive & 72.58 & 78.50 & 75.78 & 79.29 & 77.92\\
                        & Considering a threshold of \textbf{0.4} & 70.0 & 73.17 & 71.79 & 73.68 & 72.72\\
                        & Considering a threshold of \textbf{0.6} & 65.0 & 70.73 & 68.42 & 72.22 & 69.56\\
                    \midrule
                        \multirow{3}{*}{\textbf{80}} & Considering only a \textbf{single} positive & 70.56 & 77.57 & 74.46 & 78.82 & 76.67\\
                        & Considering a threshold of \textbf{0.4} & 75.0 & 68.29 & 69.76 & 65.21 & 72.22\\
                        & Considering a threshold of \textbf{0.6} & 70.0 & 63.41 & 65.11 & 60.89 & 66.66\\
                    \bottomrule
                \end{tabular}
            } 
            \label{table:edit_distance_sensitivity_results_repo_level_llama2}
        \end{table}
        
        \begin{table}
            \caption{Results of TraWiC's code inclusion detection with different edit distance thresholds on Mistral}
            \resizebox{\textwidth}{!}{
                \centering
                \begin{tabular}{c c r r r r r}
                    \toprule
                    Edit Distance Threshold & Repository Inclusion Criterion & Precision(\%) & Accuracy(\%) & F-score(\%) & Sensitivity(\%) & Specificity (\%)\\
                    \midrule
                        \multirow{3}{*}{\textbf{20}} & Considering only a \textbf{single} positive & 82.55 & 86.45 & \underline{85.37} & \underline{88.38} & 84.89\\
                        & Considering a threshold of \textbf{0.4} & 85.0 & 85.36 & 85.0 & 85.0 & \underline{85.71}\\
                        & Considering a threshold of \textbf{0.6} & 84.21 & 82.5 & 82.05 & 80.0 & 85.0\\
                    \midrule
                        \multirow{3}{*}{\textbf{40}} & Considering only a \textbf{single} positive & 80.62 & 85.52 & 84.21 & 88.13 & 83.49\\
                        & Considering a threshold of \textbf{0.4} & 85.0 & 85.36 & 85.0 & 85.0 & 85.71\\
                        & Considering a threshold of \textbf{0.6} & 84.21 & 82.05 & 82.05 & 80.0 & 84.21\\
                    \midrule
                        \multirow{3}{*}{\textbf{60}} & Considering only a \textbf{single} positive & 77.51 & 83.48 & 81.79 & 86.58 & 81.16\\
                        & Considering a threshold of \textbf{0.4} & 80.0 & 80.48 & 80.0 & 80.0 & 80.95\\
                        & Considering a threshold of \textbf{0.6} & 80.0 & 78.04 & 78.04 & 76.19 & 80.0\\
                    \midrule
                        \multirow{3}{*}{\textbf{70}} & Considering only a \textbf{single} positive & 77.51 & 82.56 & 80.97 & 84.74 & 80.85 \\
                        & Considering a threshold of \textbf{0.4} & 75.0 & 80.48 & 78.94 & 83.33 & 78.26\\
                        & Considering a threshold of \textbf{0.6} & 75.0 & 75.60 & 75.0 & 75.0 & 76.19\\
                    \midrule
                        \multirow{3}{*}{\textbf{80}} & Considering only a \textbf{single} positive & 74.74 & 83.11 & 81.31 & 86.46 & 80.64\\
                        & Considering a threshold of \textbf{0.4} & 65.0 & 70.73 & 68.42 & 72.22 & 69.56\\
                        & Considering a threshold of \textbf{0.6} & 65.0 & 63.41 & 63.41 & 61.90 & 65.0\\
                    \bottomrule
                \end{tabular}
            } 
            \label{table:edit_distance_sensitivity_results_repo_level_mistral}
        \end{table}
        
        \begin{tcolorbox}[colback=blue!5,colframe=blue!40!black]
            \textbf{Findings 1:} When analyzing a single script's inclusion, TraWiC shows high performances in detecting both true positives and true negatives. The best balance is achieved when we employ lower edit distance thresholds for considering semantic identifiers. This means that looking for similarity in natural language elements of code can be used as a strong signal for detecting dataset inclusion.
        \end{tcolorbox}
        
        Given the differences between TraWiC's specificity and sensitivity on file-level granularity in comparison to repository-level granularity, we can infer two scenarios for using TraWiC. First, if a repository contains a large number of scripts and inference on the given model is resource-intensive, using TraWiC on less than half of the repository's scripts can effectively indicate dataset inclusion. Second, in scenarios where reducing false negatives is more important (e.g., identifying every instance of dataset inclusion), applying TraWiC on file-level granularity can be an effective method to check for dataset inclusion.
        
    \subsubsection{RQ1b: How does TraWiC compare against traditional CCD approaches for dataset inclusion detection?}\label{subsubsec:results_rq1b}
        Table \ref{table:results_nicad_jplag} presents the results obtained when using NiCad and JPlag for clone detection. We used NiCad and JPlag to detect code clones across code snippets from the scripts in the LLM’s training dataset and code snippets generated by the LLM as outlined in Section \ref{subsec:methodology__comparison_with_nicad}. Given that our task is a binary classification, NiCad is unable to achieve the accuracy of a random classifier (50\%), and JPlag achieves an accuracy of 55\% which is barely higher than a random classifier indicating that code clone detection approaches are not suitable for detecting dataset inclusion. This observation is consistent with the results reported in \cite{ciniselli2022extent}. The highest accuracy, F-Score, and sensitivity are achieved by NiCad for clone detection when sampling 9\% of the LLM’s dataset. The highest accuracy, F-score, and sensitivity for JPlag are achieved when sampling 10\% of the dataset while the highest precision and specificity are achieved when sampling 5\% of the dataset. As the results show, our approach outperforms both NiCad and JPlag by a large margin in detecting whether a project was included in the training dataset of SantaCoder regardless of the edit distance threshold. Moreover, using clone detection techniques for dataset inclusion detection is expensive. In fact, before using any clone detection technique, we need to generate code snippets using the LLM and once a sufficient amount of scripts are generated, compare them in a pair-wise manner with the original scripts that were used in the LLM’s training dataset. TraWiC is much less computationally expensive compared to generating multiple snippets of code from different repositories and using NiCad/JPlag to detect similarities between them. For detecting dataset inclusion, TraWiC’s main performance constraint is inference on the LLM for generating the necessary tokens. In contrast, both NiCad and JPlag need to compare an average of 3,000 scripts per project in a pair-wise manner for inclusion detection when considering a 10\% sample of the filtered dataset on top of generating the necessary code from the LLM. As presented in Table \ref{table:edit_distance_sensitivity_results_repo_level},  the best results are achieved when we focus on identifying the inclusion of single scripts with an edit distance threshold of 20. Our analysis results presented in Table \ref{table:results_nicad_jplag} show that increasing the number of sampled repositories has only a marginal improvement on NiCad’s and JPlag’s performance.
        
        \begin{table}[ht]
            \caption{Nicad/JPlag's Performance on detecting clones across sampled repositories}
            \resizebox{\textwidth}{!}{
                \begin{tabular}{llrrrrr}
                    \toprule
                    Tool & Percentage of Dataset Sampled & Precision(\%) & Accuracy(\%) & F-Score(\%) & Sensitivity(\%) & Specificity (\%)\\
                    \midrule
                    \multirow{3}{*}{NiCad} 
                    & Sampling 5\% (1726 projects / $\approx$ 34600 scripts) & 84.81 & 43.13 & 53.60 & 38.18 & 63.63 \\
                    & Sampling 9\% (3108 projects / $\approx$ 62280 scripts) & 86.74 & 47.64 & 56.47 & 41.86 & 72.5 \\
                    & Sampling 10\% (3452 projects / $\approx$ 69200 scripts) & 80.24 & 41.22 & 49.24 & 35.51 & 64.44 \\
                    \midrule
                    \multirow{3}{*}{JPlag} 
                    & Sampling 5\% (1726 projects / $\approx$ 34600 scripts) & 83.53 & 54.49 & 58.92 & 45.51 & 77.26 \\ 
                    & Sampling 9\% (3108 projects / $\approx$ 62280 scripts) & 82.63 & 53.83 & 58.32 & 45.06 & 76.02 \\
                    & Sampling 10\% (3452 projects / $\approx$ 69200 scripts) & 81.00 & 55.00 & 59.94 & 47.57 & 73.00 \\
                    \bottomrule
                \end{tabular}
            }
            \label{table:results_nicad_jplag}
        \end{table}

        \begin{tcolorbox}[colback=blue!5,colframe=blue!40!black]
            \textbf{Findings 2:} The high cost of pair-wise clone detection alongside the LLMs' capability in generating code with different syntax makes code clone detection tools incapable of detecting dataset inclusion. This is indicated by their poor performance observed even when increasing the amount of codes to compare.
        \end{tcolorbox}

    \subsubsection{RQ1c: What is the effect of using different classification methods on TraWiC's performance?}\label{subsubsec:results_rq1c}
        Table \ref{table:different_classifier_performance_comparison} shows the F-scores of different classification methods for SantaCoder. As the results show, random forests outperform both XGB and SVM across all edit distance thresholds and repository inclusion criteria. We include a more detailed analysis of both XGB's and SVM's across different performance metrics in Section \ref{sec:appendix}.
        
        \begin{table}[ht]
            \caption{Performance comparison of different classifiers on the generated dataset}
            \centering
                \begin{tabular}{c c r r r}
                    \toprule
                    Edit Distance Threshold & Repository Inclusion Criterion & F-score XGB(\%) & F-score SVM(\%) & F-score RF(\%)\\
                    \midrule
                        \multirow{3}{*}{\textbf{20}} & Considering only a \textbf{single} positive & 86.04 & 85.53 & 88.57\\
                        & Considering a threshold of \textbf{0.4} & 83.53 & 82.30 & 84.82\\
                        & Considering a threshold of \textbf{0.6} & 80.53 & 77.95 & 80.31\\
                    \midrule
                        \multirow{3}{*}{\textbf{40}} & Considering only a \textbf{single} positive & 84.01 & 83.20 & 83.39\\
                        & Considering a threshold of \textbf{0.4} & 82.16 & 81.89 & 83.80\\
                        & Considering a threshold of \textbf{0.6} & 82.48 & 80.60 & 84.10\\
                    \midrule
                        \multirow{3}{*}{\textbf{60}} & Considering only a \textbf{single} positive & 82.49 & 82.43 & 85.46\\
                        & Considering a threshold of \textbf{0.4} & 82.18 & 82.20 & 82.95\\
                        & Considering a threshold of \textbf{0.6} & 82.02 & 82.37 & 82.89\\
                    \midrule
                        \multirow{3}{*}{\textbf{70}} & Considering only a \textbf{single} positive & 84.32 & 84.49 & 85.85\\
                        & Considering a threshold of \textbf{0.4} & 83.56 & 83.78 & 84.17\\
                        & Considering a threshold of \textbf{0.6} & 81.85 & 82.82 & 84.39\\
                    \midrule
                        \multirow{3}{*}{\textbf{80}} & Considering only a \textbf{single} positive & 82.78 & 82.22 & 85.55\\
                        & Considering a threshold of \textbf{0.4} & 82.39 & 81.91 & 82.84\\
                        & Considering a threshold of \textbf{0.6} & 81.23 & 81.24 & 82.41\\
                    \bottomrule
                \end{tabular}
            
            \label{table:different_classifier_performance_comparison}
        \end{table}

\subsection{\textbf{RQ2 [Sensitivity Analysis]}} \label{subsec:results__rq2}
    In this RQ, we assess TraWiC's robustness to data obfuscation. Afterward, we analyze the correlation between the extracted features in our dataset and discuss the feature importance of the trained classifiers.
    
    \subsubsection{RQ2a: How robust is TraWiC against data obfuscation techniques?}\label{subsubsec:results_rq2a}
        In this section, we analyze TraWiC's robustness against models trained on obfuscated datasets. Data obfuscation can be used as a defense strategy against MIAs \cite{hu2022membership}. For example, knowledge distillation \cite{shejwalkar2021membership, zheng2021resisting} or differential privacy \cite{xu2019ganobfuscator, rosenblatt2020differentially} can be used for training models that do not generate rare elements observed during their training. There also exists a myriad of code obfuscation techniques proposed for software copyright and vulnerability protection by making reverse-engineering the software more difficult. For example, Mixed Boolean Arithmetic methods \cite{david2020qsynth} obfuscate the code by replacing arithmetic operations in a code with overly complex statements. Opaque Predicates approaches \cite{xu2016generalized} obfuscate the code by setting only one direction of the code to be executed and insert dummy code in the part that never gets executed, therefore, making the code more complex without introducing any change to the code's execution. Finally, other approaches alter identifiers, variables, and string literals or remove code indentation to change the code's semantics \cite{sebastian2016study}. Approaches that use a combination of these methods \cite{kang2021obfus} have been proposed for thorough code obfuscation to protect the code as well. 
        
        As such, in this section, we examine the robustness of our approach when confronted with models trained on obfuscated datasets. Such scenarios are also similar to detecting \textit{Type-2/Type-3} code clones. It should be noted that for our sensitivity analysis, we assume that the data for training the given model has already been obfuscated before training the model, therefore making it more robust against producing textual elements similar to the original unobfuscated codes. To simulate these scenarios, we introduce noise during the hit-count process for syntactic identifiers. This noise emulates situations where the LLM produces a syntactic identifier that deviates from the original, resulting in a miss. Moreover, we test the robustness of our approach in scenarios where noise is introduced during semantic identifiers' hit count. Finally, we simulate scenarios where both syntactic and semantic identifiers have been changed in the training dataset by data obfuscation techniques. Therefore, we will be able to test TraWiC's robustness in recognizing dataset inclusion despite deliberate obfuscations in the training dataset. 
        
        Tables \ref{table:noise_sensitivity_results_santacoder}, \ref{table:noise_sensitivity_results_llama}, and \ref{table:noise_sensitivity_results_mistral} show the results of our sensitivity analysis conducted by varying the levels of noise (with noise ratios of 0.1, 0.5, and 0.9) injected during the syntactic, semantic, and their combination's hit-count process as mentioned above, for SantaCoder, Llama-2, and Mistral, respectively. Here, ``Target'' refers to the level that the noises are being applied to, a ``Noise Ratio'' of 0.1 means that during the hit-count process, a hit is ignored with a probability of 10\%. Similar to the analyses done in \textit{RQ1}, we consider different edit distance thresholds (20, 60, and 80) for detecting semantic hits. It should be noted that we conduct our analysis on single scripts instead of projects. We also conduct analyses of different ``Repository Inclusion Criterion'' (i.e., considering a project to be included in the model's training dataset if a certain amount of its corresponding scripts are detected as included in the model's training dataset) similar to the analysis done in Tables \ref{table:edit_distance_sensitivity_results_repo_level}, \ref{table:edit_distance_sensitivity_results_repo_level_llama2}, and \ref{table:edit_distance_sensitivity_results_repo_level_mistral}. We include these results in Section \ref{sec:appendix}. 
        
        As expected, increasing the noise sensitivity threshold results in a decline in precision, accuracy, F-score, and sensitivity across all edit distance thresholds, for all the models under study. This is aligned with the expectation that the introduction of noise impairs the classifier's ability to correctly identify true positives. However, the random forest's inherent robustness to overfitting and its ensemble nature appear to mitigate this effect to some extent, particularly at moderate levels of noise as also reported in \cite{ghosh2017robustness}. Finally, we observe that TraWiC's performance degrades close to that of a random classifier when there exists a lot of noise in the dataset for SantaCoder and even further for Llama-2 and Mistral. This indicates that our approach is not robust against thorough data obfuscation techniques where nearly all semantic and syntactic identifiers are changed. We will discuss possible approaches for improving our approach in Section \ref{sec:conclusion}.
        
        \begin{table}
            \caption{Results of TraWiC's dataset inclusion detection with noise injection across different levels - SantaCoder}
            \resizebox{\textwidth}{!}{
                \centering
                \begin{tabular}{c c c c r r r r r}
                    \toprule
                    Target & Edit Distance Threshold & Noise Ratio & Precision(\%) & Accuracy(\%) & F-score(\%) & Sensitivity(\%) & Specificity (\%)\\
                    \midrule
                    \multirow{9}{*}{\textbf{Syntactic}} &
                        \multirow{3}{*}{\textbf{20}} & 0.1 & 88 & 83.09 & 88.13 & 88.27 & 70.3\\
                        & & 0.5 & 86.41 & 79.26 & 85.23 & 84.08 & 67.34\\
                        & & 0.9 & 87.8 & 69.04 & 75.11 & 65.62 & 77.49\\
                    \cmidrule{2-8}
                        & \multirow{3}{*}{\textbf{60}} & 0.1 & 79.45 & 78.51 & 86.18 & 94.17 & 39.85\\
                        & & 0.5 & 79.30 & 74.89 & \underline{83.24} & 87.59 & 43.54\\
                        & & 0.9 & 81.82 & 59.73 & 66.37 & 55.83 & 69.37\\
                    \cmidrule{2-8}
                        & \multirow{3}{*}{\textbf{80}} & 0.1 & 78.92 & 78.03 & 85.94 & 94.32 & 37.82\\
                        & & 0.5 & 78.43 & 74.95 & 83.55 & 89.39 & 39.3\\
                        & & 0.9 & 79.11 & 65.69 & 74.38 & 70.18 & 54.24\\
                    \midrule
                    \multirow{9}{*}{\textbf{Semantic}} &
                        \multirow{3}{*}{\textbf{20}} & 0.1 & 87.19 & 82.29 & 87.62 & 88.04 & 68.08\\
                        & & 0.5 & 83.83 & 74.89 & 81.97 & 80.19 & 61.81\\
                        & & 0.9 & 81.7 & 51.86 & 55.22 & 41.7 & 76.94\\
                    \cmidrule{2-8}
                        & \multirow{3}{*}{\textbf{60}} & 0.1 & 79.63 & 78.67 & 86.26 & 94.1 & 40.59\\
                        & & 0.5 & 78.83 & 76.33 & 84.59 & \underline{91.26} & 39.48\\
                        & & 0.9 & 78.79 & 74.31 & 82.89 & 87.44 & 41.88\\
                    \cmidrule{2-8}
                        & \multirow{3}{*}{\textbf{80}} & 0.1 & 79.31 & 78.88 & 86.51 & 95.14 & 38.75\\
                        & & 0.5 & 78.03 & 76.76 & 85.16 & 93.72 & 34.87\\
                        & & 0.9 & 78.13 & 75.21 & 83.86 & 90.51 & 37.45\\
                    \midrule
                    \multirow{9}{*}{\textbf{Combined}} &
                        \multirow{3}{*}{\textbf{20}} & 0.1 & 87.16 & 82.07 & 87.45 & 87.74 & 68.08\\
                        & & 0.5 & 83.37 & 73.03 & 80.37 & 77.58 & 61.81\\
                        & & 0.9 & 85.25 & 50.96 & 52.18 & 37.59 & 83.95\\
                    \cmidrule{2-8}
                        & \multirow{3}{*}{\textbf{60}} & 0.1 & 79.1 & 77.87 & 85.76 & 93.65 & 38.93\\
                        & & 0.5 & 78.93 & 73.51 & \underline{82.15} & 85.65 & 43.54\\
                        & & 0.9 & 81.97 & 54.1 & 58.53 & 45.52 & 75.28\\
                    \cmidrule{2-8}
                        & \multirow{3}{*}{\textbf{80}} & 0.1 & 78.34 & 77.39 & 85.59 & 94.32 & 35.61\\
                        & & 0.5 & 78.12 & 74.36 & \underline{83.15} & 88.86 & 38.56\\
                        & & 0.9 & 82.1 & 57.45 & 63.24 & 51.42 & 72.32\\
                    \bottomrule
                \end{tabular}}
            \label{table:noise_sensitivity_results_santacoder}
        \end{table}
    
        \begin{table}
            \caption{Results of TraWiC's dataset inclusion detection with noise injection across different levels - Llama-2}
            \resizebox{\textwidth}{!}{
                \centering
                \begin{tabular}{c c c c r r r r r}
                    \toprule
                    Target & Edit Distance Threshold & Noise Ratio & Precision(\%) & Accuracy(\%) & F-score(\%) & Sensitivity(\%) & Specificity (\%)\\
                    \midrule
                    \multirow{9}{*}{\textbf{Syntactic}} &
                        \multirow{3}{*}{\textbf{20}} & 0.1 & 76.68 & 82.75 & 81.36 & 84.23 & 81.54\\
                        & & 0.5 & 68.60 & 71.80 & 69.96 & 71.37 & 72.16\\
                        & & 0.9 & 57.75 & 61.41 & 58.89 & 60.08 & 62.54\\
                    \cmidrule{2-8}
                        & \multirow{3}{*}{\textbf{60}} & 0.1 & 72.29 & 76.26 & 73.92 & 75.63 & 76.77\\
                        & & 0.5 & 62.50 & 65.92 & 63.01 & 63.52 & 67.93\\
                        & & 0.9 & 56.22 & 59.25 & 56.22 & 56.22 & 61.99\\
                    \cmidrule{2-8}
                        & \multirow{3}{*}{\textbf{80}} & 0.1 & 69.53 & 74.02 & 70.36 & 74.66 & 7357\\
                        & & 0.5 & 60.48 & 66.54 & 62.63 & 64.96 & 67.76\\
                        & & 0.9 & 51.61 & 55.14 & 51.61 & 51.61 & 58.19\\
                    \midrule
                    \multirow{9}{*}{\textbf{Semantic}} &
                        \multirow{3}{*}{\textbf{20}} & 0.1 & 70.93 & 74.40 & 72.62 & 74.39 & 74.40\\
                        & & 0.5 & 63.57 & 67.16 & 64.95 & 66.40 & 67.81\\
                        & & 0.9 & 53.10 & 56.40 & 53.83 & 54.58 & 57.99\\
                    \cmidrule{2-8}
                        & \multirow{3}{*}{\textbf{60}} & 0.1 & 67.07 & 72.34 & 69.29 & 71.67 & 72.85\\
                        & & 0.5 & 57.43 & 62.43 & 58.73 & 60.08 & 64.31\\
                        & & 0.9 & 55.42 & 57.57 & 54.87 & 54.33 & 60.50\\
                    \cmidrule{2-8}
                        & \multirow{3}{*}{\textbf{80}} & 0.1 & 66.13 & 73.46 & 69.79 & 73.87 & 73.16\\
                        & & 0.5 & 56.45 & 62.43 & 58.21 & 60.09 & 64.24\\
                        & & 0.9 & 49.60 & 54.39 & 50.20 & 50.83 & 57.43\\
                    \midrule
                    \multirow{9}{*}{\textbf{Combined}} &
                        \multirow{3}{*}{\textbf{20}} & 0.1 & 68.60 & 71.43 & 69.69 & 70.80 & 71.97\\
                        & & 0.5 & 59.30 & 63.27 & 60.71 & 62.20 & 64.16\\
                        & & 0.9 & 53.10 & 55.66 & 53.41 & 53.73 & 57.39\\
                    \cmidrule{2-8}
                        & \multirow{3}{*}{\textbf{60}} & 0.1 & 66.27 & 69.35 & 66.80 & 67.35 & 71.03\\
                        & & 0.5 & 59.04 & 64.49 & 60.74 & 62.55 & 66.00\\
                        & & 0.9 & 48.59 & 50.28 & 47.64 & 46.72 & 53.62\\
                    \cmidrule{2-8}
                        & \multirow{3}{*}{\textbf{80}} & 0.1 & 62.50 & 68.60 & 64.85 & 67.39 & 69.51\\
                        & & 0.5 & 55.65 & 60.19 & 56.44 & 57.26 & 62.59\\
                        & & 0.9 & 47.18 & 54.02 & 48.75 & 50.43 & 56.77\\
                    \bottomrule
                \end{tabular}}
            \label{table:noise_sensitivity_results_llama}
        \end{table}

        \begin{table}
            \caption{Results of TraWiC's dataset inclusion detection with noise injection across different levels - Mistral}
            \resizebox{\textwidth}{!}{
                \centering
                \begin{tabular}{c c c c r r r r r}
                    \toprule
                    Target & Edit Distance Threshold & Noise Ratio & Precision(\%) & Accuracy(\%) & F-score(\%) & Sensitivity(\%) & Specificity (\%)\\
                    \midrule
                    \multirow{9}{*}{\textbf{Syntactic}} &
                        \multirow{3}{*}{\textbf{20}} & 0.1 & 82.56 & 86.48 & 85.37 & 88.38 & 84.95\\
                        & & 0.5 & 71.32 & 73.15 & 71.73 & 72.16 & 74.04\\
                        & & 0.9 & 66.15 & 66.85 & 65.24 & 65.37 & 68.20\\
                    \cmidrule{2-8}
                        & \multirow{3}{*}{\textbf{60}} & 0.1 & 77.52 & 83.49 & 81.80 & 86.58 & 81.17\\
                        & & 0.5 & 67.83 & 71.61 & 69.58 & 71.43 & 71.77\\
                        & & 0.9 & 55.43 & 61.22 & 57.78 & 60.34 & 61.92\\
                    \cmidrule{2-8}
                        & \multirow{3}{*}{\textbf{80}} & 0.1 & 73.26 & 78.66 & 76.67 & 80.43 & 77.30\\
                        & & 0.5 & 64.34 & 71.24 & 68.17 & 72.49 & 70.32\\
                        & & 0.9 & 55.42 & 60.75 & 57.02 & 58.72 & 62.37\\
                    \midrule
                    \multirow{9}{*}{\textbf{Semantic}} &
                        \multirow{3}{*}{\textbf{20}} & 0.1 & 72.48 & 76.44 & 74.65 & 76.95 & 76.01\\
                        & & 0.5 & 67.05 & 69.02 & 67.45 & 67.84 & 70.07\\
                        & & 0.9 & 56.98 & 60.11 & 57.76 & 58.57 & 61.46\\
                    \cmidrule{2-8}
                        & \multirow{3}{*}{\textbf{60}} & 0.1 & 68.99 & 74.58 & 72.21 & 75.74 & 73.68\\
                        & & 0.5 & 61.63 & 66.42 & 63.73 & 65.98 & 66.78\\
                        & & 0.9 & 56.59 & 60.30 & 57.71 & 58.87 & 61.51\\
                    \cmidrule{2-8}
                        & \multirow{3}{*}{\textbf{80}} & 0.1 & 67.44 & 74.03 & 71.31 & 75.65 & 72.82\\
                        & & 0.5 & 59.69 & 67.16 & 63.51 & 67.84 & 66.67\\
                        & & 0.9 & 53.49 & 59.55 & 55.87 & 58.47 & 60.40\\
                    \midrule
                    \multirow{9}{*}{\textbf{Combined}} &
                        \multirow{3}{*}{\textbf{20}} & 0.1 & 72.48 & 75.88 & 74.21 & 76.02 & 75.77\\
                        & & 0.5 & 65.89 & 69.02 & 67.06 & 68.27 & 69.66\\
                        & & 0.9 & 58.91 & 61.78 & 59.61 & 69.32 & 63.07\\
                    \cmidrule{2-8}
                        & \multirow{3}{*}{\textbf{60}} & 0.1 & 69.77 & 72.36 & 70.73 & 71.71 & 72.92\\
                        & & 0.5 & 60.85 & 64.38 & 62.06 & 63.31 & 65.29\\
                        & & 0.9 & 53.88 & 54.92 & 53.36 & 52.85 & 56.88\\
                    \cmidrule{2-8}
                        & \multirow{3}{*}{\textbf{80}} & 0.1 & 67.05 & 72.73 & 70.18 & 73.62 & 72.04\\
                        & & 0.5 & 61.24 & 64.75 & 62.45 & 63.71 & 65.64\\
                        & & 0.9 & 52.71 & 57.88 & 54.51 & 56.43 & 59.06\\
                    \bottomrule
                \end{tabular}}
            \label{table:noise_sensitivity_results_mistral}
        \end{table}
        
\begin{figure}
            \centering
                \subfloat[Feature Importance (edit distance of 20)]{
                    \includegraphics[width=0.45\textwidth]{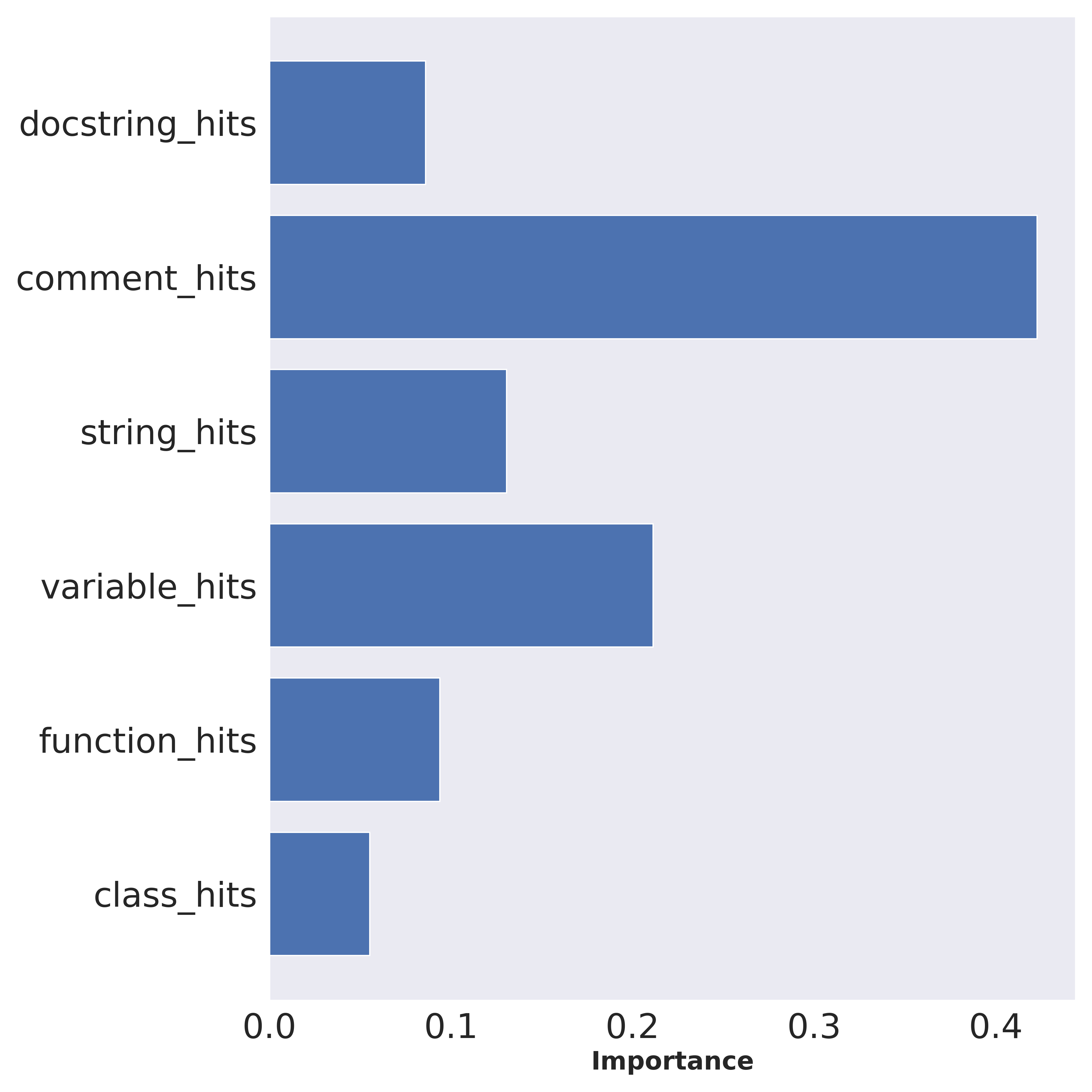}
                    \label{fig:feature_importance_sem20}
                }
                \hspace{0.05\textwidth} 
                \subfloat[Feature Importance (edit distance of 40)]{
                    \includegraphics[width=0.45\textwidth]{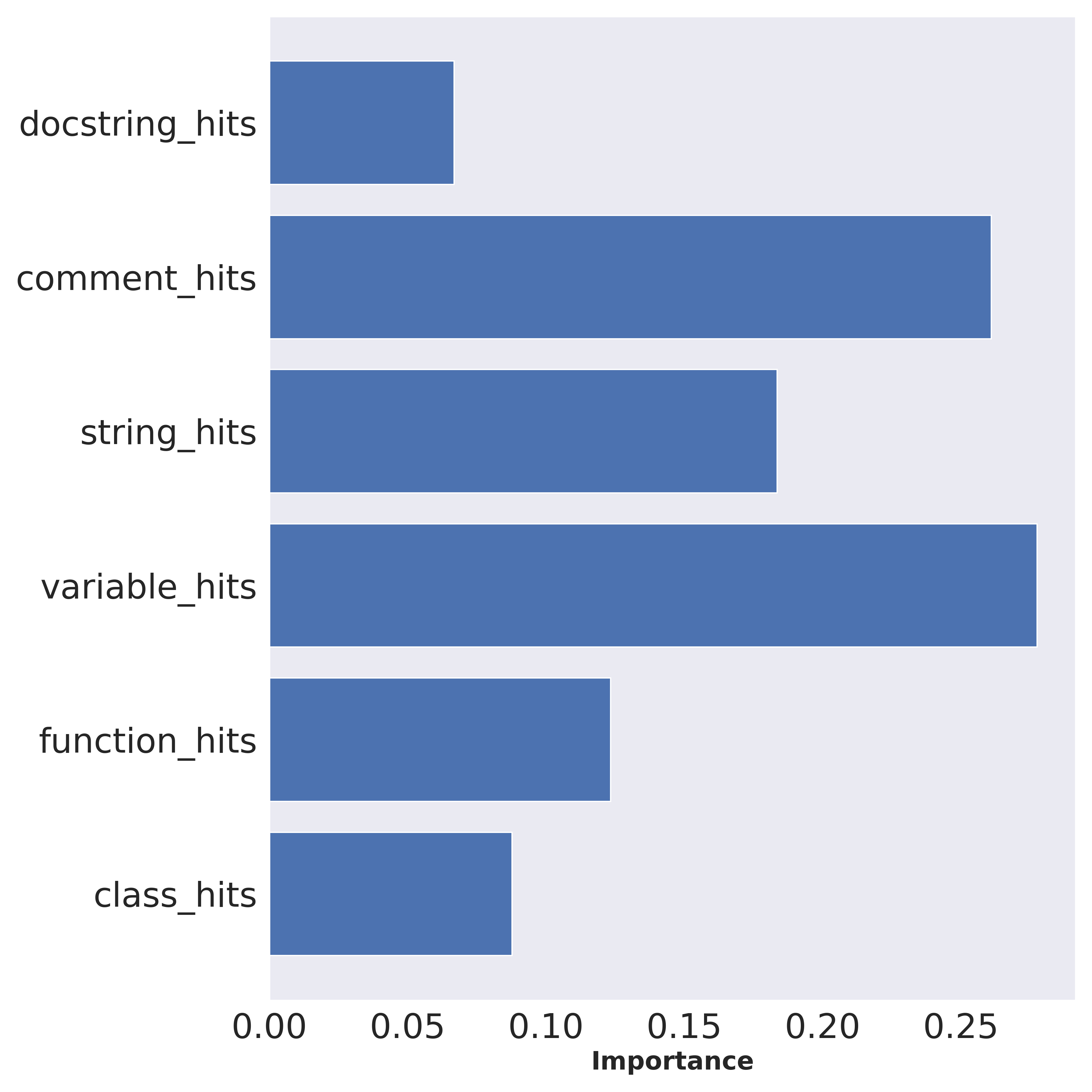}
                    \label{fig:feature_importance_sem40}
                }\\ 
                \subfloat[Feature Importance (edit distance of 60)]{
                    \includegraphics[width=0.45\textwidth]{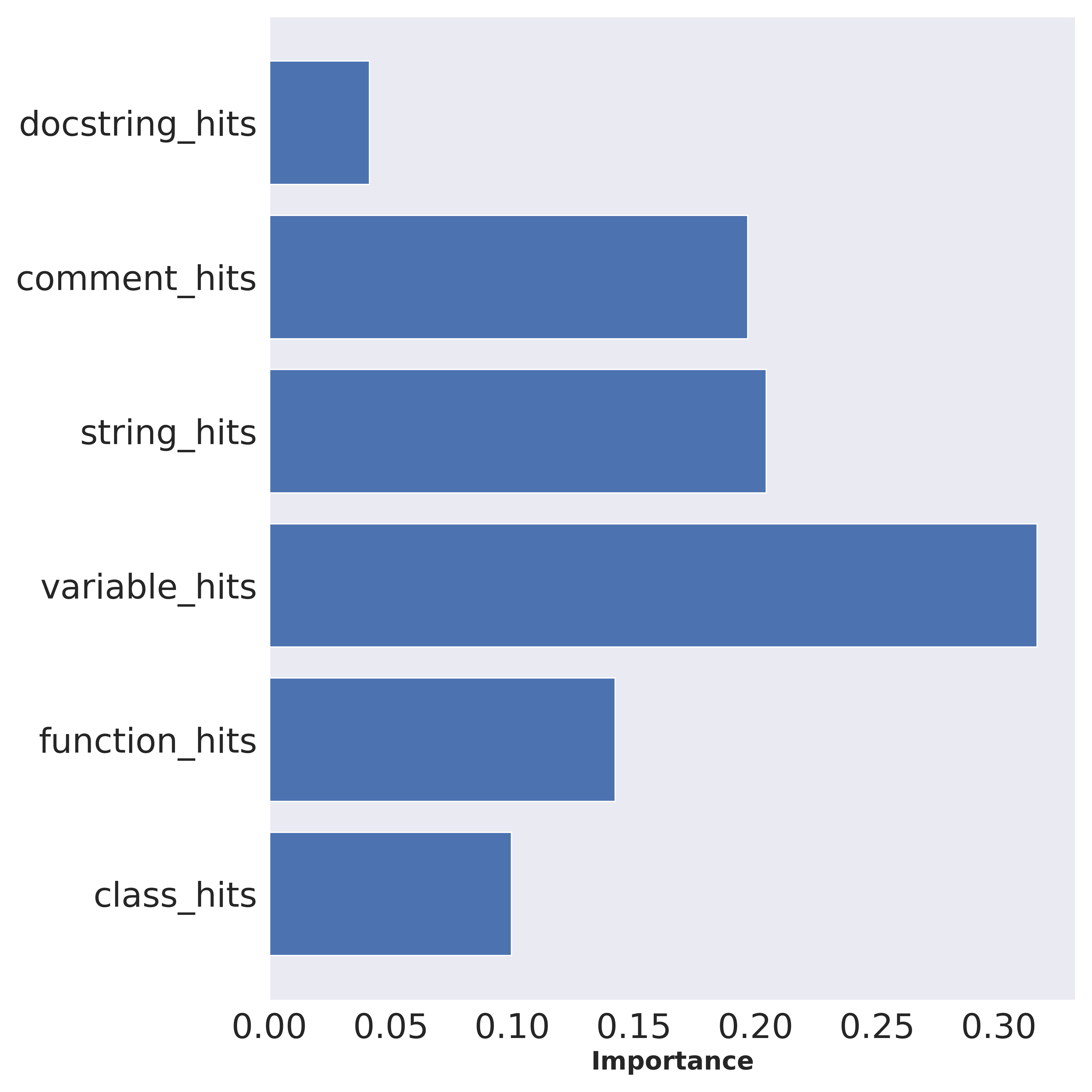}
                    \label{fig:feature_importance_sem60}
                }
                \hspace{0.05\textwidth} 
                \subfloat[Feature Importance (edit distance of 80)]{
                    \includegraphics[width=0.45\textwidth]{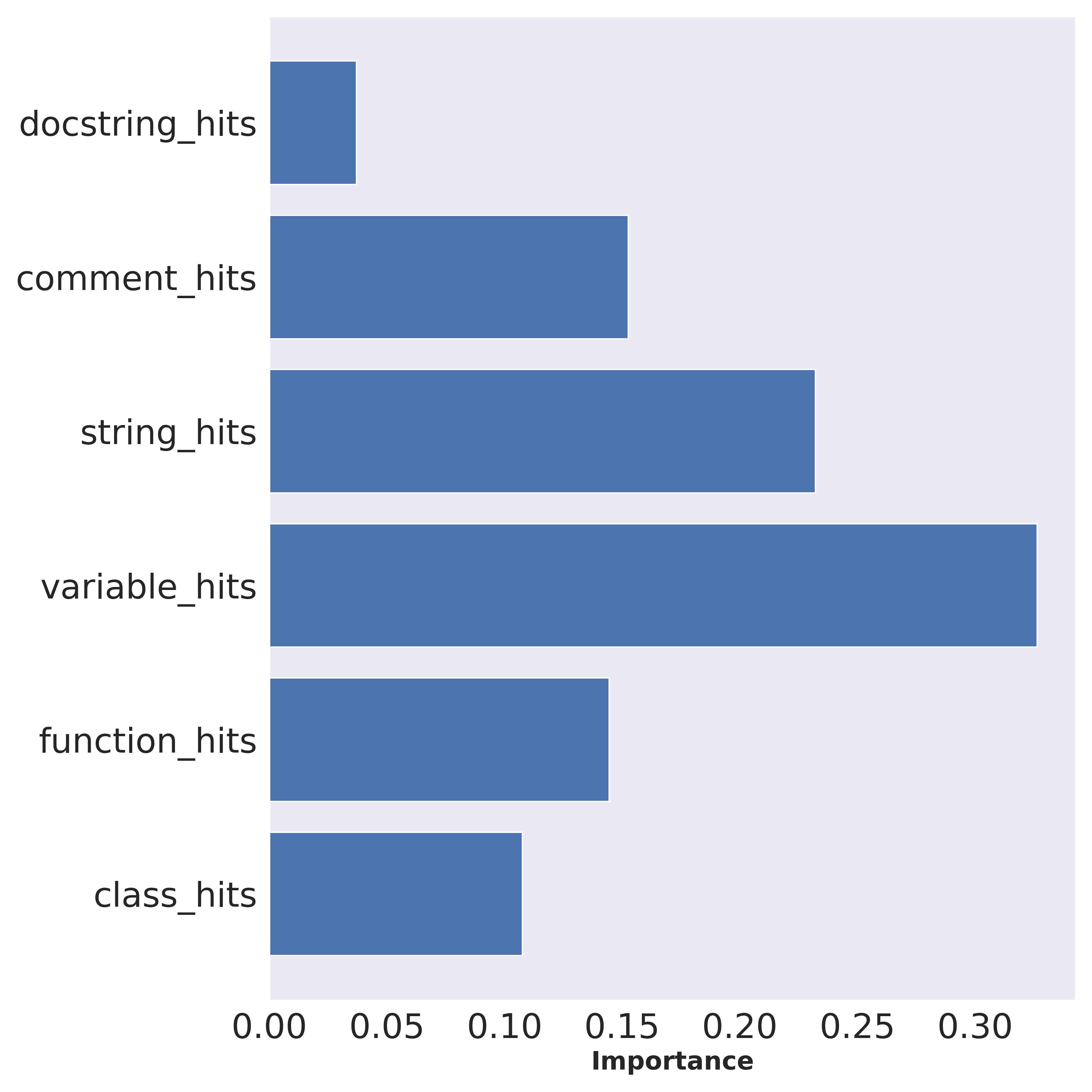}
                    \label{fig:feature_importance_sem80}
                }
            \caption{Feature importance of the final dataset with different edit distance thresholds - SantaCoder}
            \label{fig:results_feature_impact_feature_importance}
        \end{figure}

        Cross-analyzing the results of Tables \ref{table:noise_sensitivity_results_santacoder}, \ref{table:noise_sensitivity_results_llama}, and \ref{table:noise_sensitivity_results_mistral} with Tables \ref{table:edit_distance_sensitivity_results_repo_level}, \ref{table:edit_distance_sensitivity_results_repo_level_llama2}, and \ref{table:edit_distance_sensitivity_results_repo_level_mistral} which contain classifier's performance on clean data provides the following observations:    
        
        \begin{itemize}
            \item \textbf{Semantic identifiers are important in dataset inclusion detection}: We observe sharp degradation in the classifier's performance at higher edit distance thresholds. As shown in Figures \ref{fig:results_feature_impact_feature_importance_mistral_llama} and \ref{fig:results_feature_impact_feature_importance} which display the feature importance of our trained classifiers, we can observe that at lower edit distance thresholds, the classifier gives more weight to semantic identifiers compared to syntactic identifiers for all of the 3 models under study. Therefore, at lower edit distance thresholds, even though we observe performance degradation, the classifier has a high performance in detecting true positives even as the amount of noise is increased. As mentioned in Section \ref{subsec:background__code_clone_detection}, \textit{Type-2} code clones are similar to each other in syntax but have different variable/function/class names. Our results on simulating data obfuscation show that TraWiC can successfully detect dataset inclusion despite changing the syntactic identifiers. 
    
            \item \textbf{Considering a repository-level granularity for dataset inclusion detection can help against data obfuscation}: Comparing Tables \ref{table:edit_distance_sensitivity_results_repo_level}, \ref{table:edit_distance_sensitivity_results_repo_level_llama2}, and \ref{table:edit_distance_sensitivity_results_repo_level_mistral} with Tables \ref{table:noise_sensitivity_results_combined_total_santa}, \ref{table:noise_sensitivity_results_combined_total_llama}, and \ref{table:noise_sensitivity_results_combined_total_mistral}, we can observe degradation in detecting dataset inclusion in file-level granularity on higher edit distance thresholds. This decrease in performance is indicated by the decrease in sensitivity (correctly classified true positives) and the increase in specificity (correctly classified true negatives). As mentioned above, as a large number of the scripts in our constructed dataset have been included in the training dataset of the model under study, we cannot attribute the rise in specificity as an indicator of TraWiC's better performance for SantaCoder, and as we can observe from Tables \ref{table:noise_sensitivity_results_llama} and \ref{table:noise_sensitivity_results_mistral}, for both Mistral and Llama the performance of the classifier degrades as the amount of noise is increased. However, even though we observe degradation in performance while considering repository-level granularity as well, we observe that analyzing all the scripts in a project can help us detect dataset inclusion even in the presence of high levels of noise for all of the 3 models under study. Depending on the approach used for code obfuscation, obfuscating the entire project is either too costly or adds unnecessary performance overhead to the project's operation \cite{zhuang2018performance, bunse2018impact, sebastian2016study} and many of the proposed approaches focus on obfuscating blocks of code inside the script as opposed to its entirety \cite{kang2021obfus, zhang2023khaos}. As such, even though the identifiers in one script in the project may be changed, other scripts in the same project may remain unchanged or undergo lower amounts of obfuscation, therefore, aiding in dataset inclusion detection.

        \end{itemize}
        
        \begin{tcolorbox}[colback=blue!5,colframe=blue!40!black]
            \textbf{Findings 3:} Data obfuscation techniques can be used to prevent MIAs from detecting dataset inclusion. Our results show that TraWiC is capable of detecting dataset inclusion when a moderate level of noise is applied to the training data. Moreover, we show that by analyzing all the scripts in a project, we can detect dataset inclusion with an F-score of up to 83.15\%, even when half of semantic and syntactic identifiers in code are obfuscated.
        \end{tcolorbox}
    
    \subsubsection{\textbf{RQ2b}: What is the importance of each feature in detecting dataset inclusion?}\label{subsubsec:results_rq2b}
        We first analyze the constructed dataset for training the classification models using Spearman's rank correlation coefficient \cite{zar2005spearman} to assess the relationship between the constructed features of the dataset. 
        Then, we analyze the classification models' feature importance using the gini importance criterion \cite{breiman2017classification}.
        
        \begin{figure}[h]
            \centering
            \subfloat[\centering SantaCoder \label{fig:correlation_matrix__santacoder}]{{\includegraphics[width=0.45\textwidth, valign=c]{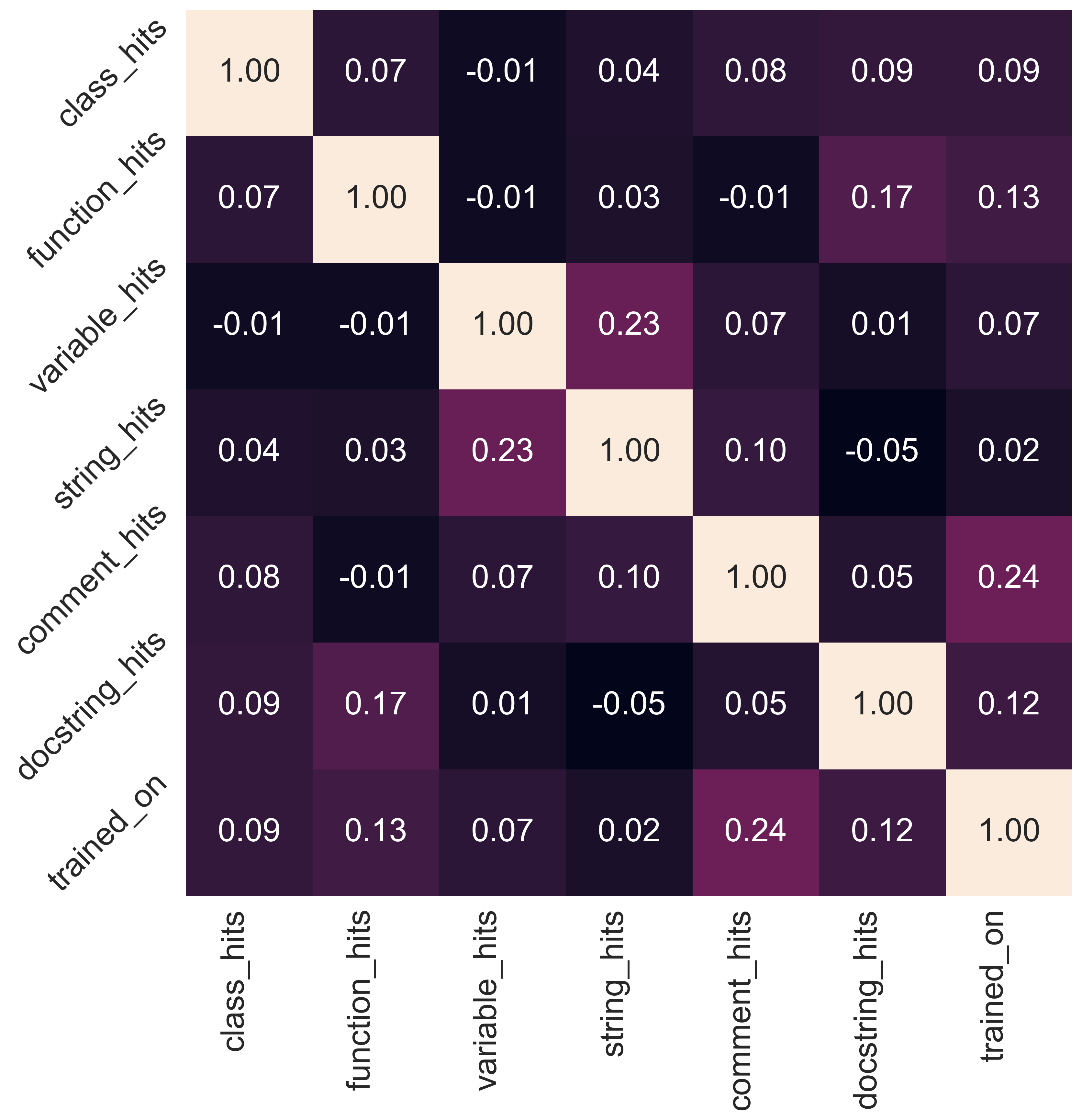} }}%
            \qquad
            \subfloat[\centering Llama-2 \label{fig:correlation_matrix__llama}]{{\includegraphics[width=0.45\textwidth, valign=c]{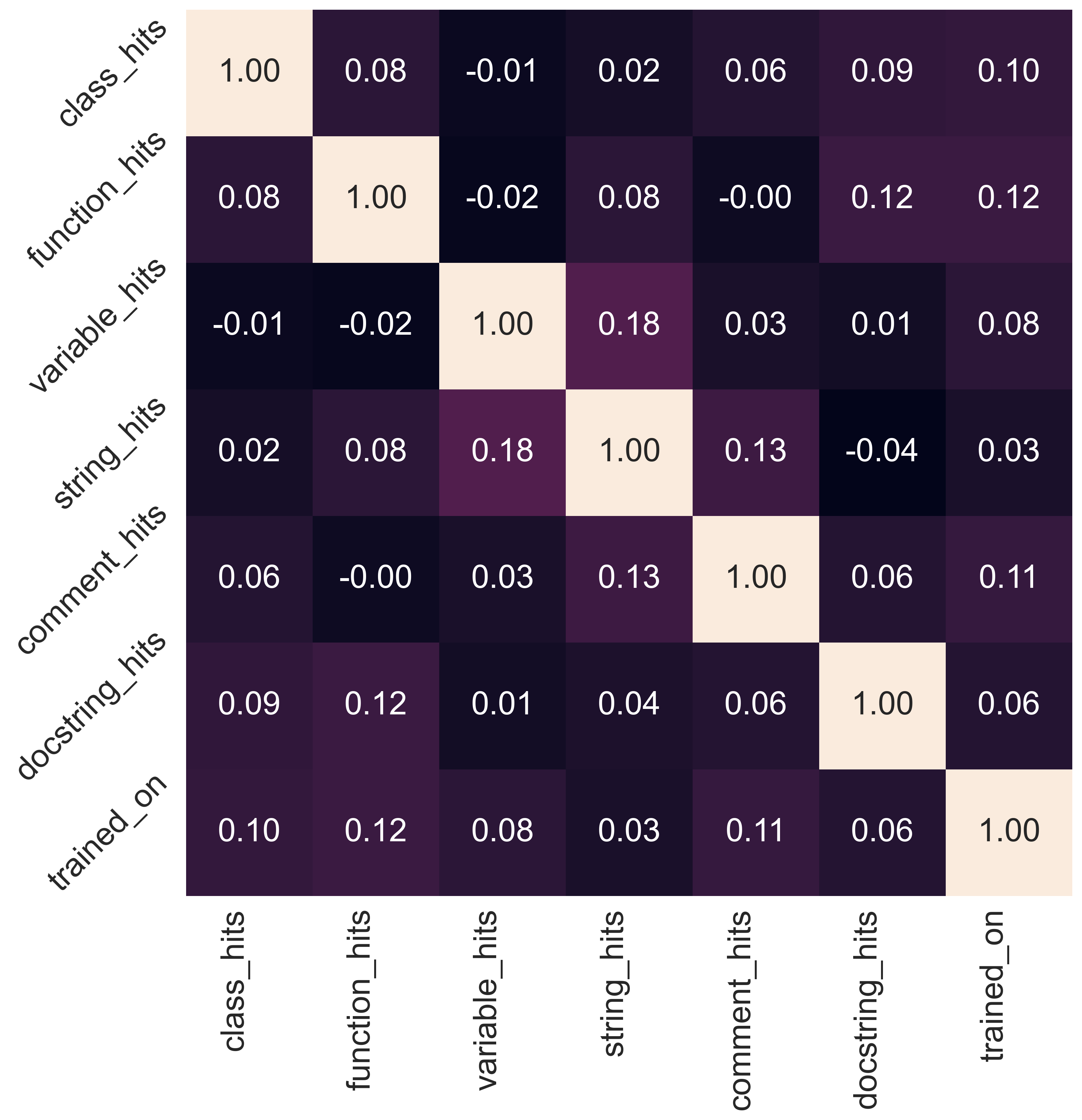}}}%
            \qquad
            \subfloat[\centering Mistral \label{fig:correlation_matrix__mistral}]{{\includegraphics[width=0.45\textwidth, valign=c]{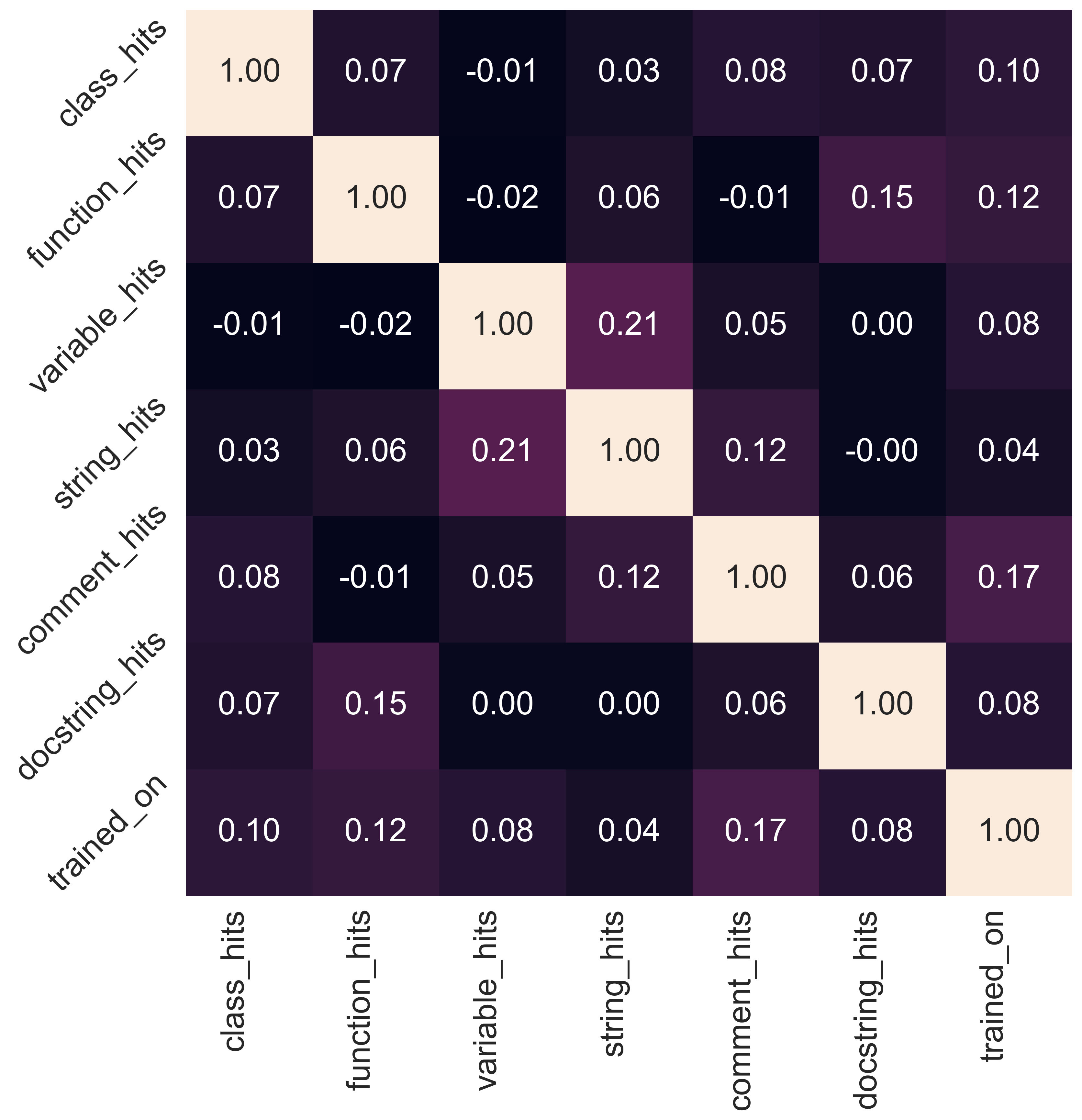}}}%
            \caption{Correlation of different features in the generated dataset}%
            \label{fig:correlation_matrix_all_models}%
        \end{figure}
        
        Figure \ref{fig:correlation_matrix_all_models} displays the correlation analysis of the extracted features with \underline{\texttt{trained\_on}} being the dependent variable indicating a script's inclusion in the LLM's training dataset for each of the models under study. We observe weak to no correlation between different features which indicates low dependency among features and that each feature provides unique information for the classification task. 
        As displayed in Tables \ref{table:edit_distance_sensitivity_results_repo_level}, \ref{table:edit_distance_sensitivity_results_repo_level_llama2}, and \ref{table:edit_distance_sensitivity_results_repo_level_mistral}, we experiment with multiple edit distance thresholds for counting semantic hits. Even though the performances of models for different edit distance thresholds are similar, there are variations in the classifiers' feature importance. Figure \ref{fig:results_feature_impact_feature_importance} displays different feature importance distributions for detecting dataset inclusion for SantaCoder for edit distances of 20, 40, 60, and 80, respectively. As displayed in this figure, the lower the edit distance threshold is, the more the importance of \textit{docstring} and \textit{comment} features for deciding code inclusion increases. Comments and docstrings are specifically designed to explain the code's functionality, as they are written to explain pieces of code in natural language, which makes them inherently unique. Therefore, with lower edit distances, which lowers the matching criterion between the original tokens and what the model generates, the number of hits for semantic elements increases, and as a result, the importance of these features increases.

        Figure \ref{fig:feature_importance_sem60} displays the feature importance of the best-performing classifier (i.e., the random forest with an edit distance of 60 and repository inclusion criterion of 0.4) for SantaCoder. The number of variable hits is the most significant feature, followed by the number of comment hits and function hits. The number of docstring hits has the least importance, which suggests that the information within the docstrings of the code has a minimal impact on the classifier's decisions. It should be noted that compared to other semantic identifiers, docstrings are longer (as they explain the functionality of a class alongside its inputs and outputs) and less prevalent (one per function/class). Therefore, there exists a smaller number of hits for docstrings compared to the other semantic identifiers.

        \begin{figure}
            \centering
                \subfloat[Feature Importance (edit distance of 20 - Llama)]{
                    \includegraphics[width=0.45\textwidth]{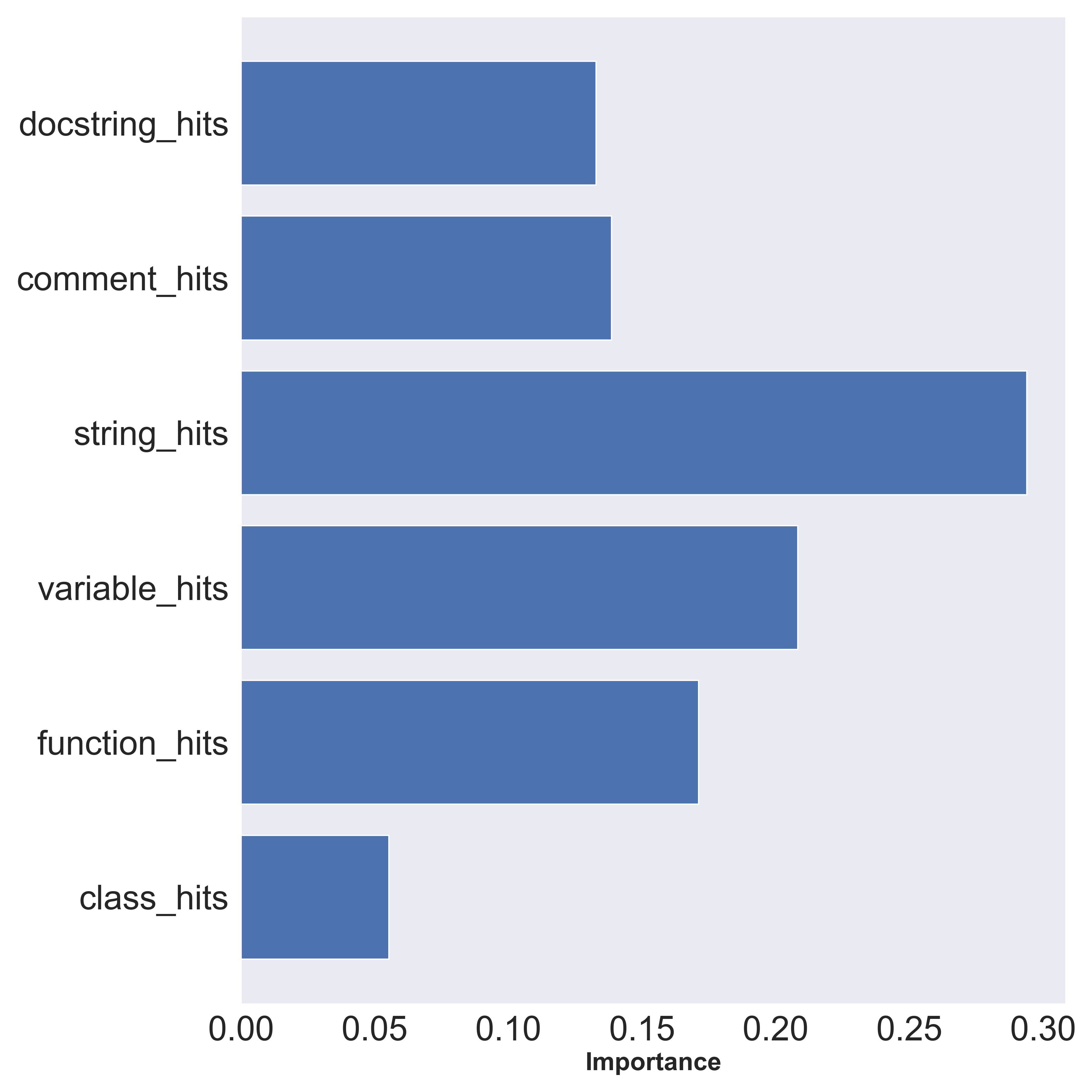}
                    \label{fig:feature_importance_sem20_llama}
                }
                \hspace{0.05\textwidth} 
                \subfloat[Feature Importance (edit distance of 20) - Mistral]{
                    \includegraphics[width=0.45\textwidth]{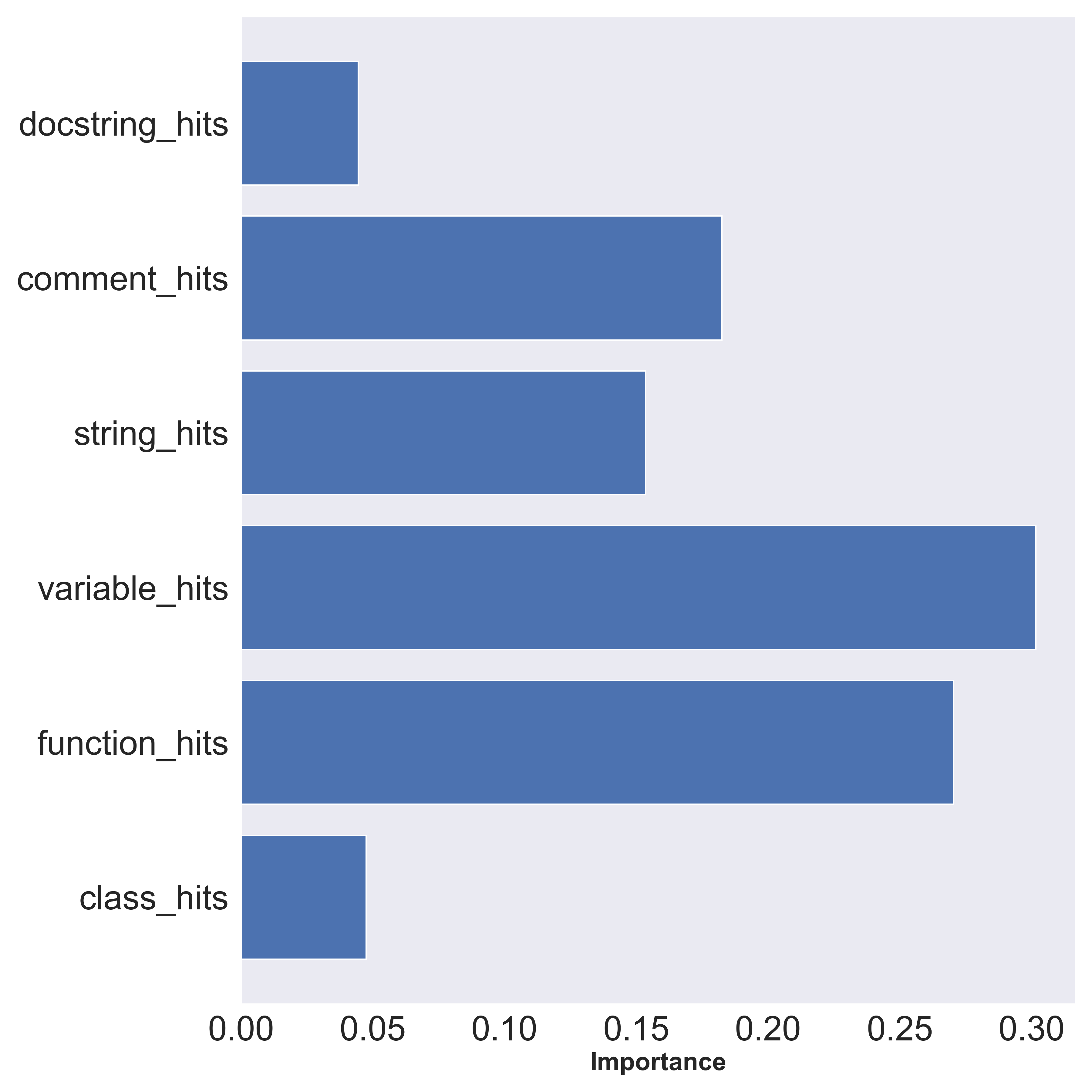}
                    \label{fig:feature_importance_sem20_mistral}
                }\\ 
                \subfloat[Feature Importance (edit distance of 60) - Llama]{
                    \includegraphics[width=0.45\textwidth]{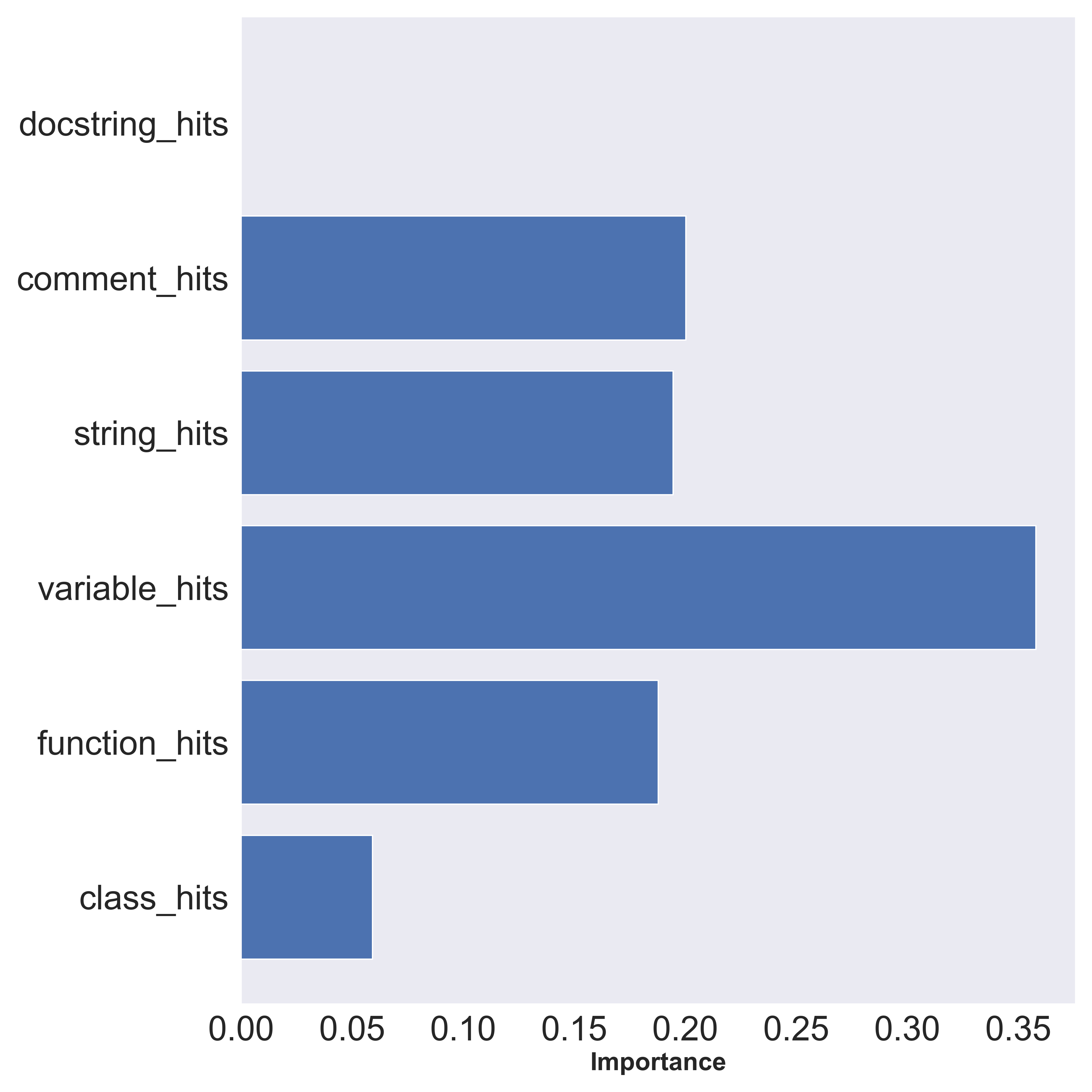}
                    \label{fig:feature_importance_sem60_llama}
                }
                \hspace{0.05\textwidth} 
                \subfloat[Feature Importance (edit distance of 60) - Mistral]{
                    \includegraphics[width=0.45\textwidth]{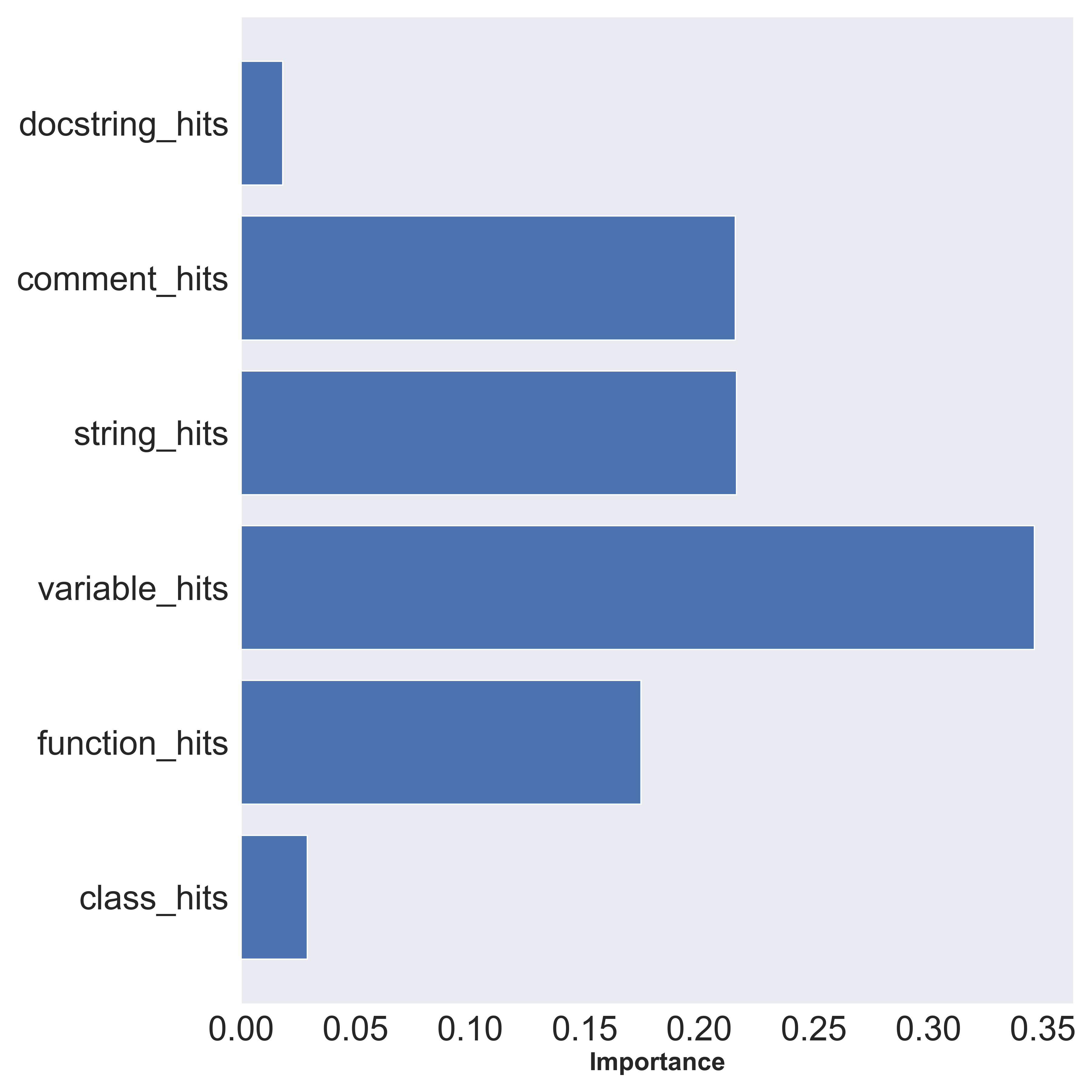}
                    \label{fig:feature_importance_sem60_mistral}
                }
                
            \caption{Feature importance of Llama-2 and Mistral with different edit distance thresholds}
            \label{fig:results_feature_impact_feature_importance_mistral_llama}
        \end{figure}
        
        We observe different feature importance and correlations of features for the fine-tuned models of Mistral and Llama compared to SantaCoder, as displayed in Figure \ref{fig:results_feature_impact_feature_importance_mistral_llama}. Figure \ref{fig:feature_importance_sem20_llama} shows that for an edit distance of 20 for Llama-2, the number of string hits is the most significant feature, followed by variable hits and function hits. In Figure \ref{fig:feature_importance_sem60_llama}, which presents the feature importance for an edit distance of 60 for Llama-2, the number of variable hits becomes the most significant feature, followed by comment hits and string hits. It should be noted that the feature importance of comment, string, and function hits is relatively the same at this edit distance. However, at both edit distances, the number of class hits has the least importance (excluding the docstring hits for an edit distance of 60). This is due to the relatively fewer occurrences of classes in scripts within the analyzed dataset, which is similar to the analysis carried out for SantaCoder. Figure \ref{fig:feature_importance_sem20_mistral} and \ref{fig:feature_importance_sem60_mistral} display the feature importance of the classifier at edit distances of 20 and 60 for Mistral, respectively. At an edit distance of 20, the number of variable hits is the most significant feature, followed by function hits and comment hits. The number of class hits is the least important, suggesting that the information within class definitions has minimal impact on the classifier’s decisions, similar to Llama-2 and SantaCoder. At an edit distance of 60, the number of variable hits remains the most significant feature, followed by string hits and comment hits. The consistency across different semantic thresholds for all three models shows that these features can be used as strong indicators for dataset inclusion in LLMs trained on code. An important observation across all three models is the relatively low importance of docstring hits in the classifier’s ability to detect dataset inclusion. We observe that as we increase the edit distance threshold, the number of docstring hits decreases, indicating that this low contribution results from low edit distance scores associated with docstrings. We discuss the reasons behind this in RQ3.

\subsection{\textbf{RQ3 [Error Analysis]}}\label{subsec:results__rq3}
    In this RQ, we investigate how the models under study make mistakes in predicting the masked element, and how these mistakes can be used for the task of dataset inclusion detection. Here, we define a ``mistake'' as the model's inability to generate the \textbf{exact} same token as what was present in the script. There exists a difference between the mistakes made by SantaCoder and our fine-tuned models, as such we will explain each as follows.
    
    \subsubsection{SantaCoder}
        In Python, variable/function/class names cannot contain any white spaces between them and are considered a single token; therefore, we focus on these elements as the model's task is predicting a single token. We observe that in most cases when the model makes a mistake, it generates the correct variable/function/class name however it also generates some extra tokens as well. Listing \ref{listing:results_analysis__mistake_example} shows an example of such cases for the variable name \underline{\texttt{product\_to\_encode}}. When the model makes a mistake for syntactic identifiers, it generates an average of 23 tokens regardless of whether the underlying script was in its training dataset or not. Therefore, to understand these mistakes better, we first remove the tokens that are either a part of the programming language's syntax or too frequent. To do so, we count the repetition frequency of each token and normalize them by dividing their repetition frequency by the total number of observed tokens. Afterward, we remove the tokens that are repeated too many times across the scripts in the dataset from the generated mistakes.
        
        The results of this analysis are presented in Table \ref{table:results__error_analysis__prediction_token_type} with the ``Filtration Threshold'' column outlining the threshold for keeping a token. A filtration threshold of 0.5 means that only tokens appearing with a frequency less than or equal to 50\% of all the observed tokens are retained, while tokens with a higher frequency are filtered out during the analysis.

        \begin{lstlisting}[style=pythonstyle, caption={A case of model's predictions for scripts not in its training data}, label={listing:results_analysis__mistake_example}]
            product_to_encode = pickle.load(open('./data/product_1.pickle', 'rb'))
            mock = OpenDataCubeRecordsProviderMock()
            encoded_product = mock._encode_dataset_type_properties(product_to_encode)
            assert 'id' in encoded_product.keys()
            id = encoded_product.get('id')
            assert id is not None
            assert id == product_to_encode.name
            
            def test__encode_dataset_type_properties():
                product_to_encode
        \end{lstlisting}
        
            A notable variance is observed in the class names generated by the model, with an approximate difference of around 6\%, depending on whether a script was part of its training dataset or not. It is important to note that classes are less recurrent than functions and variables within code. This difference (regardless of filtration criterion) shows that MIAs, particularly when directed at rare textual elements, can serve as a means to discern the data used in a model's training.
            
            \begin{table}
                \caption{Analysis of Model's outputs by token type for scripts that are included in its training dataset}
                \resizebox{\textwidth}{!}{
                    \begin{tabular}{c c c c}
                    \toprule
                    Filtration Threshold & Token Type & Output Contains Token (Script in Dataset)(\%) & Output Contains Token (Script \textit{\textbf{not}} in Dataset)(\%) \\
                    \midrule
                        \multirow{3}{*}{\textbf{0.5}} & Variable Names & 60.71 & 61\\
                        & Function Names & 14.08 & 11.77\\
                        & Class Names & 11.18 & 5.02\\
                    \midrule
                        \multirow{3}{*}{\textbf{0.3}} & Variable Names & 60.89 & 61.38\\
                        & Function Names & 15.37 & 12.9\\
                        & Class Names & 11.3 & 5.08\\
                    \midrule
                        \multirow{3}{*}{\textbf{0.1}} & Variable Names & 60.89 & 60\\
                        & Function Names & 15.37 & 11.99\\
                        & Class Names & 11.3 & 5 \\
                    \bottomrule
                    \end{tabular}
                    }
                \label{table:results__error_analysis__prediction_token_type}
            \end{table}
            
            \begin{tcolorbox}[colback=blue!5,colframe=blue!40!black]
            \textbf{Findings 4:} Even in cases where the model's output is not adhering to our \textbf{exact} matching criteria, in roughly 60\% of cases its outputs contain elements from the code that it was trained on. These results complement Finding 1 and show that MIAs directed at parts of code that are not grammatical (syntax of the programming language itself), can help identify dataset inclusion.
            \end{tcolorbox}

    \subsubsection{Llama-2 and Mistral}
        As described in Section \ref{subsubsec:results_rq2b}, for both fine-tuned versions of Llama-2 and Mistral, we observe that the number of docstring hits decreases as we increase the edit distance threshold. This indicates that there is already a low similarity between the models’ generated content for these elements and their original contents. To understand this better, we investigate the outputs of our fine-tuned models against the original targets. When it comes to generating long sequences such as docstrings, the models often fail to produce coherent outputs. They either generate sequences that are close to the target but contain additional tokens, reducing the similarity score, or they generate unrelated tokens. Listing \ref{listing:results_analysis__mistake_example_fim} shows an example of cases where the model generates special tokens. As described in Section \ref{subsubsec:results__finetuning}, Llama-2 and Mistral do not natively support the FIM objective, requiring us to fine-tune the models for this purpose. As shown in Listing \ref{listing:prompt_format}, \texttt{<fim\_prefix>} is one of the special tokens added to the models’ tokenizers for this task. During the early stages of fine-tuning, these special tokens appear frequently in the models’ outputs. However, as training progresses, their occurrence decreases, suggesting that further fine-tuning on a larger dataset would minimize these instances. Nevertheless, as shown in Listing \ref{listing:results_analysis__mistake_example_fim}, the presence of such extra tokens can result in higher edit distance scores, leading to more mistakes.

        \begin{lstlisting}[style=pythonstyle, caption={A case of fine-tuned models generating special tokens}, label={listing:results_analysis__mistake_example_fim}]
                    Target: 'SEARCH_ITERATIONS'
                    Model_output: '<fim-prefix> SEARCH_ITERATIONS'
        \end{lstlisting}

        On the other hand, in some cases, the developers have added some comments to the code in order to address or reference some of the maintenance activities such as URLs linking to related issues in the repository, as displayed in Listing \ref{listing:results_analysis__mistake_example_address}. In such cases, both models generated somewhat similar URLs but failed to generate the rest of the docstrings as intended. It is important to note that the URL generated in Listing \ref{listing:results_analysis__mistake_example_address} is incorrect and does not exist. However, "Surya" is one of the repositories included in our fine-tuning dataset\footnote{https://github.com/VikParuchuri/surya}, and we observe that the model generates a URL which, although invalid, includes the repository’s name, indicating that the model has encountered this repository during training. Our findings line up with the results reported by Carlini et al. \cite{carlini2022quantifying} where they show that MIAs on URLs, phone numbers, and names can aid in identifying memorization in LLMs.

        \begin{lstlisting}[style=pythonstyle, caption={A case of fine-tuned models generating the target with extra tokens}, label={listing:results_analysis__mistake_example_address}]
                Target: 'You should install the dependencies'
                Model_output: 
                    print("You probably should install the dependencies.") 
                    print("https://github.com/surya-ai/surya/blob/master/docs/installation.md")
        \end{lstlisting}

        Finally, similar to SantaCoder, we observe that in some cases, the models generate the target we are looking for, as shown in Listing \ref{listing:results_analysis__mistake_example_long}. However, due to the presence of additional tokens, the edit distance score is high. Consequently, by increasing the edit distance threshold, such cases are filtered out, resulting in few or no hits for docstrings.

        \begin{lstlisting}[style=pythonstyle, caption={A case of fine-tuned models generating the target with high edit distance}, label={listing:results_analysis__mistake_example_long}]
                Target : 'Action {action} is illegal'
                Model_output:
                    def step_from_state(self, action):    
                        if self.state[action] != 0:
                            raise ValueError(f"Action {action} is illegal")
                            self.state[action] = self.turn
                            self.actions_stack
        \end{lstlisting}

        The analysis of the mistakes made by TraWiC for predicting the inclusion of such occurrences highlights some limitations of our proposed approach. Therefore, our future work aims to design a more robust similarity objective for semantic elements. This would enable the detection of cases such as these, ultimately improving the accuracy of dataset inclusion detection for models trained on code.

\section{Related Works}\label{sec:related_works}
    In this section, we review the relevant literature. Specifically, we present related works that focus on determining dataset inclusion detection for code models. 

The most relevant work to our approach is the work done by Choi et al. \cite{choi2023tools} which proposes an approach for verifying whether a particular model was trained on a particular dataset. Their approach consists of investigating multiple checkpoints produced during a model's training procedure. They show that if a model was trained on a dataset as reported by the model's creators, then the model's weights would ``overfit'' on the training dataset in the next checkpoint compared to a dataset that was not included in the training data. Even though their approach shows promising results, it requires having access to the model's architecture, weights, hyperparameters, and training data. In comparison, TraWiC only requires query access to the given model which is the case for most of the code LLMs offered by enterprises.

Zhang et al. \cite{zhang2023code} proposed a framework for dataset inclusion detection on Code Pre-trained Language Models (CLPMs). In this work, the authors consider multiple levels of access to a CLPM, namely: 
\begin{enumerate}
    \item Complete access to the model and a large fraction of its training dataset (i.e., white-box access).
    \item Complete access to 
    the model's architecture and a small fraction of its training dataset (i.e., gray-box access).
    \item No access to 
    either the model or the training dataset (i.e., black-box access).
\end{enumerate} 
For each level of access, they provide different detection mechanisms and their black-box level access is the setting that is comparable with TraWiC. 
For dataset inclusion detection given only black-box access to the model, the authors provide two methods, namely, unimodal and bimodal calibration. In unimodal calibration, the code is perturbed and the differences in the model's output representations for different versions (e.g., uppercase vs lowercase) of the code snippet are observed. A calibration model, based on these differences, is used to infer the membership status of the code. In bimodal calibration, the calibration model's outputs are compared to the CPLM's outputs for the same input. This comparison is focused on identifying disparities between the two outputs, which are indicative of whether a given code snippet was used in training the CPLM or not. This method relies on the unique interaction between the natural language and programming language aspects of CPLMs, to make inferences about membership. In comparison to TraWiC, the approach proposed by Zhang et al. is focused on CLPMs that produce embeddings for downstream tasks and require expensive training of multiple large ML components (i.e., calibration models) while TraWiC is focused on LLMs that generate code. Furthermore, TraWiC is not limited to pre-trained models and does not contain any resource-intensive ML components outside of inference on the LLM itself.

Finally, Yang et al. \cite{yang2023gotcha} examined the risk of membership leakage in LLMs trained on code. In their approach, they assume that a fraction of the data used to train an LLM on code is available. They train a surrogate model on both said fraction and another dataset that was not included in the original model's training. The surrogate model is supposed to mimic the behavior of the original model. Afterward, they train an MIA classifier based on the outputs of the surrogate model as an input to detect dataset inclusion. They study the results of varying degrees of membership leakage risks depending on factors such as the attacker's knowledge of the victim model, the model's training epochs, and the fraction of the training data that is known to the attacker. In their approach, they consider that at least a part of the training data of a model is available and it involves resource-intensive training of a surrogate model. In comparison, the main resource-intensive component of TraWiC is inference on the LLM and it does not require re-training of a surrogate model or access to the original training dataset. 

\section{Limitations}\label{sec:limitations}
    In this section, we describe and discuss the limitations of our approach in detecting dataset inclusion and possible approaches for overcoming them.

As discussed in Section \ref{subsec:results__rq1}, we observe that for larger models, namely Mistral and Llama-2, TraWiC's performance degrades when the edit distance threshold for considering semantic similarity between the models' outputs and the target tokens is increased. According to our sensitivity analysis in Section \ref{subsec:results__rq2}, TraWiC’s performance degrades to the level of a random classifier when the dataset is heavily obfuscated (obfuscation of more than half of syntactic and semantic variables). By analyzing the causes behind these failures in Section \ref{subsec:results__rq3}, we conclude that the main issue lies in TraWiC's ability to detect semantic similarity. This is not surprising, as discussed in Section \ref{subsubsec:results_rq2b}, since TraWiC focuses on the number of semantic element hits to detect dataset inclusion. Listings \ref{listing:results_analysis__mistake_example_fim} and \ref{listing:results_analysis__mistake_example_long} show that most failures are due to high edit distances between the targets and the models’ predictions, even when the element of interest is present in the predictions. Therefore, we believe that by designing more robust and complex methods for similarity detection on semantic elements, we could detect these instances, reduce false negatives, and improve TraWiC’s performance in such cases.

Another limitation of our approach is that it requires some ground truths to train the initial classifier for dataset inclusion detection. As discussed in Sections \ref{sec:introduction} and \ref{subsubsec:results__model_under_study},  developers of LLMs rarely provide complete information about the training datasets of their models, opting instead for general descriptions \cite{touvron2023llama, jiang2023mistral}. However, based on the specific model under study, we can infer some priors about the dataset used to train the model from the developers' descriptions and reports or use other MIA approaches in order to construct the initial dataset for training the classifier.

Finally, there is a requirement for the models under study to support the Fill-in-the-Middle (FIM) objective. The FIM task is a common step in the pre-training procedures of LLMs \cite{allal2023santacoder, touvron2023llama, jiang2023mistral, devlin2018bert, anil2023palm}. However, as shown with some of the models under study, the released versions by the developers may not natively support this task. Therefore, studying such models for dataset inclusion using TraWiC would require either fine-tuning these models to support the FIM task or having access to a version that already does.
    
\section{Threats to Validity}\label{sec:threats_to_validity}
    In this section, we report on threats to the validity of our study. 

\textbf{Threats to internal validity} concern factors internal to our study, that could have influenced our study. Our reliance on MIAs as the primary mechanism for detecting code inclusion can be considered a potential threat to the validity of our research. MIAs have been successfully used in other contexts, and our results show that leveraging MIAs is also an effective approach for detecting code inclusion. Additionally, our choice of comparison criteria for syntactic/semantic elements might not encompass all aspects of the code inclusion detection task. We mitigate this threat by experimenting with multiple values of edit distance and conducting a rigorous sensitivity analysis to show the effectiveness of our approach. Finally, the bias introduced by the randomness in sampling a small portion of the original datasets of the LLMs under study can be considered as a potential threat. While we conducted multiple experimental runs and fixed hyperparameters to ensure consistency and reproducibility, we did not perform statistical significance tests given the large size of the original dataset and the resources available to us. We mitigate this threat by conducting our study on multiple LLMs and separate datasets. However, future work should include multiple experimental runs and statistical tests to better account for and mitigate the effects of random bias.

\textbf{Threats to construct validity} concern the relationship between theory and observation. The first threat to our construct validity involves our experimental design. To mitigate this threat and to ensure that the results of code inclusion detection were not only valid for certain scripts and projects we ran our approach using random sampling of both scripts and projects from the generated dataset to validate the results. We acknowledge that the results of detecting code inclusion using NiCad and JPlag may vary as the underlying dataset is extremely large and a pair-wise comparison across the entire dataset is not feasible. We mitigate this issue by increasing the percentage of the sampled dataset to have a comprehensive analysis and use two popular token-based code clone detection approaches. Finally, there exists the threat that our fine-tuning of the models is selective based on the model under study. We address this threat by fine-tuning two models on the same dataset until the training loss on both models stops improving. We include the details of our fine-tuning regiment in Section \ref{sec:appendix}. 

\textbf{Threats to External Validity} concern the generalization of our findings. We identify the replicability of our results as the most important external threat. Given that replicating a certain output on LLMs is difficult, we release the generated dataset, alongside TraWiC's code \cite{reppackage}. Moreover, there exists the possibility that our approach may not perform as reported on other code models. Since SantaCoder's introduction, many larger code models with higher performance have been introduced. There exist larger and more capable models such as other versions of Llama-2 with higher numbers of parameters and Llama-3. It should be noted that the majority of these models are so large that they require computing resources that are not available outside of the enterprise. Furthermore, the training datasets of these large models are not available. Given such limitations, SantaCoder, Mistral 7B, and Llama-2 7B were the only feasible options for this study. Considering that as a model's capacity increases so does its internal ability to remember instances from its training dataset \cite{carlini2022quantifying} and that our approach is model-agnostic, it is very likely that our approach would also perform well even on larger models.

\section{Conclusion}\label{sec:conclusion}
    In this study, we introduced TraWiC, a model-agnostic, interpretable approach for detecting code inclusion. TraWiC exploits the memorization ability of LLMs trained on code to detect whether codes from a project (collection of codes) were included in a model’s training
dataset. Evaluation results show that 
%
TraWiC can detect whether a project was included in an LLM's training dataset with a recall of up to 99.19\%. Our results also show that TraWiC 
significantly outperforms code clone detection approaches in identifying dataset inclusion and is robust to noise. 
In the future, we plan to test TraWiC on more capable LLMs trained on code. We also plan to investigate what other aspects of code can be used for conducting MIAs on a code model. We aim to use deep reinforcement learning to train a model that can detect dataset inclusion based on the outputs of a code model by comparing the AST representation of the generated output to the AST representation of the input. Doing so would considerably lower TraWiC's performance bottleneck and allow for constructing an end-to-end solution for dataset inclusion detection which is more robust against code obfuscation. 

\section{Acknowledgments}\label{sec:acknowledgments}
    This work is partially supported by the Fonds de Recherche du Quebec (FRQ), the Canadian Institute for Advanced Research (CIFAR), and the Natural Sciences and Engineering Research Council of Canada (NSERC).

\bibliographystyle{ACM-Reference-Format}
\bibliography{sample-base}

\section{Appendix}\label{sec:appendix}
    \subsection{Grid search parameters for classifiers}
        Here, we present the grid search parameters that were used to train the best classifiers. It should be noted that the best classifier was selected based on the highest F-score achieved on the test set. All scripts for training these models are available at \cite{reppackage}.  
        \subsubsection{Random forest}
            \begin{itemize}
                \item Number of estimators: This hyperparameter is used to indicate the number of trees in the forest. We used values of 50, 100, and 200.
                \item Number of features for best split: The number of features to consider when looking for the best split. We used the values ``sqrt'' indicating the square root of the number of features in the input; and ``log2'' indicating the binary logarithm of the number of features in the input.
                \item Max tree depth: Maximum depth allowed for a single tree in the forest. we used the values of 10, 20, and 30.
                \item Split criterion: The function to measure the quality of a split. With ``gini'' indicating Gini impurity and ``entropy'' indicating Shannon information gain.
            \end{itemize}

        \subsubsection{SVM}
            \begin{itemize}
                \item C: The regularization parameter with the strength of regularization being inversely proportional to C. We used values of 0.1, 1, 10, 100.
                \item Kernel: Specifies the kernel type to be used in the algorithm. We used radial basis and linear functions.
                \item Kernel Coefficient: The kernel coefficient is a parameter that determines the shape of the kernel function used to map the input data into a higher-dimensional feature space. We used values of 1 , 0.1, 0.01, and 0.001. 
            \end{itemize}

        \subsubsection{XGB}
            \begin{itemize}
                \item Learning rate: Also denoted as ``eta'' is the step size shrinkage used in the update. We used values of 0.01, 0.1, and 0.5.
                \item Max depth: maximum depth of a tree. We used values of 3, 5, and 7.
                \item Number of estimators: Number of boosting rounds or trees to build. We used values of 50, 100, and 200.
                \item Subsample: Subsample ratio of the training instances. For example, a subsample of 0.5 means that we randomly sample half of the training data before growing the trees. We used values of 0.6, 0.8, and 1.0.
                \item Subsample ratio of columns: Subsample ratio of columns when constructing each tree. Subsampling occurs once for every tree constructed. We used values of 0.6, 0.8, and 1.0.
            \end{itemize}

\textbf{}
    \subsection{Performance of XGB and SVM}
        Both SVM and XGB classifiers were trained on the same dataset as the random forest model. Tables \ref{table:edit_distance_sensitivity_results_repo_level_svm} and \ref{table:edit_distance_sensitivity_results_repo_level_xgb} show their complete performance metrics, respectively.
        
        \begin{table} 
            \caption{Results of TraWiC's code inclusion detection with different edit distance thresholds - SVM}
            \centering
            \resizebox{\textwidth}{!}{
                \begin{tabular}{c c r r r r r}
                    \toprule
                    Edit Distance Threshold & Repository Inclusion Criterion & Precision(\%) & Accuracy(\%) & F-score(\%) & Sensitivity(\%) & Specificity (\%)\\
                    \midrule
                            \multirow{3}{*}{\textbf{20}} & Considering only a \textbf{single} positive & 84.80 & 79.41 & 85.53 & 86.27 & 63.00\\
                            & Considering a threshold of \textbf{0.4} & 77.66 & 73.60 & 82.30 & 87.53 & 40.88\\
                            & Considering a threshold of \textbf{0.6} & 80.81 & 70.13 & 77.95 & 75.29 & 58.01\\
                        \midrule
                            \multirow{3}{*}{\textbf{40}} & Considering only a \textbf{single} positive & 74.98 & 73.67 & 83.20 & 93.45 & 27.99\\
                            & Considering a threshold of \textbf{0.4} & 70.66 & 70.30 & 81.89 & 97.37 & 10.11\\
                            & Considering a threshold of \textbf{0.6} & 72.69 & 69.97 & 80.60 & 90.43 & 24.47\\
                        \midrule
                            \multirow{3}{*}{\textbf{60}} & Considering only a \textbf{single} positive & 74.12 & 72.93 & 82.43 & 92.85 & 29.80\\
                            & Considering a threshold of \textbf{0.4} & 71.11 & 70.72 & 82.20 & 97.39 & 10.22\\
                            & Considering a threshold of \textbf{0.6} & 74.29 & 72.53 & 82.37 & 92.42 & 27.42\\
                        \midrule
                            \multirow{3}{*}{\textbf{70}} & Considering only a \textbf{single} positive & 76.60 & 75.00 & 84.49 & 94.19 & 24.95\\
                            & Considering a threshold of \textbf{0.4} & 73.41 & 72.67 & 83.78 & 97.56 & 7.56\\
                            & Considering a threshold of \textbf{0.6} & 74.87 & 72.19 & 82.82 & 92.67 & 18.60\\
                        \midrule
                            \multirow{3}{*}{\textbf{80}} & Considering only a \textbf{single} positive & 72.83 & 72.07 & 82.22 & 94.40 & 23.74\\
                            & Considering a threshold of \textbf{0.4} & 70.79 & 69.97 & 81.91 & 97.17 & 6.59\\
                            & Considering a threshold of \textbf{0.6} & 72.16 & 69.97 & 81.24 & 92.92 & 16.48\\
                    \bottomrule
                \end{tabular}
        }
        \label{table:edit_distance_sensitivity_results_repo_level_svm}
    \end{table}

        \begin{table}
            \caption{Results of TraWiC's code inclusion detection with different edit distance thresholds - XGB}
            \centering
            \resizebox{\textwidth}{!}{
                \begin{tabular}{c c r r r r r}
                        \toprule
                        Edit Distance Threshold & Repository Inclusion Criterion & Precision(\%) & Accuracy(\%) & F-score(\%) & Sensitivity(\%) & Specificity (\%)\\
                        \midrule
                                \multirow{3}{*}{\textbf{20}} & Considering only a \textbf{single} positive & 84.94 & 80.05 & 86.04 & 87.18 & 63.00\\
                                & Considering a threshold of \textbf{0.4} & 77.85 & 75.08 & 83.53 & 90.12 & 39.78\\
                                & Considering a threshold of \textbf{0.6} & 82.31 & 73.27 & 80.53 & 78.82 & 60.22\\
                        \midrule
                            \multirow{3}{*}{\textbf{40}} & Considering only a \textbf{single} positive & 75.23 & 74.73 & 84.01 & 95.12 & 27.64\\
                            & Considering a threshold of \textbf{0.4} & 70.69 & 70.63 & 82.16 & 98.09 & 9.57\\
                            & Considering a threshold of \textbf{0.6} & 73.46 & 72.44 & 82.48 & 94.02 & 24.47\\
                        \midrule    
                            \multirow{3}{*}{\textbf{60}} & Considering only a \textbf{single} positive & 73.29 & 72.61 & 82.49 & 94.32 & 25.59\\
                            & Considering a threshold of \textbf{0.4} & 70.58 & 70.39 & 82.18 & 98.34 & 6.99\\
                            & Considering a threshold of \textbf{0.6} & 72.71 & 71.38 & 82.02 & 94.08 & 19.89\\
                        \midrule                    
                            \multirow{3}{*}{\textbf{70}} & Considering only a \textbf{single} positive & 76.61 & 74.79 & 84.32 & 93.75 & 25.34\\
                            & Considering a threshold of \textbf{0.4} & 73.32 & 72.35 & 83.56 & 97.11 & 7.56\\
                            & Considering a threshold of \textbf{0.6} & 74.59 & 70.90 & 81.85 & 90.67 & 19.19\\
                        \midrule    
                            \multirow{3}{*}{\textbf{80}} & Considering only a \textbf{single} positive & 73.38 & 72.98 & 82.78 & 94.95 & 25.42\\
                            & Considering a threshold of \textbf{0.4} & 71.26 & 70.79 & 82.39 & 97.64 & 8.24\\
                            & Considering a threshold of \textbf{0.6} & 71.87 & 69.80 & 81.23 & 93.40 & 14.84\\
                        \bottomrule
                    \end{tabular}
            }
            \label{table:edit_distance_sensitivity_results_repo_level_xgb}
        \end{table}
        
    \subsection{TraWiC's performance on different noise levels}
        Here, we include the complete results of TraWic's inclusion detection against various levels of noise on script-level granularity and project-level granularity with different inclusion thresholds. Tables \ref{table:noise_sensitivity_results_semantic_total_santa} - \ref{table:noise_sensitivity_results_combined_total_mistral} display our results on noise injected at semantic, syntactic, and both levels, respectively.
        
        \begin{table}
            \caption{Results of TraWiC's code inclusion detection with different edit distance thresholds and different noise ratio - Semantic - SantaCoder} 
            \centering
            \resizebox{0.9\textwidth}{!}{
            \begin{tabular}{c c c r r r r r}
                \toprule
                Edit Distance Threshold & Noise Ratio & Repository Inclusion Criterion & Precision(\%) & Accuracy(\%) & F-score(\%) & Sensitivity(\%) & Specificity (\%)\\
                \midrule
                    \multirow{9}{*}{\textbf{20}} & \multirow{3}{*}{0.1} & Considering only a \textbf{single} positive & 87.19 & 82.29 & 87.62 & 88.04 & 68.08\\
                    & & Considering a threshold of \textbf{0.4} & 79.56 & 76.82 & 86.17 & 93.97 & 20.00\\
                    & & Considering a threshold of \textbf{0.6} & 84.55 & 73.51 & 82.30 & 80.17 & 51.43\\
                    & \multirow{3}{*}{0.5} & Considering only a \textbf{single} positive & 83.83 & 74.89 & 81.97 & 80.19 & 61.81\\
                    & & Considering a threshold of \textbf{0.4} & 79.14 & 76.82 & 86.27 & 94.83 & 17.14\\
                    & & Considering a threshold of \textbf{0.6} & 85.32 & 74.17 & 82.67 & 80.17 & 54.29\\
                    & \multirow{3}{*}{0.9} & Considering only a \textbf{single} positive & 81.70 & 51.86 & 55.22 & 41.70 & 76.94\\
                    & & Considering a threshold of \textbf{0.4} & 80.15 & 75.50 & 85.02 & 90.52 & 25.71\\
                    & & Considering a threshold of \textbf{0.6} & 87.18 & 61.59 & 70.10 & 58.62 & 71.43\\
                \midrule
                    \multirow{9}{*}{\textbf{40}} & \multirow{3}{*}{0.1} & Considering only a \textbf{single} positive & 80.47 & 79.79 & 86.94 & 94.54 & 43.36\\
                    & & Considering a threshold of \textbf{0.4} & 77.70 & 77.48 & 87.12 & 99.14 & 5.71\\
                    & & Considering a threshold of \textbf{0.6} & 81.29 & 80.79 & 88.63 & 97.41 & 25.71\\
                    & \multirow{3}{*}{0.5} & Considering only a \textbf{single} positive & 79.07 & 76.86 & 84.95 & 91.78 & 40.04\\
                    & & Considering a threshold of \textbf{0.4} & 77.18 & 76.82 & 86.79 & 99.14 & 2.86\\
                    & & Considering a threshold of \textbf{0.6} & 81.16 & 80.13 & 88.19 & 96.55 & 25.71\\
                    & \multirow{3}{*}{0.9} & Considering only a \textbf{single} positive & 77.96 & 70.27 & 79.53 & 81.17 & 43.36\\
                    & & Considering a threshold of \textbf{0.4} & 78.23 & 78.15 & 87.45 & 99.14 & 8.57\\
                    & & Considering a threshold of \textbf{0.6} & 81.62 & 80.13 & 88.10 & 95.69 & 28.57\\
                \midrule
                    \multirow{9}{*}{\textbf{80}} & \multirow{3}{*}{0.1} & Considering only a \textbf{single} positive & 79.31 & 78.88 & 86.51 & 95.14 & 38.75\\
                    & & Considering a threshold of \textbf{0.4} & 77.70 & 77.48 & 87.12 & 99.14 & 5.71\\
                    & & Considering a threshold of \textbf{0.6} & 80.71 & 80.13 & 88.28 & 97.41 & 22.86\\
                    & \multirow{3}{*}{0.5} & Considering only a \textbf{single} positive & 78.03 & 76.76 & 85.16 & 93.72 & 34.87\\
                    & & Considering a threshold of \textbf{0.4} & 77.33 & 77.48 & 87.22 & 100.00 & 2.86\\
                    & & Considering a threshold of \textbf{0.6} & 80.71 & 80.13 & 88.28 & 97.41 & 22.86\\
                    & \multirow{3}{*}{0.9} & Considering only a \textbf{single} positive & 78.13 & 75.21 & 83.86 & 90.51 & 37.45\\
                    & & Considering a threshold of \textbf{0.4} & 77.85 & 78.15 & 87.55 & 100.00 & 5.71\\
                    & & Considering a threshold of \textbf{0.6} & 81.43 & 81.46 & 89.06 & 98.28 & 25.71\\
                \bottomrule
            \end{tabular}
            }
            \label{table:noise_sensitivity_results_semantic_total_santa}
        \end{table}
    
        \begin{table}
            \caption{Results of TraWiC's code inclusion detection with different edit distance thresholds and different noise ratio - Syntactic - SantaCoder}
            \centering
            \resizebox{0.9\textwidth}{!}{
                \begin{tabular}{c c c r r r r r}
                    \toprule
                    Edit Distance Threshold & Noise Ratio & Repository Inclusion Criterion & Precision(\%) & Accuracy(\%) & F-score(\%) & Sensitivity(\%) & Specificity (\%)\\
                    \midrule
                        \multirow{9}{*}{\textbf{20}} & \multirow{3}{*}{0.1} & Considering only a \textbf{single} positive & 88.00 & 83.09 & 88.13 & 88.27 & 70.30\\
                        & & Considering a threshold of \textbf{0.4} & 79.56 & 76.82 & 86.17 & 93.97 & 20.00\\
                        & & Considering a threshold of \textbf{0.6} & 80.73 & 67.55 & 78.22 & 75.86 & 40.00\\
                        & \multirow{3}{*}{0.5} & Considering only a \textbf{single} positive & 86.41 & 79.26 & 85.23 & 84.08 & 67.34\\
                        & & Considering a threshold of \textbf{0.4} & 78.68 & 74.83 & 84.92 & 92.24 & 17.14\\
                        & & Considering a threshold of \textbf{0.6} & 81.13 & 66.89 & 77.48 & 74.14 & 42.86\\
                        & \multirow{3}{*}{0.9} & Considering only a \textbf{single} positive & 87.80 & 69.04 & 75.11 & 65.62 & 77.49\\
                        & & Considering a threshold of \textbf{0.4} & 78.52 & 74.17 & 84.46 & 91.38 & 17.14\\
                        & & Considering a threshold of \textbf{0.6} & 83.51 & 66.23 & 76.06 & 69.83 & 54.29\\
                    \midrule
                        \multirow{9}{*}{\textbf{40}} & \multirow{3}{*}{0.1} & Considering only a \textbf{single} positive & 80.34 & 79.41 & 86.68 & 94.10 & 43.17\\
                        & & Considering a threshold of \textbf{0.4} & 78.23 & 78.15 & 87.45 & 99.14 & 8.57\\
                        & & Considering a threshold of \textbf{0.6} & 81.16 & 80.13 & 88.19 & 96.55 & 25.71\\
                        & \multirow{3}{*}{0.5} & Considering only a \textbf{single} positive & 79.52 & 75.00 & 83.26 & 87.37 & 44.46\\
                        & & Considering a threshold of \textbf{0.4} & 77.70 & 77.48 & 87.12 & 99.14 & 5.71\\
                        & & Considering a threshold of \textbf{0.6} & 81.48 & 79.47 & 87.65 & 94.83 & 28.57\\
                        & \multirow{3}{*}{0.9} & Considering only a \textbf{single} positive & 83.74 & 64.15 & 70.97 & 61.58 & 70.48\\
                        & & Considering a threshold of \textbf{0.4} & 77.93 & 76.82 & 86.59 & 97.41 & 8.57\\
                        & & Considering a threshold of \textbf{0.6} & 83.61 & 77.48 & 85.71 & 87.93 & 42.86\\
                    \midrule
                        \multirow{9}{*}{\textbf{80}} & \multirow{3}{*}{0.1} & Considering only a \textbf{single} positive & 78.92 & 78.03 & 85.94 & 94.32 & 37.82\\
                        & & Considering a threshold of \textbf{0.4} & 77.85 & 78.15 & 87.55 & 100.00 & 5.71\\
                        & & Considering a threshold of \textbf{0.6} & 80.71 & 80.13 & 88.28 & 97.41 & 22.86\\
                        & \multirow{3}{*}{0.5} & Considering only a \textbf{single} positive & 78.43 & 74.95 & 83.55 & 89.39 & 39.30\\
                        & & Considering a threshold of \textbf{0.4} & 77.33 & 77.48 & 87.22 & 100.00 & 2.86\\
                        & & Considering a threshold of \textbf{0.6} & 82.01 & 82.12 & 89.41 & 98.28 & 28.57\\
                        & \multirow{3}{*}{0.9} & Considering only a \textbf{single} positive & 79.11 & 65.59 & 74.38 & 70.18 & 54.24\\
                        & & Considering a threshold of \textbf{0.4} & 78.38 & 78.81 & 87.88 & 100.00 & 8.57\\
                        & & Considering a threshold of \textbf{0.6} & 82.71 & 80.79 & 88.35 & 94.83 & 34.29\\
                    \bottomrule
                \end{tabular}
                }
            \label{table:noise_sensitivity_results_syntactic_total_santa}
        \end{table}
    
        \begin{table}
            \caption{Results of TraWiC's code inclusion detection with different edit distance thresholds and different noise ratio - Combined - SantaCoder}
            \centering
            \resizebox{0.9\textwidth}{!}{
                \begin{tabular}{c c c r r r r r}
                    \toprule
                    Edit Distance Threshold & Noise Ratio & Repository Inclusion Criterion & Precision(\%) & Accuracy(\%) & F-score(\%) & Sensitivity(\%) & Specificity (\%)\\
                    \midrule
                        \multirow{9}{*}{\textbf{20}} & \multirow{3}{*}{0.1} & Considering only a \textbf{single} positive & 87.16 & 82.07 & 87.45 & 87.74 & 68.08\\
                        & & Considering a threshold of \textbf{0.4} & 79.71 & 77.48 & 86.61 & 94.83 & 20.00\\
                        & & Considering a threshold of \textbf{0.6} & 83.93 & 73.51 & 82.46 & 81.03 & 48.57\\
                        & \multirow{3}{*}{0.5} & Considering only a \textbf{single} positive & 83.37 & 73.03 & 80.37 & 77.58 & 61.81\\
                        & & Considering a threshold of \textbf{0.4} & 79.71 & 77.48 & 86.61 & 94.83 & 20.00\\
                        & & Considering a threshold of \textbf{0.6} & 85.45 & 74.83 & 83.19 & 81.03 & 54.29\\
                        & \multirow{3}{*}{0.9} & Considering only a \textbf{single} positive & 85.25 & 50.96 & 52.18 & 37.59 & 83.95\\
                        & & Considering a threshold of \textbf{0.4} & 82.11 & 75.50 & 84.52 & 87.07 & 37.14\\
                        & & Considering a threshold of \textbf{0.6} & 87.84 & 60.26 & 68.42 & 56.03 & 74.29\\
                    \midrule
                        \multirow{9}{*}{\textbf{40}} & \multirow{3}{*}{0.1} & Considering only a \textbf{single} positive & 79.51 & 78.30 & 86.00 & 93.65 & 40.41\\
                        & & Considering a threshold of \textbf{0.4} & 78.23 & 78.15 & 87.45 & 99.14 & 8.57\\
                        & & Considering a threshold of \textbf{0.6} & 80.71 & 80.13 & 88.28 & 97.41 & 22.86\\
                        & \multirow{3}{*}{0.5} & Considering only a \textbf{single} positive & 78.94 & 73.99 & 82.57 & 86.55 & 42.99\\
                        & & Considering a threshold of \textbf{0.4} & 77.70 & 77.48 & 87.12 & 99.14 & 5.71\\
                        & & Considering a threshold of \textbf{0.6} & 81.16 & 80.13 & 88.19 & 96.55 & 25.71\\
                        & \multirow{3}{*}{0.9} & Considering only a \textbf{single} positive & 84.54 & 55.21 & 59.05 & 45.37 & 79.52\\
                        & & Considering a threshold of \textbf{0.4} & 79.43 & 78.15 & 87.16 & 96.55 & 17.14\\
                        & & Considering a threshold of \textbf{0.6} & 80.95 & 66.23 & 76.92 & 73.28 & 42.86\\
                    \midrule
                        \multirow{9}{*}{\textbf{80}} & \multirow{3}{*}{0.1} & Considering only a \textbf{single} positive & 78.34 & 77.39 & 85.59 & 94.32 & 35.61\\
                        & & Considering a threshold of \textbf{0.4} & 77.70 & 77.48 & 87.12 & 99.14 & 5.71\\
                        & & Considering a threshold of \textbf{0.6} & 82.01 & 82.12 & 89.41 & 98.28 & 28.57\\
                        & \multirow{3}{*}{0.5} & Considering only a \textbf{single} positive & 78.12 & 74.36 & 83.15 & 88.86 & 38.56\\
                        & & Considering a threshold of \textbf{0.4} & 77.70 & 77.48 & 87.12 & 99.14 & 5.71\\
                        & & Considering a threshold of \textbf{0.6} & 81.29 & 80.79 & 88.63 & 97.41 & 25.71\\
                        & \multirow{3}{*}{0.9} & Considering only a \textbf{single} positive & 82.10 & 57.45 & 63.24 & 51.42 & 72.32\\
                        & & Considering a threshold of \textbf{0.4} & 79.17 & 78.81 & 87.69 & 98.28 & 14.29\\
                        & & Considering a threshold of \textbf{0.6} & 82.20 & 73.51 & 82.91 & 83.62 & 40.00\\
                    \bottomrule
            \end{tabular}
            }
            \label{table:noise_sensitivity_results_combined_total_santa}
        \end{table}

        \begin{table}
            \caption{Results of TraWiC's code inclusion detection with different edit distance thresholds and different noise ratio - Syntactic - Llama-2}
            \centering
            \resizebox{0.9\textwidth}{!}{
                \begin{tabular}{c c c r r r r r}
                    \toprule
                    Edit Distance Threshold & Noise Ratio & Repository Inclusion Criterion & Precision(\%) & Accuracy(\%) & F-score(\%) & Sensitivity(\%) & Specificity (\%)\\
                    \midrule
                        \multirow{9}{*}{\textbf{20}} & \multirow{3}{*}{0.1} & Considering only a \textbf{single} positive & 76.68 & 82.57 & 81.36 & 84.23 & 81.54\\
                        & & Considering a threshold of \textbf{0.4} & 80.00 & 82.93 & 82.05 & 84.21 & 81.82\\
                        & & Considering a threshold of \textbf{0.6} & 70.00 & 66.67 & 62.22 & 56.00 & 76.92\\
                        & \multirow{3}{*}{0.5} & Considering only a \textbf{single} positive & 68.60 & 71.80 & 69.96 & 71.37 & 72.16\\
                        & & Considering a threshold of \textbf{0.4} & 70.00 & 73.17 & 71.79 & 73.68 & 72.73\\
                        & & Considering a threshold of \textbf{0.6} & 70.00 & 64.71 & 60.87 & 53.85 & 76.00\\
                        & \multirow{3}{*}{0.9} & Considering only a \textbf{single} positive & 57.75 & 61.41 & 58.89 & 60.08 & 62.54\\
                        & & Considering a threshold of \textbf{0.4} & 60.00 & 60.98 & 60.00 & 60.00 & 61.90\\
                        & & Considering a threshold of \textbf{0.6} & 60.00 & 56.86 & 52.17 & 46.15 & 68.00\\
                    \midrule
                        \multirow{9}{*}{\textbf{40}} & \multirow{3}{*}{0.1} & Considering only a \textbf{single} positive & 73.23 & 77.01 & 75.15 & 77.18 & 76.87\\
                        & & Considering a threshold of \textbf{0.4} & 75.00 & 75.61 & 75.00 & 75.00 & 76.19\\
                        & & Considering a threshold of \textbf{0.6} & 61.90 & 65.00 & 65.00 & 68.42 & 61.90\\
                        & \multirow{3}{*}{0.5} & Considering only a \textbf{single} positive & 57.09 & 59.44 & 57.20 & 57.31 & 61.37\\
                        & & Considering a threshold of \textbf{0.4} & 75.00 & 75.61 & 75.00 & 75.00 & 76.19\\
                        & & Considering a threshold of \textbf{0.6} & 57.14 & 55.00 & 57.14 & 57.14 & 52.63\\
                        & \multirow{3}{*}{0.9} & Considering only a \textbf{single} positive & 43.31 & 45.05 & 42.80 & 42.31 & 47.64\\
                        & & Considering a threshold of \textbf{0.4} & 70.00 & 60.98 & 63.64 & 58.33 & 64.71\\
                        & & Considering a threshold of \textbf{0.6} & 47.62 & 57.50 & 54.05 & 62.50 & 54.17\\
                    \midrule
                        \multirow{9}{*}{\textbf{80}} & \multirow{3}{*}{0.1} & Considering only a \textbf{single} positive & 69.53 & 74.02 & 70.36 & 74.66 & 73.57\\
                        & & Considering a threshold of \textbf{0.4} & 75.00 & 65.85 & 68.18 & 62.50 & 70.59\\
                        & & Considering a threshold of \textbf{0.6} & 65.00 & 60.98 & 61.90 & 59.09 & 63.16\\
                        & \multirow{3}{*}{0.5} & Considering only a \textbf{single} positive & 60.48 & 66.54 & 62.63 & 64.96 & 67.76\\
                        & & Considering a threshold of \textbf{0.4} & 60.00 & 56.10 & 57.14 & 54.55 & 57.89\\
                        & & Considering a threshold of \textbf{0.6} & 65.00 & 60.98 & 61.90 & 59.09 & 63.16\\
                        & \multirow{3}{*}{0.9} & Considering only a \textbf{single} positive & 51.61 & 55.14 & 51.61 & 51.61 & 58.19\\
                        & & Considering a threshold of \textbf{0.4} & 60.00 & 53.66 & 55.81 & 52.17 & 55.56\\
                        & & Considering a threshold of \textbf{0.6} & 60.00 & 53.66 & 55.81 & 52.17 & 55.56\\
                    \bottomrule
            \end{tabular}
            }
            \label{table:noise_sensitivity_results_syntactic_total_llama}
        \end{table}

        \begin{table}
            \caption{Results of TraWiC's code inclusion detection with different edit distance thresholds and different noise ratio - Semantic - Llama2}
            \centering
            \resizebox{0.9\textwidth}{!}{
                \begin{tabular}{c c c r r r r r}
                    \toprule
                    Edit Distance Threshold & Noise Ratio & Repository Inclusion Criterion & Precision(\%) & Accuracy(\%) & F-score(\%) & Sensitivity(\%) & Specificity (\%)\\
                    \midrule
                        \multirow{9}{*}{\textbf{20}} & \multirow{3}{*}{0.1} & Considering only a \textbf{single} positive & 70.93 & 74.40 & 72.62 & 74.39 & 74.40\\
                        & & Considering a threshold of \textbf{0.4} & 75.00 & 80.49 & 78.95 & 83.33 & 78.26\\
                        & & Considering a threshold of \textbf{0.6} & 60.00 & 64.71 & 57.14 & 54.55 & 72.41\\
                        & \multirow{3}{*}{0.5} & Considering only a \textbf{single} positive & 63.57 & 67.16 & 64.95 & 66.40 & 67.81\\
                        & & Considering a threshold of \textbf{0.4} & 70.00 & 73.17 & 71.79 & 73.68 & 72.73\\
                        & & Considering a threshold of \textbf{0.6} & 60.00 & 60.78 & 54.55 & 50.00 & 70.37\\
                        & \multirow{3}{*}{0.9} & Considering only a \textbf{single} positive & 53.10 & 56.40 & 53.83 & 54.58 & 57.99\\
                        & & Considering a threshold of \textbf{0.4} & 70.00 & 60.98 & 63.64 & 58.33 & 64.71\\
                        & & Considering a threshold of \textbf{0.6} & 50.00 & 47.06 & 42.55 & 37.04 & 58.33\\
                    \midrule
                        \multirow{9}{*}{\textbf{40}} & \multirow{3}{*}{0.1} & Considering only a \textbf{single} positive & 69.29 & 72.15 & 70.26 & 71.26 & 72.92\\
                        & & Considering a threshold of \textbf{0.4} & 70.00 & 68.29 & 68.29 & 66.67 & 70.00\\
                        & & Considering a threshold of \textbf{0.6} & 61.90 & 67.50 & 66.67 & 72.22 & 63.64\\
                        & \multirow{3}{*}{0.5} & Considering only a \textbf{single} positive & 78.94 & 73.99 & 82.57 & 86.55 & 42.99\\
                        & & Considering a threshold of \textbf{0.4} & 75.98 & 80.93 & 79.10 & 82.48 & 79.73\\
                        & & Considering a threshold of \textbf{0.6} & 52.38 & 60.00 & 57.89 & 64.71 & 56.52\\
                        & \multirow{3}{*}{0.9} & Considering only a \textbf{single} positive & 52.76 & 56.07 & 53.28 & 53.82 & 58.04\\
                        & & Considering a threshold of \textbf{0.4} & 65.00 & 60.98 & 61.90 & 59.09 & 63.16\\
                        & & Considering a threshold of \textbf{0.6} & 42.86 & 52.50 & 48.65 & 56.25 & 50.00\\
                    \midrule
                        \multirow{9}{*}{\textbf{80}} & \multirow{3}{*}{0.1} & Considering only a \textbf{single} positive & 66.13 & 73.46 & 69.79 & 73.87 & 73.16\\
                        & & Considering a threshold of \textbf{0.4} & 70.00 & 63.41 & 65.12 & 60.87 & 66.67\\
                        & & Considering a threshold of \textbf{0.6} & 65.00 & 58.54 & 60.47 & 56.52 & 61.11\\
                        & \multirow{3}{*}{0.5} & Considering only a \textbf{single} positive & 56.45 & 62.43 & 58.21 & 60.09 & 64.24\\
                        & & Considering a threshold of \textbf{0.4} & 65.00 & 56.10 & 59.09 & 54.17 & 58.82\\
                        & & Considering a threshold of \textbf{0.6} & 55.00 & 48.78 & 51.16 & 47.83 & 50.00\\
                        & \multirow{3}{*}{0.9} & Considering only a \textbf{single} positive & 49.60 & 54.39 & 50.20 & 50.83 & 57.43\\
                        & & Considering a threshold of \textbf{0.4} & 50.00 & 48.78 & 48.78 & 47.62 & 50.00\\
                        & & Considering a threshold of \textbf{0.6} & 50.00 & 43.90 & 46.51 & 43.48 & 44.44\\
                    \bottomrule
            \end{tabular}
            }
            \label{table:noise_sensitivity_results_semantic_total_lamma}
        \end{table}

        \begin{table}
            \caption{Results of TraWiC's code inclusion detection with different edit distance thresholds and different noise ratio - Combined - Llama2}
            \centering
            \resizebox{0.9\textwidth}{!}{
                \begin{tabular}{c c c r r r r r}
                    \toprule
                    Edit Distance Threshold & Noise Ratio & Repository Inclusion Criterion & Precision(\%) & Accuracy(\%) & F-score(\%) & Sensitivity(\%) & Specificity (\%)\\
                    \midrule
                        \multirow{9}{*}{\textbf{20}} & \multirow{3}{*}{0.1} & Considering only a \textbf{single} positive & 68.60 & 71.43 & 69.69 & 70.80 & 71.97\\
                        & & Considering a threshold of \textbf{0.4} & 75.00 & 75.61 & 75.00 & 75.00 & 76.19\\
                        & & Considering a threshold of \textbf{0.6} & 65.00 & 62.75 & 57.78 & 52.00 & 73.08\\
                        & \multirow{3}{*}{0.5} & Considering only a \textbf{single} positive & 59.30 & 63.27 & 60.71 & 62.20 & 64.16\\
                        & & Considering a threshold of \textbf{0.4} & 65.00 & 68.29 & 66.67 & 68.42 & 68.18\\
                        & & Considering a threshold of \textbf{0.6} & 55.00 & 56.86 & 50.00 & 45.83 & 66.67\\
                        & \multirow{3}{*}{0.9} & Considering only a \textbf{single} positive & 53.10 & 55.66 & 53.41 & 53.73 & 57.39\\
                        & & Considering a threshold of \textbf{0.4} & 55.00 & 46.34 & 50.00 & 45.83 & 47.06\\
                        & & Considering a threshold of \textbf{0.6} & 45.00 & 47.06 & 40.00 & 36.00 & 57.69\\
                    \midrule
                        \multirow{9}{*}{\textbf{40}} & \multirow{3}{*}{0.1} & Considering only a \textbf{single} positive & 60.24 & 63.74 & 61.20 & 62.20 & 65.05\\
                        & & Considering a threshold of \textbf{0.4} & 65.00 & 68.29 & 66.67 & 68.42 & 68.18\\
                        & & Considering a threshold of \textbf{0.6} & 61.90 & 60.00 & 61.90 & 61.90 & 57.89\\
                        & \multirow{3}{*}{0.5} & Considering only a \textbf{single} positive & 57.09 & 61.50 & 58.47 & 59.92 & 62.80\\
                        & & Considering a threshold of \textbf{0.4} & 50.00 & 53.66 & 51.28 & 52.63 & 54.55\\
                        & & Considering a threshold of \textbf{0.6} & 52.38 & 55.00 & 55.00 & 57.89 & 52.38\\
                        & \multirow{3}{*}{0.9} & Considering only a \textbf{single} positive & 54.33 & 58.69 & 55.53 & 56.79 & 60.27\\
                        & & Considering a threshold of \textbf{0.4} & 60.00 & 48.78 & 53.33 & 48.00 & 50.00\\
                        & & Considering a threshold of \textbf{0.6} & 45.00 & 47.06 & 40.00 & 36.00 & 57.69\\
                    \midrule
                        \multirow{9}{*}{\textbf{80}} & \multirow{3}{*}{0.1} & Considering only a \textbf{single} positive & 62.50 & 68.60 & 64.85 & 67.39 & 69.51\\
                        & & Considering a threshold of \textbf{0.4} & 65.00 & 63.41 & 63.41 & 61.90 & 65.00\\
                        & & Considering a threshold of \textbf{0.6} & 65.00 & 56.10 & 59.09 & 54.17 & 58.82\\
                        & \multirow{3}{*}{0.5} & Considering only a \textbf{single} positive & 55.65 & 60.19 & 56.44 & 57.26 & 62.59\\
                        & & Considering a threshold of \textbf{0.4} & 55.00 & 56.10 & 55.00 & 55.00 & 57.14\\
                        & & Considering a threshold of \textbf{0.6} & 50.00 & 48.78 & 48.78 & 47.62 & 50.00\\
                        & \multirow{3}{*}{0.9} & Considering only a \textbf{single} positive & 47.18 & 54.02 & 48.75 & 50.43 & 56.77\\
                        & & Considering a threshold of \textbf{0.4} & 45.00 & 43.90 & 43.90 & 42.86 & 45.00\\
                        & & Considering a threshold of \textbf{0.6} & 45.00 & 48.78 & 46.15 & 47.37 & 50.00\\
                    \bottomrule
            \end{tabular}
            }
            \label{table:noise_sensitivity_results_combined_total_llama}
        \end{table}

        \begin{table}
            \caption{Results of TraWiC's code inclusion detection with different edit distance thresholds and different noise ratio - Syntactic - Mistral}
            \centering
            \resizebox{0.9\textwidth}{!}{
                \begin{tabular}{c c c r r r r r}
                    \toprule
                    Edit Distance Threshold & Noise Ratio & Repository Inclusion Criterion & Precision(\%) & Accuracy(\%) & F-score(\%) & Sensitivity(\%) & Specificity (\%)\\
                    \midrule
                        \multirow{9}{*}{\textbf{20}} & \multirow{3}{*}{0.1} & Considering only a \textbf{single} positive & 82.56 & 86.48 & 85.37 & 88.38 & 84.95\\
                        & & Considering a threshold of \textbf{0.4} & 85.00 & 82.93 & 82.93 & 80.95 & 85.00\\
                        & & Considering a threshold of \textbf{0.6} & 84.21 & 82.50 & 82.50 & 80.00 & 85.00\\
                        & \multirow{3}{*}{0.5} & Considering only a \textbf{single} positive & 71.32 & 73.15 & 71.73 & 72.16 & 74.04\\
                        & & Considering a threshold of \textbf{0.4} & 80.00 & 78.05 & 78.05 & 76.19 & 80.00\\
                        & & Considering a threshold of \textbf{0.6} & 73.68 & 72.50 & 71.79 & 70.00 & 75.00\\
                        & \multirow{3}{*}{0.9} & Considering only a \textbf{single} positive & 66.15 & 66.85 & 65.24 & 65.37 & 68.20\\
                        & & Considering a threshold of \textbf{0.4} & 75.00 & 65.85 & 68.18 & 62.50 & 70.59\\
                        & & Considering a threshold of \textbf{0.6} & 73.68 & 70.00 & 70.00 & 66.67 & 73.68\\
                    \midrule
                        \multirow{9}{*}{\textbf{40}} & \multirow{3}{*}{0.1} & Considering only a \textbf{single} positive & 78.29 & 80.71 & 79.53 & 80.80 & 80.62\\
                        & & Considering a threshold of \textbf{0.4} & 75.00 & 78.05 & 76.92 & 78.95 & 77.27\\
                        & & Considering a threshold of \textbf{0.6} & 73.68 & 71.79 & 71.79 & 70.00 & 73.68\\
                        & \multirow{3}{*}{0.5} & Considering only a \textbf{single} positive & 70.93 & 74.77 & 72.91 & 75.00 & 74.58\\
                        & & Considering a threshold of \textbf{0.4} & 60.00 & 63.41 & 61.54 & 63.16 & 63.64\\
                        & & Considering a threshold of \textbf{0.6} & 68.42 & 69.23 & 68.42 & 68.42 & 70.00\\
                        & \multirow{3}{*}{0.9} & Considering only a \textbf{single} positive & 63.57 & 65.86 & 64.06 & 64.57 & 67.02\\
                        & & Considering a threshold of \textbf{0.4} & 55.00 & 60.98 & 57.89 & 61.11 & 60.87\\
                        & & Considering a threshold of \textbf{0.6} & 52.63 & 56.41 & 54.05 & 55.56 & 57.14\\
                    \midrule
                        \multirow{9}{*}{\textbf{80}} & \multirow{3}{*}{0.1} & Considering only a \textbf{single} positive & 73.26 & 78.66 & 76.67 & 80.43 & 77.30\\
                        & & Considering a threshold of \textbf{0.4} & 65.00 & 70.73 & 68.42 & 72.22 & 69.57\\
                        & & Considering a threshold of \textbf{0.6} & 60.00 & 56.10 & 57.14 & 54.55 & 57.89\\
                        & \multirow{3}{*}{0.5} & Considering only a \textbf{single} positive & 64.34 & 71.24 & 68.17 & 72.49 & 70.32\\
                        & & Considering a threshold of \textbf{0.4} & 55.00 & 58.54 & 56.41 & 57.89 & 59.09\\
                        & & Considering a threshold of \textbf{0.6} & 55.00 & 51.22 & 52.38 & 50.00 & 52.63\\
                        & \multirow{3}{*}{0.9} & Considering only a \textbf{single} positive & 55.42 & 60.75 & 57.02 & 58.72 & 62.37\\
                        & & Considering a threshold of \textbf{0.4} & 50.00 & 48.78 & 48.78 & 47.62 & 50.00\\
                        & & Considering a threshold of \textbf{0.6} & 50.00 & 51.22 & 50.00 & 50.00 & 52.38\\
                    \bottomrule
            \end{tabular}
            }
            \label{table:noise_sensitivity_results_syntactic_total_mistral}
        \end{table}

        \begin{table}
            \caption{Results of TraWiC's code inclusion detection with different edit distance thresholds and different noise ratio - Semantic - Mistral}
            \centering
            \resizebox{0.9\textwidth}{!}{
                \begin{tabular}{c c c r r r r r}
                    \toprule
                    Edit Distance Threshold & Noise Ratio & Repository Inclusion Criterion & Precision(\%) & Accuracy(\%) & F-score(\%) & Sensitivity(\%) & Specificity (\%)\\
                    \midrule
                        \multirow{9}{*}{\textbf{20}} & \multirow{3}{*}{0.1} & Considering only a \textbf{single} positive & 72.48 & 76.44 & 74.65 & 76.95 & 76.0`\\
                        & & Considering a threshold of \textbf{0.4} & 75.00 & 78.05 & 76.92 & 78.95 & 77.27\\
                        & & Considering a threshold of \textbf{0.6} & 68.42 & 70.00 & 68.42 & 68.42 & 71.43\\
                        & \multirow{3}{*}{0.5} & Considering only a \textbf{single} positive & 67.05 & 69.02 & 67.45 & 67.84 & 70.07\\
                        & & Considering a threshold of \textbf{0.4} & 75.00 & 70.73 & 71.43 & 68.18 & 73.68\\
                        & & Considering a threshold of \textbf{0.6} & 68.42 & 65.00 & 65.00 & 61.90 & 68.42\\
                        & \multirow{3}{*}{0.9} & Considering only a \textbf{single} positive & 56.98 & 60.11 & 57.76 & 58.57 & 62.37\\
                        & & Considering a threshold of \textbf{0.4} & 60.00 & 56.10 & 57.14 & 54.55 & 57.89\\
                        & & Considering a threshold of \textbf{0.6} & 73.68 & 60.00 & 63.64 & 56.00 & 66.67\\
                    \midrule
                        \multirow{9}{*}{\textbf{40}} & \multirow{3}{*}{0.1} & Considering only a \textbf{single} positive & 80.25 & 79.77 & 78.44 & 76.71 & 82.59\\
                        & & Considering a threshold of \textbf{0.4} & 85.00 & 82.93 & 82.93 & 80.95 & 85.00\\
                        & & Considering a threshold of \textbf{0.6} & 68.42 & 71.79 & 70.27 & 72.22 & 71.43\\
                        & \multirow{3}{*}{0.5} & Considering only a \textbf{single} positive & 68.07 & 69.17 & 66.94 & 65.85 & 72.16\\
                        & & Considering a threshold of \textbf{0.4} & 70.00 & 75.61 & 73.68 & 77.78 & 73.91\\
                        & & Considering a threshold of \textbf{0.6} & 73.68 & 69.23 & 70.00 & 66.67 & 72.22\\
                        & \multirow{3}{*}{0.9} & Considering only a \textbf{single} positive & 58.82 & 60.50 & 57.73 & 56.68 & 63.97\\
                        & & Considering a threshold of \textbf{0.4} & 50.00 & 63.41 & 57.14 & 66.67 & 61.54\\
                        & & Considering a threshold of \textbf{0.6} & 63.16 & 51.28 & 55.81 & 50.00 & 53.33\\
                    \midrule
                        \multirow{9}{*}{\textbf{80}} & \multirow{3}{*}{0.1} & Considering only a \textbf{single} positive & 67.44 & 74.03 & 71.31 & 75.65 & 72.82\\
                        & & Considering a threshold of \textbf{0.4} & 65.00 & 68.29 & 66.67 & 68.42 & 68.18\\
                        & & Considering a threshold of \textbf{0.6} & 60.00 & 58.54 & 58.54 & 57.14 & 60.00\\
                        & \multirow{3}{*}{0.5} & Considering only a \textbf{single} positive & 59.69 & 67.16 & 63.51 & 67.84 & 66.67\\
                        & & Considering a threshold of \textbf{0.4} & 50.00 & 56.10 & 52.63 & 55.56 & 56.52\\
                        & & Considering a threshold of \textbf{0.6} & 45.00 & 48.78 & 46.15 & 47.37 & 50.00\\
                        & \multirow{3}{*}{0.9} & Considering only a \textbf{single} positive & 53.49 & 59.55 & 55.87 & 58.47 & 60.40\\
                        & & Considering a threshold of \textbf{0.4} & 50.00 & 60.98 & 55.56 & 62.50 & 60.00\\
                        & & Considering a threshold of \textbf{0.6} & 60.00 & 51.22 & 54.55 & 50.00 & 52.94\\
                    \bottomrule
            \end{tabular}
            }
            \label{table:noise_sensitivity_results_semantic_total_mistral}
        \end{table}

        \begin{table}
            \caption{Results of TraWiC's code inclusion detection with different edit distance thresholds and different noise ratio - Combined - Mistral}
            \centering
            \resizebox{0.9\textwidth}{!}{
                \begin{tabular}{c c c r r r r r}
                    \toprule
                    Edit Distance Threshold & Noise Ratio & Repository Inclusion Criterion & Precision(\%) & Accuracy(\%) & F-score(\%) & Sensitivity(\%) & Specificity (\%)\\
                    \midrule
                        \multirow{9}{*}{\textbf{20}} & \multirow{3}{*}{0.1} & Considering only a \textbf{single} positive & 72.48 & 75.88 & 74.21 & 76.02 & 75.77\\
                        & & Considering a threshold of \textbf{0.4} & 75.00 & 75.61 & 75.00 & 75.00 & 76.19\\
                        & & Considering a threshold of \textbf{0.6} & 84.21 & 80.00 & 80.00 & 76.19 & 84.21\\
                        & \multirow{3}{*}{0.5} & Considering only a \textbf{single} positive & 65.89 & 69.02 & 67.06 & 68.27 & 69.66\\
                        & & Considering a threshold of \textbf{0.4} & 70.00 & 73.17 & 71.79 & 73.68 & 72.73\\
                        & & Considering a threshold of \textbf{0.6} & 63.16 & 60.00 & 60.00 & 57.14 & 63.16\\
                        & \multirow{3}{*}{0.9} & Considering only a \textbf{single} positive & 58.91 & 61.78 & 59.61 & 69.32 & 63.07\\
                        & & Considering a threshold of \textbf{0.4} & 55.00 & 53.66 & 53.66 & 52.38 & 55.00\\
                        & & Considering a threshold of \textbf{0.6} & 57.89 & 55.00 & 55.00 & 52.38 & 57.89\\
                    \midrule
                        \multirow{9}{*}{\textbf{40}} & \multirow{3}{*}{0.1} & Considering only a \textbf{single} positive & 69.77 & 73.84 & 71.86 & 74.07 & 73.65\\
                        & & Considering a threshold of \textbf{0.4} & 75.00 & 78.05 & 76.92 & 78.95 & 77.27\\
                        & & Considering a threshold of \textbf{0.6} & 78.95 & 74.36 & 75.00 & 71.43 & 77.78\\
                        & \multirow{3}{*}{0.5} & Considering only a \textbf{single} positive & 63.18 & 66.05 & 64.05 & 64.94 & 67.01\\
                        & & Considering a threshold of \textbf{0.4} & 70.00 & 70.73 & 70.00 & 70.00 & 71.43\\
                        & & Considering a threshold of \textbf{0.6} & 73.68 & 66.67 & 68.29 & 63.64 & 70.59\\
                        & \multirow{3}{*}{0.9} & Considering only a \textbf{single} positive & 51.55 & 58.63 & 54.40 & 57.58 & 59.42\\
                        & & Considering a threshold of \textbf{0.4} & 55.00 & 58.54 & 56.41 & 57.89 & 59.09\\
                        & & Considering a threshold of \textbf{0.6} & 73.68 & 66.67 & 68.29 & 63.64 & 70.59\\
                    \midrule
                        \multirow{9}{*}{\textbf{80}} & \multirow{3}{*}{0.1} & Considering only a \textbf{single} positive & 67.05 & 72.73 & 70.18 & 73.62 & 72.04\\
                        & & Considering a threshold of \textbf{0.4} & 65.00 & 65.85 & 65.00 & 65.00 & 66.67\\
                        & & Considering a threshold of \textbf{0.6} & 60.00 & 58.54 & 58.54 & 57.14 & 60.00\\
                        & \multirow{3}{*}{0.5} & Considering only a \textbf{single} positive & 61.24 & 64.75 & 62.45 & 63.71 & 65.64\\
                        & & Considering a threshold of \textbf{0.4} & 50.00 & 56.10 & 52.63 & 55.56 & 56.52\\
                        & & Considering a threshold of \textbf{0.6} & 55.00 & 53.66 & 53.66 & 52.38 & 55.00\\
                        & \multirow{3}{*}{0.9} & Considering only a \textbf{single} positive & 52.71 & 57.88 & 54.51 & 56.43 & 59.06\\
                        & & Considering a threshold of \textbf{0.4} & 40.00 & 39.02 & 39.02 & 38.10 & 40.00\\
                        & & Considering a threshold of \textbf{0.6} & 40.00 & 41.46 & 40.00 & 40.00 & 42.86\\
                    \bottomrule
            \end{tabular}
            }
            \label{table:noise_sensitivity_results_combined_total_mistral}
        \end{table}
        
\subsection{Details of Fine-tuning}

\begin{itemize}
    \item LoRA attention dimension: 64
    \item LoRA dropout: 0.1
    \item Max gradient norm: 0.3
    \item Learning rate: 0.002
    \item Weight decay: 0.001
    \item Warmup ratio: 0.03
\end{itemize}

\begin{figure}
    \centering
    \includegraphics[width=0.5\linewidth]{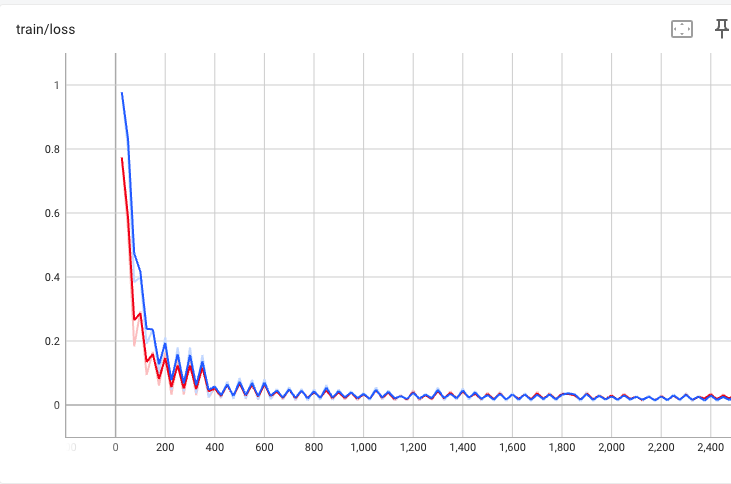}
    \caption{Training loss of the fine-tuning process}
    \label{fig:train_loss}
\end{figure}

Figure \ref{fig:train_loss} displays the training loss of both models during fine-tuning. The blue line indicates the loss of Llama-2 and the red line indicates Mistral's loss during the fine-tuning process.

\end{document}